%% file: SED.tex
\documentclass[fleqn,usenatbib]{mnras}

\usepackage{booktabs,caption}
\usepackage[flushleft]{threeparttable}

\usepackage[T1]{fontenc}

\DeclareRobustCommand{\VAN}[3]{#2}
\let\VANthebibliography\thebibliography
\def\thebibliography{\DeclareRobustCommand{\VAN}[3]{##3}\VANthebibliography}

\usepackage{graphicx}	
\usepackage{amsmath}	
\usepackage{amssymb}	
\usepackage{multirow}
\usepackage{newtxtext,newtxmath}

\title[A survey of stellar populations in quiescent UDGs]{The stellar populations of quiescent ultra-diffuse galaxies from optical to mid-infrared spectral energy distribution fitting}

\author[M. L. Buzzo et al.]{
Maria Luisa {Buzzo}$^{1}$\thanks{E-mail: lgomesbuzzo@swin.edu.au},
Duncan A. Forbes$^{1}$,
Jean P. Brodie$^{1,2}$,
Aaron J. Romanowsky$^{2,3}$,
Michelle E. Cluver$^{1,4}$,
\newauthor
Thomas H. Jarrett$^{5}$,
Seppo Laine$^{6}$,
Warrick J. Couch$^{1}$,
Jonah S. Gannon$^{1}$, 
Anna {Ferré-Mateu}$^{7,1}$ and
\newauthor
Nobuhiro Okabe$^{8}$
\\ \\
$^{1}$ Centre for Astrophysics and Supercomputing, Swinburne University, John Street, Hawthorn VIC 3122, Australia \\
$^{2}$ University of California Observatories, 1156 High Street, Santa Cruz, CA 95064, USA \\
$^{3}$ Department of Physics and Astronomy, San José State University, One Washington Square, San Jose, CA 95192, USA \\
$^{4}$ Department of Physics and Astronomy, University of the Western Cape, Robert Sobukwe Road, Cape Town, 7535, South Africa \\
$^{5}$ Department of Astronomy, University of Cape Town, Private Bag X3, Rondebosch, 7701, South Africa \\
$^{6}$ IPAC, Mail Code 314-6, Caltech, 1200 E. California Blvd., Pasadena, CA 91125, USA \\
$^{7}$ Instituto de Astrof\'isica de Canarias, Calle V\'ia L\'actea S/N, E-38205, La Laguna, Tenerife, Spain \\
$^{8}$ Department of Physics, Hiroshima University, 1-3-1 Kagamiyama, Higashi-Hiroshima, Hiroshima 739-8526, Japan
}

\date{Accepted 2022 August 24. Received 2022 July 27; in original form 2022 June 17}

\pubyear{2022}

\begin{document}
\label{firstpage}
\pagerange{\pageref{firstpage}--\pageref{lastpage}}
\maketitle

\begin{abstract}
We use spectral energy distribution (SED) fitting to place constraints on the stellar population properties of 29 quiescent ultra-diffuse galaxies (UDGs) across different environments. We use the fully Bayesian routine \texttt{PROSPECTOR} coupled with archival data in the optical, near, and mid-infrared from \textit{Spitzer} and \textit{WISE} under the assumption of an exponentially declining star formation history. We recover the stellar mass, age, metallicity, dust content, star formation time scales and photometric redshifts (photo-zs) of the UDGs studied. 
Using the mid-infrared data, we probe the existence of dust in UDGs. Although its presence cannot be confirmed, we find that the inclusion of small amounts of dust in the models brings the stellar populations closer to those reported with spectroscopy. Additionally, we fit the redshifts of all galaxies. We find a high accuracy in recovering photo-zs compared to spectroscopy, allowing us to provide new photo-z estimates for three field UDGs with unknown distances. 
We find evidence of a stellar population dependence on the environment, with quiescent field UDGs being systematically younger than their cluster counterparts. Lastly, we find that all UDGs lie below the mass--metallicity relation for normal dwarf galaxies. Particularly, the globular cluster (GC)-poor UDGs are consistently more metal-rich than GC-rich ones, suggesting that GC-poor UDGs may be puffed-up dwarfs, while most GC-rich UDGs are better explained by a failed galaxy scenario. As a byproduct, we show that two galaxies in our sample, NGC 1052-DF2 and NGC 1052-DF4, share equivalent stellar population properties, with ages consistent with 8 Gyr. This finding supports formation scenarios where the galaxies were formed together.
\end{abstract}

\begin{keywords}
galaxies: formation – galaxies: stellar content – galaxies: fundamental parameters 
\end{keywords}



\section{Introduction}
\label{sec:introduction}
The existence of extremely faint and diffuse galaxies has been known since the mid-1980s \citep{Sandage_84,Bothun_87,Impey_88,Dalcanton_97,Conselice_03}. However, the eagerness to understand this underlying low surface brightness universe resurfaced again only recently, when \cite{vanDokkum_15} unexpectedly found a large population of these galaxies, dubbed ultra-diffuse galaxies (UDGs), in the Coma cluster. 

UDGs are tipically characterised by a central surface brightness of $\mu_{(g,0)}~>~24$ mag. arcsec$^{-2}$. One of the most stunning properties of these galaxies is that although they are very faint, they have the effective radii of giants, e.g., (${R}_\mathrm{e}$) $\geq 1.5$ kpc.

UDGs have been found in various environments, including clusters  (e.g., \citealt{vanDokkum_15,Yagi_16,Mihos_15,Venhola_17,Venhola_21,Wittmann_17,Gannon_22,Janssens_19,ManceraPina_19}), groups \citep[e.g.,][]{vanDokkum_18,Roman_Trujillo_17, Forbes_20b,Forbes_19b}, the field \citep[e.g.,][]{Leisman_17,Papastergis_17}, filaments \citep[e.g.,][]{MartinezDelgado_16}, and even in voids \citep[e.g.,][]{Roman_19}. While most of these studies focused on pointed observations to find these galaxies, studies relying on machine- and deep-learning and/or exploration of large-area imaging have systematically found thousands of new UDG candidates across a wide range of environments \citep{Greco_18,Zaritsky_19,Zaritsky_21, Tanoglidis_21,Greene_22}. These discoveries have increased substantially the number of UDGs known and thus provide richer statistics for inferences about their formation histories as a class. 

One downside of these photometric searches is that they provide only the positions and photometric properties of the sources, but not their distances, leaving the true size of these galaxies unknown. 
For the galaxies that do not have spectroscopy available, their size estimates are only possible through estimating their photometric redshifts. These estimates, in turn, necessitate a comprehensive wavelength coverage, so that different spectral features can be correctly identified and fitted. This has been pursued by \cite{Barbosa_20}, who estimated the distance and size of 100 UDG candidates in the field. 

Although the number of known UDGs has been growing steeply in the past few years, a consensus about how they were formed has not yet been reached. Various formation pathways have been proposed for UDGs, with most of them suggesting that they could either be ``failed galaxies'' \citep[][]{vanDokkum_15,Peng_Lim_16} or ``puffed-up dwarfs''. \citep{Burkert_17,Jiang_19,Amorisco_16,diCintio_17,Chan_18}. Under the failed galaxy scenario, UDGs would have started their lifetimes like normal galaxies, but faced a sudden halt in star formation at early epochs ($z>2$). This could happen, for example, due to early infall into a cluster environment \citep{Yozin_15}, where processes such as ram pressure stripping and harassment would prevent these galaxies from continuing to form stars and resulting in old stellar populations. Another explanation would be early quenching combined with an early, violent and fast star formation episode that would naturally create many globular clusters (GCs). As a result, these GCs would contribute a disproportionately large fraction of the main stellar light of the galaxy \citep{Danieli_22}. This could explain the high number of GCs found in many UDGs. The scenario proposed by \cite{Danieli_22} is capable of explaining the presence of UDGs both in high- and low-density environments, but it requires that the galaxies are GC-rich, and we may expect that they would be extremely metal-poor as these would be made up of mainly (now disrupted) GCs.

The puffed-up dwarf scenario is based on the assumption that UDGs are simply dwarf galaxies that have undergone some process capable of increasing their effective radii. These processes could be externally- or internally-driven. Some external explanations include dwarfs undergoing tidal stripping and heating \citep{Carleton_19} or mergers \citep{Wright_21}. Alternatively, internal processes could include having high-spins \citep{Amorisco_16,Rong_17,Amorisco_18} or stellar feedback \citep{diCintio_17,Chan_18}, capable of quenching star formation in the galaxies. This scenario would allow, unlike the failed galaxy one, the presence of gas in UDGs, and thus fit well the observation of bluer colours in isolated UDGs \citep{Roman_Trujillo_17,ManceraPina_19b}. 
None of these puffed-up dwarf scenarios is capable of explaining the existence of red isolated UDGs, which may instead have originated as a result of backsplash orbits \citep{Benavides_21}.

Recently, there have been attempts to probe these scenarios using a combination of photometric and spectroscopic data. These studies have mostly focused on the kinematic and dynamical properties of these galaxies and their globular cluster systems \citep[e.g., ][Gannon et al. 2022b, submitted]{Beasley_16,BeasleyTrujillo_16,vanDokkum_19,Emsellem_19,Forbes_20a,Gannon_20,Gannon_21,Gannon_22,ManceraPina_19b,ManceraPina_22}. Interestingly, there is evidence to support both scenarios, indicating that UDGs may not be formed by a single pathway, but rather by a variety of them \citep[][Gannon et al. 2022b, submitted]{Ferre-Mateu_18,Forbes_20a}.

The effort to better understand their formation histories, nonetheless, has not taken sufficient advantage of the knowledge of their stellar populations. The reason is clear; in order to study in detail the stellar population properties of UDGs (or any other galaxy, for that matter), one needs statistically large samples of them. 

To date, only about a dozen works have dedicated to the study of the stellar populations of UDGs. In what follows, we briefly describe some of them. 
Using a variety of spectroscopic data, \cite{Kadowaki_17}, \cite{Ferre-Mateu_18}, \cite{Ruiz-Lara_18}, \cite{Gu_18} and \cite{Villaume_22} focused on UDGs in the Coma cluster. All of them found evidence that UDGs in clusters mostly host intermediate-age ($\sim$6-8 Gyr) stellar populations and are metal-poor. Additionally, \cite{Ferre-Mateu_18} and  Ferre-Mateu et al. (2022, in prep.) have studied the alpha enhancement of quiescent UDGs, finding that on average they have [$\alpha$/Fe]$\sim$0.3 dex. \cite{Fensch_19} and \cite{Muller_19} focused on the stellar populations of group UDGs (NGC 1052-DF2 and NGC5846\_UDG1 , respectively). They both found that the studied UDGs host stellar populations with $\sim$8 Gyr and slightly higher metallicities than most of the ones studied in clusters. \cite{MartinNavarro_19} studied the field UDG DGSAT I, finding evidence that it hosts a stellar population of $\sim$8 Gyr, with very high alpha abundances ([Mg/Fe] = +1.5 dex). Focused on star forming UDGs, \cite{Rong_20}, using the stacked spectra of 28 UDGs from SDSS, have found that these are much more metal-rich ([$Z$/H]$\sim -0.8$ dex) and younger ($\sim 5$ Gyr) than the quiescent ones.

Although these studies have advanced a lot our understanding of the stellar populations of UDGs, obtaining spectra with high enough signal-to-noise (S/N) ratios for such low surface brightness sources requires extremely long integration times at the world's largest telescopes. Thus, alternative less-expensive methods, such as spectral energy distribution (SED) fitting, must be explored to increase our knowledge on this front. Yet another advantage of SED fitting techniques is that they allow reaching lower surface brightnesses than with spectroscopy.
To date, many studies have used imaging to try to investigate the colours and photometric properties of UDGs, but only two focused on recovering stellar populations via SED fitting techniques. \cite{Pandya_18} studied 2 UDGs, one in the field (DGSAT I) and one in the Virgo cluster (VCC1287), providing the first comparison of UDGs' stellar populations obtained with the same method across different environments. Similarly to the findings with spectroscopy, they found that the cluster UDG was older than the field one. \cite{Pandya_18} have also found the first evidence of interstellar diffuse dust in UDGs.
\cite{Barbosa_20} have studied 100 field UDGs and found on average intermediate age ($\sim$7 Gyr) stellar populations, with some of the UDGs being metal-poor and others metal-rich.
\cite{Barbosa_20} also inferred dust in all of their studied UDGs, with an average reddening of $A_{V}$ = 0.1 mag, consistent with the findings of \cite{Pandya_18}.

The finding of dust in UDGs raises many questions, e.g., is this dust component real? If so, is its presence expected from any formation scenario? Is there any correlation between the environment that the galaxies reside in and their dust content? Regardless of the answers to these questions, these findings are surprising, especially in UDGs in clusters, since such galaxies are expected to have quenched long ago and thus mechanisms of dust destruction, such as supernovae-driven shock waves, would have destroyed all of the dust out of the galaxy by now \citep{Jones_04,Jones_11}. 
Therefore, a more thorough exploration of this finding using appropriate data is necessary to better understand the role of dust in the formation history of UDGs.

In this work, we employ SED fitting techniques to explore the stellar population properties and photometric redshifts of UDGs. We do this for a moderate sample of 29 galaxies distributed across a variety of environments.
We use data from the optical to the mid-infrared to better constrain the shape of the SED, and also probe for the presence of dust in UDGs.
We use the stellar populations recovered to understand how UDGs fit into known scaling relations for both dwarf and giant galaxies. Finally, we seek to test different formation scenarios for them.

The paper is structured as follows: in Section \ref{sec:data} we present a summary of the UDG sample studied in this work and the data available for each UDG. We also discuss how we measured the photometry in each different dataset and the difficulties of doing this for such faint sources as UDGs. In Section \ref{sec:analysis} we describe our SED fitting methodology. In Section \ref{sec:results} we provide our results. In Section \ref{sec:discussion} we discuss the implications of our results within the theoretical predictions for UDGs and as compared to the literature. In Section \ref{sec:conclusions} we present the summary and the conclusions of the paper. In Appendix \ref{sec:appendix_stamps} we show the processed postage stamps of our sample of galaxies. In Appendix \ref{sec:appendixA} we show all of the SED fits and resulting corner plots. In Appendix \ref{sec:appendixB} we analyse the impact of excluding the infrared bands in the recovered stellar populations. In Appendix \ref{sec:Appendix_Table}, we show our SED fitting results without the inclusion of dust attenuation in the models.

\section{Data sample and photometry}
\label{sec:data}

In this work, we use data from the optical to mid-infrared to study the stellar population properties of twenty-nine quiescent ultra-diffuse galaxies. Below we present the data used for each galaxy, along with how the photometry was measured in each band. The data sample is summarised in Table \ref{tab:properties}.
Our sample consists of UDGs with central surface brightnesses brighter than 26 mags. arcsec$^{-2}$ and is based on two \textit{Spitzer}-IRAC programs (P.I. Romanowsky, with program IDs 13125 and 14114). 
These galaxies were selected to be quiescent, e.g., from red colours, but we note that some of them (e.g., LSBG-044, Hayes et al. in prep.) showed emission lines when followed up spectroscopically.
The Coma cluster galaxies were chosen to span a range of sizes, magnitudes, globular cluster specific frequencies and clustercentric radii.
The other UDGs were drawn from the fairly rare discoveries of other nearby non-Coma UDGs until early 2018, with an emphasis on including those with a diversity of properties and environments. We did not target any gas-rich, star forming UDGs \citep[e.g.,][]{Leisman_17}.
We also include in the sample two regular dwarf galaxies for control. These are the dwarf elliptical VCC1122 and local group dwarf irregular DDO 190.

The GC-richness classification of our sample of UDGs is made primarily based on GC numbers. GC-rich UDGs are those with more or exactly 20 GCs, while GC-poor ones are those with less than 20 GCs. The GC numbers for the UDGs in our sample come from a combination of the studies of \cite{vanDokkum_17,Forbes_20a,Lim_20,Gannon_21} and \cite{Saifollahi_22}. 

\begin{table*}
\centering
\scalebox{0.78}{
\begin{threeparttable}
\caption{Data sample properties.}
\begin{tabular}{c|c|c|c|c|c|c|c|c|c|l} \hline
Galaxy & RA & Dec & Environment & GC richness & V$_{r}$ & Distance  & $\mu_{\rm 0}$  &  $r_{\rm eff}$ & $r_{\rm eff}$ & \multirow{3}{*}{Comments} \\ 
 & (deg) & (deg) & & & [km s$^{-1}$] & [Mpc] & [mag arcsec$^{-2}$] & [kpc] & [arcsec] \\
 (1) & (2) & (3) & (4) & (5) & (6) & (7) & (8) & (9) & (10) & (11) \\ \hline
PUDG-R24 & 49.648 & 41.809 & Field$^{\rm c}$ [Perseus] & Poor & 7784 & 75  & 23.6 ($g^{\rm a}$) & 3.6 & 10.1 & Recent infall into Perseus? \\
NGC 1052-DF4  & 39.813 & -8.116 & Group [NGC 1052] & Poor & 1445 & 20  & 23.7 (V$_{606}$) & 1.6 & 16.5 & \\
LSBG-490   & 181.933 & 01.172 & Field & -- & --  & -- & 23.7 ($g^{\rm b}$) & -- & 4.7 & \\
Dragonfly X1 (DFX1) & 195.316  & 27.210 & Cluster [Coma] & Rich & 8223 & 100  & 24.1 ($g$) &  3.5 & 7.2  & \\
Dragonfly 26  (DF26) &  195.086 & 27.787 & Cluster [Coma] & Rich & 6611 & 100   & 24.1 ($g$) &  3.3 & 6.8 & Tidally disrupting (S-shaped tails)\\
LSBG-378  & 137.273 & 04.995 & Field &  -- & --   & -- & 24.3 ($g^{\rm b}$) & -- & 10.3 & \\
NGC 1052-DF2 & 40.445 & -8.403 & Group [NGC 1052] & Poor & 1803 & 22.1  & 24.4 (V$_{606}$) & 2.2 & 22.6 & \\
Dragonfly 02 (DF02) & 194.790 & 29.007  & Cluster [Coma] & Poor & -- & 100  & 24.4 ($g$) & 2.1 & 4.5 & \\
Dragonfly 07 (DF07) & 194.257 & 28.390  & Cluster [Coma] & Rich & 6587 & 100  & 24.4 ($g$) & 4.3 & 9.1 & \\
Dragonfly 03 (DF03) & 195.569 & 28.955 & Group & Poor & 10150 &  145 & 24.5 ($g$) & 2.9 & 6.2 & Disrupting galaxy in group behind Coma\\
Yagi358 (Y358) & 194.810 & 28.038 & Cluster [Coma] & Rich & 7967 & 100   & 24.5 ($r$) & 2.3 & 4.7 & \\
Dragonfly 44 (DF44) & 195.242 & 26.976 & Cluster [Coma] & Rich & 6280 &   100   & 24.5 ($g$) &  4.7 & 9.4 & Recent infall into Coma?\\
Dragonfly X2 (DFX2) & 195.272 & 27.160 & Cluster [Coma] & -- & 6473 & 100 & 24.5 ($g$) & 1.7 & 3.6 & Recent infall into Coma? \\
PUDG-R16  &  49.652 & 41.192 & Cluster [Perseus] & Poor & 4679 & 75  & 24.5 ($g^{\rm a}$) & 4.2 & 11.7 & \\
Dragonfly 40  (DF40) & 194.505 & 27.191 & Cluster [Coma] & Poor &  7792 & 100  & 24.6 ($g$)  & 2.9 & 5.9 & \\
LSBG-044 & 237.847 & 43.306 & Field & -- & -- & -- & 24.7 ($g^{\rm b}$) & -- &  6.4 & \\
DGSAT I  & 19.398 & 33.528 & Field & -- & 5439 & 70  & 24.8 & 4.7 & 12 & \\
Dragonfly 23 (DF23) & 194.849 & 27.791 & Cluster [Coma] & Rich & 7068 & 100  & 24.8 ($g$) & 2.3 & 4.9 & \\
M-161-1 & 202.473 & 46.372 & Field & -- & 5600 &  81  &  24.8 ($V$) & 4.1 & 10.6 & \\
Yagi436 (Y436) & 195.122 & 27.990 & Cluster [Coma] & Rich &  -- & 100 & 24.9 ($r$) &  1.7 & 3.5 & \\
Yagi534 (Y534) & 194.254 & 27.532 & Cluster [Coma] & Rich & -- &  100  & 25.1 ($r$) & 1.9 & 3.9 & \\
Dragonfly 17 (DF17) & 195.494 & 27.836 & Cluster [Coma] & Rich & 8315 & 100  & 25.1 ($g$) & 3.3 & 9.0 & \\
Dragonfly 25 (DF25) & 194.952 & 27.778 & Cluster [Coma] & Poor & 6959 & 100  & 25.2 ($g$) & 4.4 & 9.3 & \\
Dragonfly 08 (DF08) & 195.377 & 28.374  & Cluster [Coma] & Rich & 7051 & 100  & 25.4 ($g$) & 4.4 & 9.3 & \\
Dragonfly 46 (DF46) & 195.197 & 26.783 & Cluster [Coma] & Poor & -- & 100  & 25.4 ($g$) & 3.4 & 7.2 & \\
VCC1287  & 187.602 & 13.982 & Cluster [Virgo] & Rich & 1116 & 16.5  & 25.4 ($g^{\rm a}$) & 3.4 & 45.8 & \\
Dragonfly 06 (DF06) & 194.124 & 28.444  & Cluster [Coma] & Rich & -- & 100  & 25.5 ($g$) & 4.4 & 9.3 & \\
VCC1884 & 190.414 & 9.208 &  Cluster [Virgo] & Poor &  -- & 16.5 & 25.5 ($g^{\rm a}$) & 3.1 & 24.0 & \\
VCC1052  & 186.980 & 12.369 & Cluster [Virgo] & Poor & -- & 16.5 &  25.8 ($g^{\rm a}$) & 3.7 & 25.0 & Peculiar morphology  \\ \hline
VCC1122  & 187.174 & 12.916 & Cluster [Virgo] & -- & 465.1 & 16.5 &  22.4 ($B$) & 1.3 & 17.3 & Virgo dwarf elliptical  \\ 
DDO 190  & 216.181 & 44.526 & Local Group & -- & 69$^{\rm d}$ & 2.8 &  23.6 ($V$) & 0.7 & 54.6$^{\rm d}$ & Local group dwarf irregular  \\ 
 \hline
\end{tabular}
\begin{tablenotes}
      \small
      \item \textbf{Note.} Columns stand for: (1) Galaxy ID. (2-3) Coordinates in degrees. (4) Environment that the galaxies reside in; (5) Globular cluster richness (Rich: more or equal to 20 GCs, Poor: less than 20 GCs); (6) Radial velocity; (7) Distance to group/cluster in megaparsecs; (8) Central surface brightness with band in parenthesis; (9) Effective radius in kiloparsecs; (10) Effective radius in arcseconds; (11) Comments about the galaxies. (a) These are approximate values, as we convert values assuming that the central surface brightness is three times brighter than the mean surface brightness within one effective radius (i.e., 1.2 mag arcsec$^{-2}$ brighter) \protect\citep{Graham_Driver_05}. (b) Converted to $g$ band using the $g-i$ colour provided by \protect\cite{Greco_18}. (c) Although this galaxy is in the Perseus cluster, its radial velocity \protect\citep{Gannon_22} indicates that it is at the very outskirts of the cluster, thus being better classified as a recently accreted field galaxy than a cluster one. (d) Measurements from \cite{McConnachie_12, Cook_14}.
\end{tablenotes}
\label{tab:properties}
\end{threeparttable}}
\end{table*}
\subsection{\textit{WISE} near-IR and mid-IR imaging}
\label{sec:WISE_data}

\begin{figure}
    \centering
    \includegraphics[width=0.9\columnwidth]{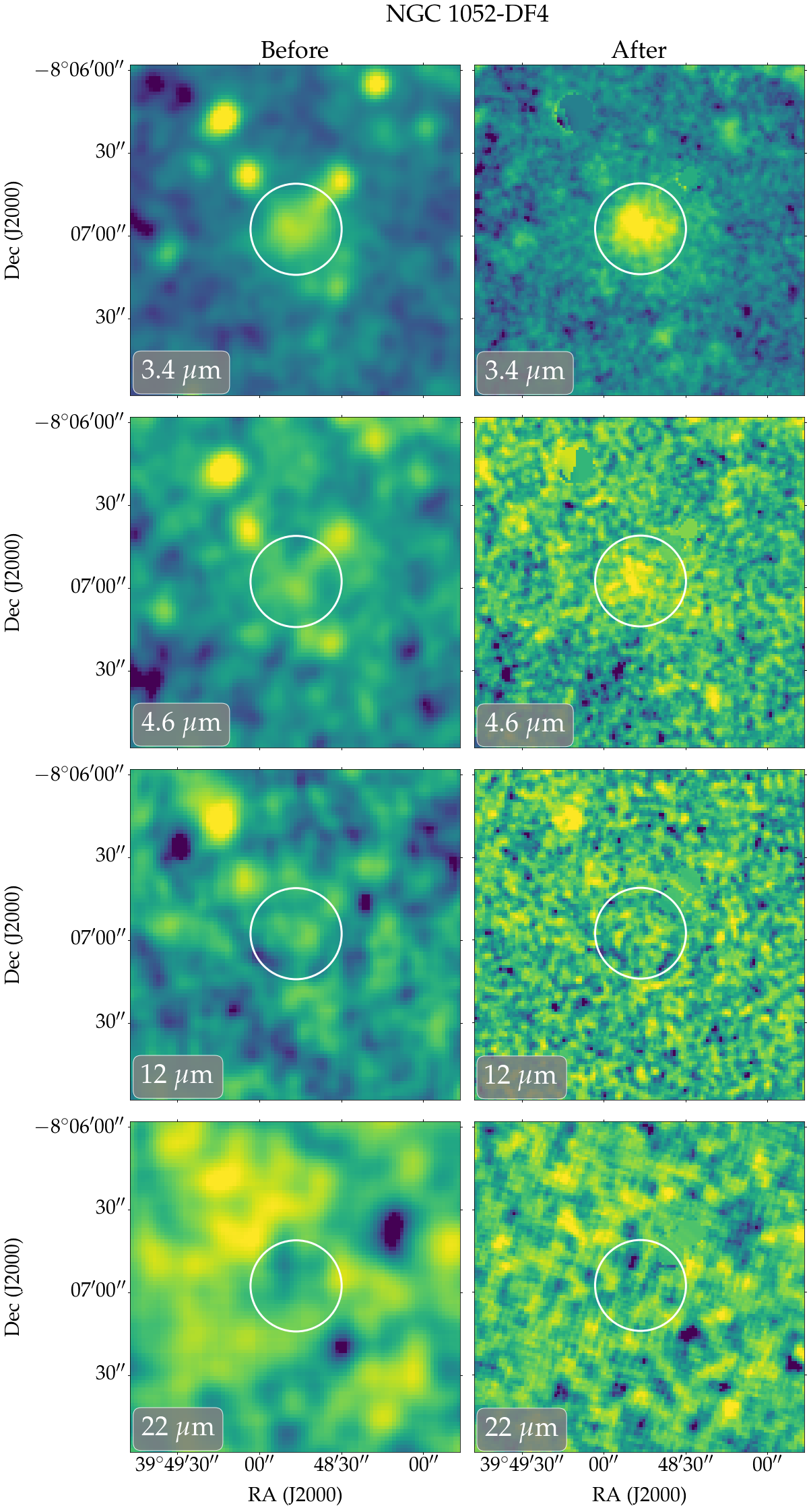}
    \caption{\textit{WISE} images of NGC 1052-DF4 before and after correcting for sensitivity issues and removing stellar contaminants. This technique follows the procedure described in \protect\cite{Jarrett_19}. Rows stand for the 3.4, 4.6, 12 and 22 $\mu$m bands, respectively. The 3.4 and 4.6 $\mu$m bands were measured for this galaxy, while the 12 and 22 $\mu$m bands just provided upper limits. The white circle represents the effective radius of the galaxy.}
    \label{fig:comparison_WISE_corrected}
\end{figure}

The \textit{Wide-field Infrared Survey Explorer} \citep[\textit{WISE}, ][]{Wright_10} is a space telescope that has imaged the entire sky in four filters with effective wavelengths of 3.368, 4.618, 12.082 and 22.194 $\mu$m (near to mid-infrared). For this work, we gathered \textit{WISE} data for all the galaxies in our sample. These data are a mix of archival ALLWISE data and bespoke data construction and analysis, including custom mosaic construction from \textit{WISE} single frames. These are done using the \texttt{ICORE} software developed by the \textit{WISE} team and IPAC \citep{Masci_13}. The frames include both classic \textit{WISE} mission data and the follow-up NEOWISE mission (W1 and W2 only), creating very deep mosaics. Native angular resolution is preserved in all four bands \citep[see][for further details]{Jarrett_12}. We corrected the frames by removing stars and background galaxies using a combination of
PSF profile-fitting and masking, while brighter (or resolved) sources required aperture masking. Masked pixels are replaced with the local background, thus preserving the integrated flux of the target galaxy \citep[more details are given in][]{Jarrett_13}.
In Fig. \ref{fig:comparison_WISE_corrected}, we show one example (NGC 1052-DF4) of \textit{WISE} images before and after the corrections described in \cite{Jarrett_19} to highlight the power of the technique.

For two galaxies of our sample which had the brightest detections, NGC 1052-DF2 and NGC 1052-DF4, axisymmetric radial profiles were constructed. We fitted a double-S\'ersic function, attempting to model the galaxies, as well as extrapolate to determine total fluxes. 
For the remaining galaxies, where little emission was detected beyond the beam (and hence radial profiles were not constructed), we instead carried out aperture photometry. We apply an aperture correction that at the very least accounts for the point spread function (PSF) emission that is not detected \citep[for further details, see][]{Cluver_20}.

The uncertainties in the \textit{WISE} photometry include mostly contributions from the Poisson errors and background subtraction errors. Additional uncertainties coming from colour and calibration corrections are also applied. An uncertainty in the zero point flux-to-magnitude conversion, corresponding to 1.5\%, is added to all \textit{WISE} bands. Additionally, in the case of aperture photometry, an uncertainty of 1\% coming from aperture correction is added to all bands. We reiterate that UDGs push the boundaries of what is possible with \textit{WISE} and just moving the background annulus around them can have a large (10-20\%) effect on the integrated flux. Thus, \textit{WISE} photometric uncertainties may be underestimated given that the technique \citep{Jarrett_19} was not designed to deal with such faint sources. 

A summary of the photometric measurement method used for each galaxy is given in Table \ref{tab:photometry}.

\subsection{\textit{Spitzer}-IRAC NIR imaging}

\textit{Spitzer}-IRAC \citep{Fazio_04,Werner_04} observations of our sample of galaxies were taken over the years of 2017 and 2018 (P.I. Romanowsky, with program IDs 13125 and 14114). 
3.6 and 4.5 $\mu$m band observations were taken for six galaxies of the sample: DFX1, DF44, M-161-1, DF17, DGSAT I and VCC1287. The remaining 23 UDGs were observed only in the 3.6 $\mu$m band. 
The reduction process applied to the \textit{Spitzer} data of the galaxies in our sample is thoroughly described in \cite{Pandya_18}.

To extract the photometry in the \textit{Spitzer}-IRAC images, we started by masking the compact foreground/background sources in both bands. We did this by defining a brightness threshold above which sources should be masked. We then applied Gaussian kernel smoothing ($\sigma = 2$ pixels) to our masks in order to remove any persisting halo features. 
Some galaxies (M-161-1, PUDG-R24, DF26, DF44) were especially tricky to mask because they had compact sources within the measured aperture. For these galaxies, we replaced the pixels within the masked area with their median value and then performed the photometry.
We note that the photometry in the \textit{Spitzer}-IRAC bands for DGSAT I, VCC1287 and VCC1122 comes from \cite{Pandya_18}, as we used the same aperture as they did in all our photometric bands.
We then measured sky backgrounds in five empty areas of the sky around the galaxies in each band, and from the results we estimated an average sky background to be subtracted at the position of the galaxies. 

As explained above, the \textit{WISE} pipeline will iteratively find the best way to measure the photometry of a galaxy (i.e., aperture photometry or isophotal). Thus, in order to have consistent magnitudes, we used the same photometry extraction method found by \textit{WISE} in our \textit{Spitzer}-IRAC images (see column 2 of Table \ref{tab:photometry}). Independently of the measurement type, we used the \texttt{Astropy} Python library commands after masking the foregrond/background sources. We corrected the results with the aperture correction factors appropriate for the aperture used, according to the \href{https://irsa.ipac.caltech.edu/data/SPITZER/docs/irac/iracinstrumenthandbook/}{IRAC instrument handbook}. 

The uncertainty in aperture photometry was estimated by performing aperture photometry on several empty positions in the sky and taking the rms scatter in these measurements.
We estimated the uncertainty due to masking by doing random variations of the masks and reperforming the photometry. We added the standard deviation of these measurements to our final uncertainties.
The sky background subtraction uncertainty ranges from 0.01 to 0.04 mag in the 3.6 $\mu$m band. This was estimated by taking the maximum difference in sky background measurements in three empty sky areas around the galaxies, adding this difference to all the pixels within the aperture and performing aperture photometry again.
The calibration uncertainty was estimated to be 0.02 mag in the 3.6 $\mu$m and 4.5 $\mu$m bands. There is an additional uncertainty due to the aperture correction applied. This ranged from 0.02--0.09 mag in the 3.6 $\mu$m band and 0.01--0.04 mag in the 4.5 $\mu$m band.  
We added all uncertainties quadratically, resulting in the values reported in Table \ref{tab:photometry}. 

\begin{figure}
    \centering
    \includegraphics[width=\columnwidth]{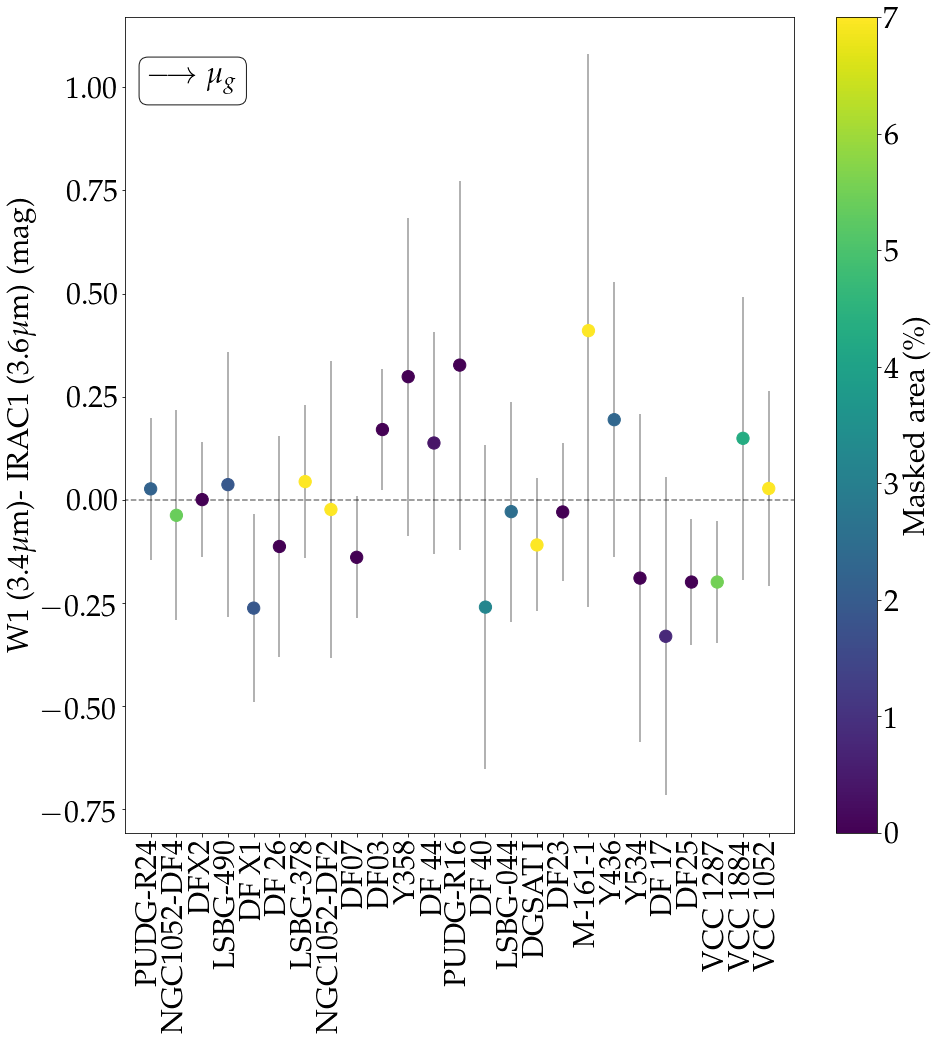}
    \caption{Photometric difference in magnitudes between measurements in the \textit{WISE} 3.4 $\mu$m band and the \textit{Spitzer}-IRAC 3.6 $\mu$m band. UDGs are ordered by their central surface brightness as shown by the arrow at the upper left corner (see Table \ref{tab:properties}). The colorbar shows the percentage of the galaxies that were masked by the threshold-masking technique applied to the \textit{Spitzer}-IRAC images. Results are consistent within the uncertainties and no systematic errors are observed as a function of surface brightness or masked area.}
    \label{fig:difference_w1_irac1}
\end{figure}

For consistency, we compared the photometry measured in the \textit{WISE} 3.4 $\mu$m and the \textit{Spitzer}-IRAC 3.6 $\mu$m bands in order to check if there was any bias correlated with the surface brightness of the galaxies or the ``threshold-masking'' applied to the \textit{Spitzer}-IRAC data. We show this comparison in Fig. \ref{fig:difference_w1_irac1}.
This figure clearly shows that, even if the \textit{WISE} image is much shallower than the \textit{Spitzer}-IRAC 3.4 $\mu$m, the results are consistent within the quoted uncertainties. Additionally, as shown in Table \ref{tab:photometry}, the difference between the \textit{WISE} 4.6 $\mu$m and the \textit{Spitzer}-IRAC 4.5 $\mu$m bands ranges from 0.06 to 0.36 mag and the measurements from the two bands are always consistent within the uncertainties.

\subsection{Optical imaging}
The optical data used in this work comes from several telescopes and surveys. Below we briefly describe where the data for each galaxy comes from.

The optical data for the two Perseus cluster galaxies in our sample, PUDG-R16 and PUDG-R24, come from observations with the Hyper Suprime-Cam (HSC) taken on the 8.2m Subaru Telescope on the night of 2014 September 24 (P.I. Okabe). Observations were carried out with a field-of-view (FoV) of 1.5 degree diameter covering the entire Perseus cluster in the $g$, $r$ and $i$ bands, with a seeing of 0.8\arcsec. The data reduction followed the standard HSC pipeline \citep[][refer to \cite{Gannon_22} for further details]{Bosch_18}.

Archival Gemini Multi-Object Spectrometer (GMOS) North data for DF44, DFX1 and DFX2 were obtained in the $g$ and $i$ bands. The observations and the reduction process are described in \cite{vanDokkum_16}.

Additionally, archival optical data in the $g$ and $r$ bands were obtained for the UDG M-161-1 from the Canada--France--Hawaii Telescope (CFHT) MegaCam archive. Astrometric and photometric calibrations were first performed on the individual exposures, and the backgrounds adjusted. The images were then combined using SWARP.

For the remaining galaxies (including DDO 190), archival optical data in the $g$, $r$ and $z$ bands were obtained from the Dark Energy Camera Legacy Survey \citep[DECaLS,][]{Dey_19}. The reduction and calibration of the DECaLS data are described by \cite{Dey_19}. Images from DECaLS have shallower depths and more uncertain sky subtractions than other optical surveys that focused on low surface brightness galaxies. For this reason, we selected three galaxies in our sample (VCC1287, VCC1052 and VCC1884) to test how accurate our photometric measurements were. We compared our results to those obtained by \cite{Lim_20} and \cite{Pandya_18} using deeper data reduced with a pipeline developed specifically to deal with low surface brightness galaxies. We find that our measurements are on average 0.1 magnitudes fainter than the ones in the literature.
We attribute this difference to the sky subtraction applied to the DECaLS images that may have considered a fraction of the UDGs as part of the background. This finding is incorporated into our uncertainties.

Photometric measurements in the optical were carried out similarly to those in the infrared regime, i.e., matching the method to perform the photometry (aperture photometry or isophotal) to those obtained with \textit{WISE} and \textit{Spitzer}-IRAC.  
The photometry in the optical for VCC1122 comes from \cite{Pandya_18}, since we used the same aperture as they did to measure the photometry in our other photometric bands. Aperture photometry in the optical for DGSAT I comes from private communication with S. Janssens \citep[with aperture colours equivalent to the total ones in][]{Janssens_22}. We note that we do not use \cite{Pandya_18} photometry for DGSAT I because they found an optical colour for it inconsistent with that found by \cite{MartinezDelgado_16} and with \cite{Janssens_22}.
Masks in the optical were applied for only a few of the galaxies in the sample (PUDG-R24, NGC 1052-DF4, NGC 1052-DF2, DF44 and M-161-1). Galaxies that did not appear to have foreground stars in the measured aperture were not masked. The uncertainty due to masking is on average 0.01 mag in the $g$ band, 0.03 mag in the $r$ band, 0.01 mag in the $i$ band and 0.04 mag in the $z$ band for the five galaxies where masks were applied. The sky background uncertainty was calculated as described above for the \textit{Spitzer}-IRAC data. Images from DECaLS ($5\sigma$ depth = 24.7 mag; \citealt{Dey_19}) had overall higher background uncertainties than images from the HSC, GMOS and MegaCam given the shallower depth and uncertain sky subtraction, resulting in higher uncertainties.

\begin{figure*}
    \centering
    \includegraphics[width=\textwidth]{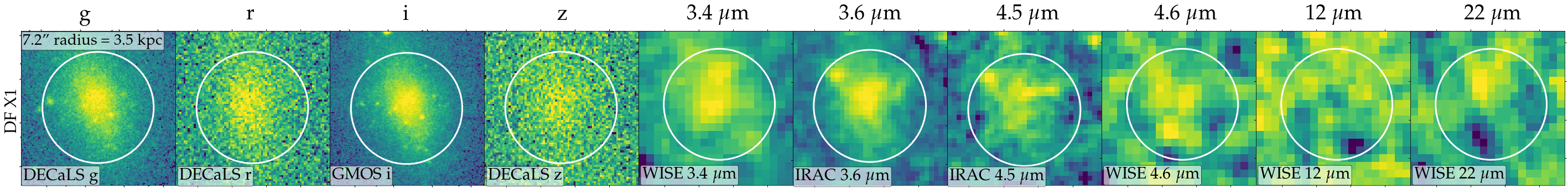}
    \includegraphics[width=\textwidth]{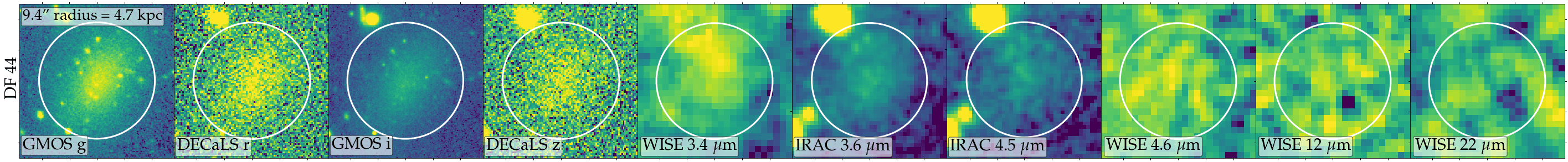}
    \includegraphics[width=\textwidth]{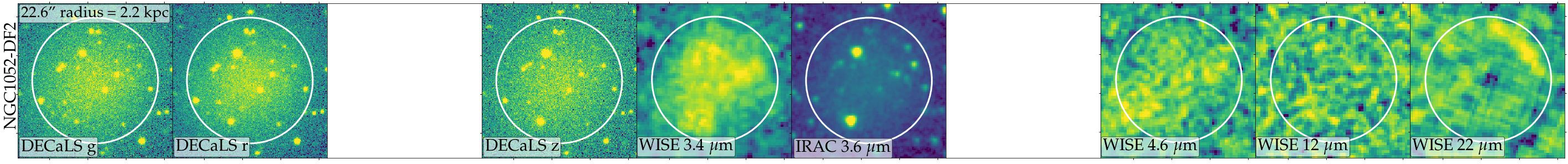}
    \includegraphics[width=\textwidth]{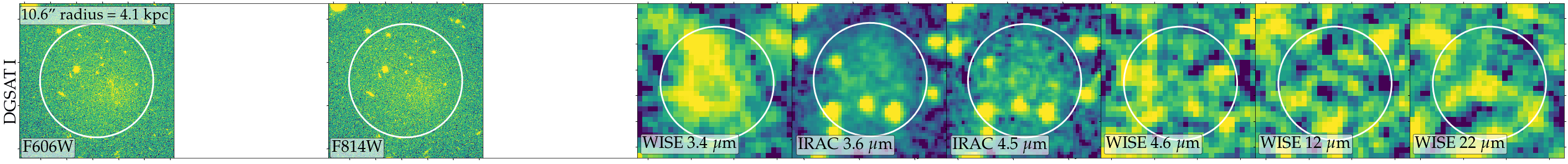}
    \includegraphics[width=\textwidth]{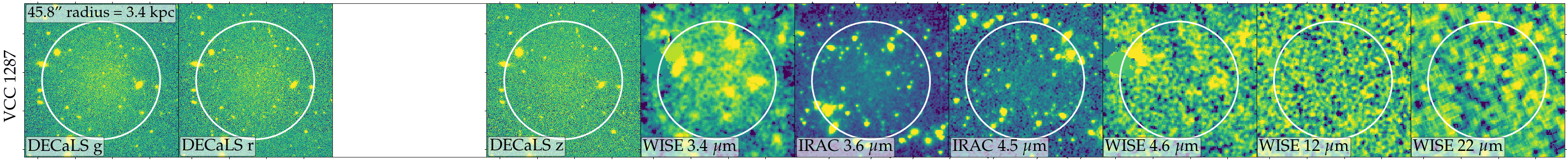}
    \caption{Final processed postage stamp images of five of the UDGs analysed in this work (the remaining 24 can be found in Appendix \ref{sec:appendix_stamps}). Rows are ordered by optical surface brightness (see Table \ref{tab:properties}). The stamps are constructed using a logarithmic normalization. Columns show the respective $g$ (F606W for DGSAT I), $r$, $i$ (F814W for DGSAT I), $z$, \textit{WISE} 3.4 $\mu$m, IRAC 3.6 $\mu$m, IRAC 4.5 $\mu$m, \textit{WISE} 4.6 $\mu$m, \textit{WISE} 12 $\mu$m, and \textit{WISE} 22 $\mu$m band images. Blank squares stand for unavailable data. White circles show the effective radius of the UDGs (see Table \ref{tab:properties}), also highlighted in the $g$ band image (first column) in arcsecs and kiloparsecs (if radial velocity known). Each stamp is annotated with the band and instrument that they come from.}
    \label{fig:UDGs_stamps1}
\end{figure*}

Fig. \ref{fig:UDGs_stamps1} shows the final processed postage stamp images of five UDGs in our sample, including all the bands used. The remaining 24 processed postage stamp images can be found in Appendix \ref{sec:appendix_stamps}. We note that the \textit{WISE} images included in these figures are the ones already corrected for stellar contaminants, as described in Section \ref{sec:WISE_data}. The \textit{Spitzer}-IRAC and optical images included are not the ones masked for contaminants.

The magnitude uncertainties were added quadratically, resulting in the values shown in Table \ref{tab:photometry}, containing a summary of the photometry measured for all the galaxies in every band.
All the magnitude measurements are in AB magnitudes and were corrected for Galactic extinction using the two-dimensional dust maps of \citealt{SFD} (recalibrated by \citealt{Schlafly_11}) and the extinction law of \cite{Calzetti_00}.

\begin{table*}
\centering
\scalebox{0.72}{
\begin{threeparttable}
\caption{Photometric measurements in the optical, near-IR and mid-IR for our sample of 29 UDGs.}
\begin{tabular}{c|c|c|c|c|c|c|c|c|c|cc} \hline
Galaxy & Method & \textit{g} & \textit{r} & \textit{i} & \textit{z} & \textit{WISE} 3.4 $\mu$m &  IRAC 3.6 $\mu$m & IRAC 4.5 $\mu$m & \textit{WISE} 4.6 $\mu$m & \textit{WISE} 12 $\mu$m & \textit{WISE} 22 $\mu$m \\
(1) & (2) & (3) & (4) & (5) & (6) & (7) & (8) & (9) & (10) & (11) & (12) \\ \hline
PUDG-R24 & Aper \{10\} & $19.63 \pm 0.12$ & $19.03 \pm 0.07$ & $18.68 \pm 0.11$ & -- & $19.37 \pm 0.10$ & $19.34 \pm 0.14$ & -- & $20.19 \pm 0.39$ & $> 16.68$ & $> 16.22$ \\ 
NGC1052-DF4 & Isophotal & $17.77 \pm 0.19$ & $17.10 \pm 0.15$ & -- & $16.67 \pm 0.16$ & $17.34 \pm 0.05$ & $17.68 \pm 0.25$ & -- & $18.25 \pm 0.30$ & $> 18.27$ & $> 16.55$ \\
LSBG-490  & Aper \{5\} & $20.56 \pm 0.23$ & $19.90 \pm 0.23$ & -- & $19.52 \pm 0.13$  & $19.94 \pm 0.02$ & $19.90 \pm 0.22$ & -- &  -- & $> 17.74$ & $> 16.17$ \\ 
DFX1 & Aper \{7\} & $20.15 \pm 0.15$ & $19.47 \pm 0.17$ & $19.28 \pm 0.22$ & $19.18 \pm 0.15$ & $20.04 \pm 0.06$ & $20.30 \pm 0.22$ & $20.99 \pm 0.25$  & $20.93 \pm 0.22$ & $> 18.25$ & $> 16.63$ \\
DF26 & Aper \{7\} & $20.20 \pm 0.18$ & $19.44 \pm 0.18$ & -- & $19.08 \pm 0.13$ & $19.81 \pm 0.19$ & $19.92 \pm 0.19$ & -- & $20.55 \pm	0.82 $ & $> 18.09$ & $> 16.41$ \\
LSBG-378 & Aper \{10\} & $19.16 \pm 0.23$ & $18.58 \pm 0.24$ & -- & $18.24 \pm 0.17$ & $18.84 \pm 0.11$ & $18.82 \pm 0.15$ & -- & $19.49 \pm 0.43$ &	$> 17.85$ & $> 16.21$ \\ 
NGC1052-DF2 & Isophotal & $17.09 \pm 0.12$ & $16.45 \pm 0.22$ & -- &  $16.07 \pm 0.23$ & $16.87 \pm 0.27$ & $17.15 \pm 0.24$ & -- & $17.42 \pm 0.93$ & $> 18.76$ & $> 16.57$ \\ 
DF02 & Aper \{5\} & $20.74 \pm 0.17$ & $20.13 \pm 0.21$ & -- & $19.74 \pm 0.21$ & -- & $> 18.25$ & -- & -- & -- & -- \\
DF07 & Aper \{9\} & $19.71 \pm 0.13$ & $19.13 \pm 0.12$ & -- & $18.69 \pm 0.15$ & $19.37 \pm 0.09$ & $19.51 \pm 0.11$ & -- & $19.16 \pm 0.17$ & $> 18.09$ & $> 16.59$\\
DF03 & Aper \{6\} & $20.73 \pm 0.17$ & $20.03 \pm 0.22$ & -- & $19.74 \pm 0.24$ & $21.08 \pm 0.45$ & $20.91 \pm 0.19$ & -- & -- & $> 18.11$ & $> 16.44$\\
Y358 & Aper \{5\} & $20.78 \pm 0.16$ & $20.11 \pm 0.14$ & -- & $19.71 \pm 0.18$ & $20.42 \pm 0.13$ & $19.94 \pm 0.16$ & --  & -- & $> 17.98$ & $> 16.34$  \\ 
DF44 & Aper \{12\} & $20.05 \pm 0.14$ & $19.15 \pm 0.15$ &$19.35 \pm 0.18$ & $18.77 \pm 0.16$ & $19.76 \pm 0.23$ & $19.62 \pm 0.14$ & $19.93 \pm 0.27$ & $20.08 \pm	0.10$ & $> 18.12$ & $> 16.45$ \\ 
DFX2 & Aper \{4\} & $18.95 \pm 0.07$ & $18.81 \pm 0.15$ & $18.75 \pm 0.06$ & $18.61 \pm 0.22$ & $19.01 \pm 0.05$ & $19.01 \pm 0.13$ & -- & $19.39 \pm 0.20$ & $> 18.32$ & $> 16.70$\\
PUDG-R16 & Aper \{12\} & $20.01 \pm 0.08$ & $19.19 \pm 0.10$ & $18.66 \pm 0.08$ & -- & $ 19.94 \pm 0.41$ & $19.61 \pm 0.18$ & -- & $20.21 \pm 0.64$ & 	$> 17.93$ & $> 16.13$ \\
DF40 & Aper \{6\} & $20.80 \pm 0.17$ & $20.35 \pm 0.18$ & -- & $19.96 \pm 0.21$ & $19.94 \pm 0.15$ & $20.20 \pm 0.18$ & -- & $20.72 \pm 0.96$ &	$> 18.04$ & $> 16.23$ \\ 
LSBG-044 &  Aper \{6\} & $20.63 \pm 0.19$ & $20.07 \pm 0.12$ & -- & -- & $20.21 \pm 0.06$ & $20.24 \pm 0.26$ & --  &	$ 21.25 \pm 0.19$ &	$> 18.72$ & $> 16.87$ \\ 
DGSAT I &  Aper \{15\} & $18.71 \pm 0.02$** & -- & $18.24 \pm 0.03$** & --  & $19.01 \pm 0.14$ & $19.12 \pm 0.07$ & $19.43 \pm 0.08$ &	$19.07 \pm 0.14$ & $> 18.50$ & $> 16.63$ \\
DF23 &  Aper \{5\} & $20.73 \pm 0.21$ & $20.09 \pm 0.20$ & -- & $19.76 \pm 0.23$ & $21.12\pm 0.50$ & $21.15 \pm 0.26$ & -- & -- & $> 17.97$ & $> 16.44$  \\
M-161-1 &  Aper \{11\} & $20.56 \pm 0.10$ & $20.13 \pm 0.17$ & -- & $19.49 \pm 0.17$ & $21.53 \pm 0.63$ & $21.12 \pm 0.23$ & $21.45 \pm 0.15$ & -- & $> 18.15$ & $> 16.56$ \\ 
Y436 & Aper \{4\} &  $21.69 \pm 0.18$ & $21.14 \pm 0.15$ & -- & $20.66 \pm 0.22$ & $20.77 \pm 0.26$ & $20.58 \pm 0.21$ & -- & $ 20.49 \pm 0.49$ & $> 18.13$ & $> 16.43$ \\ 
Y534 &  Aper \{4\} & $21.60 \pm 0.25$ & $21.02 \pm 0.17$ & -- & $20.54 \pm 0.21$ & $21.23 \pm 0.35$ & $21.42 \pm 0.19$ & -- & $21.08 \pm 0.85$ & $> 18.03$ & $> 16.33$ \\ 
DF17 &  Aper \{10\} & $20.27 \pm 0.14$ & $19.59 \pm 0.15$ & -- & $19.33 \pm 0.19$ & $20.36 \pm 0.36$ & $20.69 \pm 0.14 $ & $21.15 \pm 0.36$ & -- & $> 18.29$ & $> 16.34$ \\ 
DF25 &  Aper \{9\} &  $20.80 \pm 0.35$ & $20.29 \pm 0.32$ & -- & $19.95 \pm 0.44$ & $20.23 \pm 0.30$ & $20.43 \pm 0.15$ & -- & -- & $> 17.88$ & $> 16.25$ \\ 
DF08 &  Aper \{9\} & $20.90 \pm 0.22$ & $20.24 \pm 0.16$ & -- & $19.83 \pm 0.18$ &  -- & $> 21.12$ & -- & -- & -- & --  \\
DF46 &  Aper \{7\} & $21.41 \pm 0.25$ & $20.79 \pm 0.17$ & -- & $20.50 \pm 0.21$ &  -- & $> 21.80$ & -- & -- & -- & --  \\
VCC1287 & Aper \{30\} &  $17.17 \pm 0.21$ & $16.51 \pm 0.17$ & -- & $16.18 \pm 0.23$  & $17.22 \pm 0.13$ & $17.42 \pm 0.07$ & $17.72 \pm 0.07$ &	$17.95 \pm 0.17$ & $> 19.30$ & $> 16.20$ \\
DF06 & Aper \{6\} &  $21.16 \pm 0.31$ & $20.71 \pm 0.27$ & -- & $20.46 \pm 0.30$ & -- & $19.29\pm 0.17$ & -- & -- & -- & -- \\
VCC1884 & Aper \{22\} &  $16.95 \pm 0.15$ & $16.39 \pm 0.18$ & -- & $15.98 \pm 0.13$  & $17.38 \pm 0.05$ & $17.23 \pm 0.14$ & -- &	$19.28 \pm 0.47$ & $> 18.13$ & $> 16.33$ \\
VCC1052 &  Aper \{20\} & $18.21 \pm 0.15$ & $17.80 \pm 0.29$ & -- & $17.39 \pm 0.16$ & $18.54 \pm 0.09$ & $18.51 \pm 0.22$ & -- & $18.88 \pm 0.43$ & $> 17.94$ & $> 16.22$ \\ 
 \hline
\end{tabular}
\begin{tablenotes}
      \small
      \item \textbf{Note.} UDGs are ordered by surface brightness (see Table \ref{tab:properties}). Columns are: (1) Galaxy ID; (2) Method used for photometric measurement. The aperture radius used is indicated in curly brackets, in arcsec. (3--6) Optical photometry in the $g$, $r$, $i$ and $z$ bands, respectively; (7,10--12) Near- and mid-IR \textit{WISE} photometry in the 3.4, 4.6, 12 and 22 $\mu$m bands, respectively; (8,9) Near-IR \textit{Spitzer}-IRAC photometry in the 3.6 and 4.5 $\mu$m bands. '--' stands for unavailable or unmeasurable data. * Data in F606W band instead of g band, and F814W instead of i band. '>' stands for the 3$\sigma$ upper limits.
\end{tablenotes}
\label{tab:photometry}
\end{threeparttable}}
\end{table*}

\section{Analysis}
\label{sec:analysis}

\label{sec:SEDfitting}

For the SED fitting, we run the fully Bayesian Markov Chain Monte Carlo (MCMC) based inference code \texttt{PROSPECTOR} (version 1.1) \citep{Leja_17,Johnson_21}. Models with \texttt{PROSPECTOR} are generated on the fly, thus allowing more flexible model specifications with a larger number of parameters, since these will not be as computationally heavy as typical grid-based searches. Additionally, by including nested sampling of the Bayesian posterior probability distribution, \texttt{PROSPECTOR} is able to fully account for possible degeneracies in the stellar population parameters. As a downside, as with any Bayesian-based routine, the dependency of \texttt{PROSPECTOR} results on the assumed priors is strong and the question of whether one is actually learning anything from fitting the data or if the results are simply dominated by the prior assumptions is hard to disentangle. Thus, it is always important to feed the code with informative data and to run models with different prior assumptions to test and understand the dependencies so that plausible interpretations can be made.

The flexibility of \texttt{PROSPECTOR} is complemented by the Flexible Stellar Population Synthesis package \citep[FSPS;][version 0.4.2]{Conroy_09,Conroy_10a,Conroy_10b}, which in turn allows for all stellar population parameters to be potentially free, depending on the user's choice. To sample the posteriors, differently from \cite{Pandya_18} which used the \texttt{emcee} package, we used the dynamic nestled sampling \citep{Skilling_04,Higson_19} algorithm \texttt{dynesty} \citep{Speagle_20}. 

We use the MILES stellar spectral library \citep{SanchezBlazquez_06,Vazdekis_15}, and the Padova isochrones \citep{Marigo_07,Marigo_08} to construct our stellar population models. These models allow us to explore stellar metallicities in the range $-2.0 <$ [$Z$/H] $< 0.2$ dex. The FSPS models used in \texttt{PROSPECTOR} assume solar-scaled abundances (i.e., [$\alpha$/Fe]=0 and [$Z$/X]=[Fe/H]). We note the caveat that the alpha abundances can have a great impact on the colours of galaxies \citep[for a further discussion see][]{Byrne_22}, and thus this assumption may affect our results. Additionally, a \cite{Kroupa_01} initial mass function was assumed for all fits. 

We account for internal dust emission using the \cite{Draine_Li_07} models. However, we do not account for dust emission from AGB stars, differently from \cite{Pandya_18}. We fit the interstellar diffuse dust attenuation using the \cite{Gordon_03} attenuation curve. This dust attenuation law choice was based on \cite{Salim_20} and references therein which suggested that a steeper Small Magellanic Cloud (SMC)-like extinction curve is better suited for dwarf and lower mass galaxies. No significant difference was found in the posteriors using a Milky Way dust law or the one from \cite{Calzetti_00}. We note that FSPS specifically fits for the dust optical depth at 551 nm ($\tau_{551}$), which is the normalisation of the attenuation curve (i.e., $ {\rm I}(\lambda) = {\rm I}_0(\lambda) \times  {\rm exp}(-\tau_{\lambda})$). For simplicity, here we report dust reddening in $A_{V}$, as this is a more commonly used parameter. To do so, we use the optical depth to $A_{V}$ conversion: $A_{V} = 1.086 \times \tau_{551}$ \citep{Spitzer_98,Remy_18}.

For our fits of most galaxies, we include upper limit fluxes coming from the 12 and 22 $\mu$m bands from \textit{WISE} (see Table \ref{tab:photometry}). \texttt{Prospector} implements upper limits by requiring that the SEDs cannot surpass these limits, but these points do not necessarily need to be fitted by the best-fit model \citep[for further description, see Appendix A of][]{Sawicki_12}.

We place very strong priors on the form of the star formation history (SFH). Specifically, we assume an exponentially declining SFH, as this was shown to be better suited for UDGs and low-surface brightness galaxies by \cite{Greco_18} when compared to simple stellar populations.
The assumption of an exponentially declining SFH (or smooth SFH) is accompanied by an assumption of a long star formation timescale and thus does not allow for bursty or stochastic SFHs. However, it does allow for single stellar population (SSP) models in the limit where the e-folding timescale (i.e., the time over which the star formation decreases by a factor of \textit{e}) goes to 0 Gyr. This SFH is similar to those in the literature which use minimization techniques that impose regularization, since these penalize sharp transitions \citep[thus not allowing bursty SFHs, ][]{Ferre-Mateu_18,Ruiz-Lara_18}. A smooth SFH preferentially returns ages of half the age of the universe or younger, and is biased against ancient ages, which are preferred by bursty SFHs.
Because of that, an exponentially declining SFH may be a poor assumption for bursty episodes of star formation expected in supernova feedback scenarios, for example. 
As a final caveat, we note that our exponentially declining SFH does not include metallicity evolution unlike some other SFH models \citep[e.g., CIGALE,][]{Burgarella_05,Noll_09}.

For our particular case, we use four different configurations in \texttt{PROSPECTOR}. 

\begin{enumerate}
    \item A$_V \neq 0; z = z_{\rm spec}$: five free parameters. Stellar mass (log(M$_{\star}$/M$_{\odot}$)), metallicity ([$Z$/H]), age (t$_{\rm age}$), star formation time scale ($\tau$) and diffuse interstellar dust ($A_{V}$). Redshifts ($z$) are fixed to the spectroscopic value in this scenario. The spectroscopic redshifts used are based on the radial velocities (assuming $V_r=cz$, where $c$ is the speed of light) shown in Table \ref{tab:properties} for the galaxies where these are available. If not, we use the distance to the group/cluster. If none of those are available, this configuration is not carried out for that particular galaxy.\\
    
    \item A$_V = 0; z = z_{\rm spec}$: four free parameters (log(M$_{\star}$/M$_{\odot}$), [$Z$/H], t$_{\rm age}$ and $\tau$). We assume no dust ($A_{V}$ fixed to zero) and redshift fixed to the spectroscopic value.\\
    
    \item A$_V \neq 0; z \neq z_{\rm spec}$: six free parameters (log(M$_{\star}$/M$_{\odot}$), [$Z$/H], t$_{\rm age}$, $\tau$, $A_{V}$ and $z$). In this case, we use the galaxies with spectroscopic redshifts to test our ability to estimate their distances and then extrapolate this for the galaxies to which we do not know the distance to.\\
    
    \item A$_V = 0; z \neq z_{\rm spec}$: five free parameters (log(M$_{\star}$/M$_{\odot}$), [$Z$/H], t$_{\rm age}$, $\tau$ and $z$). $A_{V}$ is fixed to zero.
\end{enumerate}

For these scenarios, unless stated otherwise, we placed linearly uniform priors on our free parameters. Those are:  log(M$_{\star}$/M$_{\odot}$) = $6$ -- $10$, [$Z$/H] = $-$2.0 to 0.2 dex, $\tau$ = 0.1--10 Gyr, t$_{\rm age}$ = 0.1--14 Gyr, $\tau_{551}$ = 0--4 (A$_V = 0 - 4.344$ mag) and redshift $z =$ 0--0.045. Assuming a flat $\Lambda$CDM model with H$_0$=70.5 km s$^{-1}$ Mpc$^{-1}$ \citep{Komatsu_09}, this redshift range translates to a luminosity distance range of $0 < D_L < 200$ Mpc. We note that different prior assumptions do not significantly alter our results or conclusions, with the exception of the prior assumption on the shape of the SFH. The effect of this assumption is thoroughly discussed by \cite{Webb_22}.

The aperture stellar masses obtained from \texttt{Prospector} were converted to total stellar masses by using total magnitudes present in the literature and the mass-to-light (M$_{\star}$/L) ratios found by \texttt{Prospector} in the aperture that we performed the photometry. The total magnitudes come from a combination of studies in literature. For the galaxies in the Coma cluster, the total magnitudes come from \cite{Yagi_16} and \cite{Forbes_20a}. For the ones in the Virgo cluster, magnitudes come from \cite{Lim_20}. For the three field LSBG galaxies, magnitudes come from \cite{Greco_18}. For M-161-1, the total magnitude comes from \cite{Dalcanton_97}. For DGSAT I and VCC1122, total magnitudes come from \cite{Pandya_18}. For the galaxies in the NGC 1052 group, total magnitudes are from \cite{Muller_19}. For the ones in the Perseus cluster, magnitudes are from \cite{Gannon_22}. For DDO 190, the total magnitude comes from \cite{DDO_06}. These total stellar masses are further explored in Section \ref{sec:MZR} and presented in Tables \ref{tab:prospector_results} and \ref{tab:prospector_results_redshift} (with dust) and Table \ref{tab:prospector_results_nodust} and \ref{tab:prospector_results_redshift_nodust} (without dust).

\section{Results}
\label{sec:results}

Here we present our results on the stellar population properties of the 29 UDGs studied in this work. Additionally, in Appendix \ref{sec:appendixB} we report results of SED model fitting without the \textit{WISE} and/or the \textit{Spitzer}-IRAC bands, to investigate the impact of adding these infrared bands in the fitting. In Appendix \ref{sec:Appendix_Table}, we report our SED fitting results for the models that do not include dust attenuation.

\subsection{Comparing \texttt{Prospector} configurations (i) and (ii): the effect of having the dust as a free parameter}
\label{sec:results_1_2}

For all the galaxies with confirmed spectroscopic redshifts or that reside in groups/clusters that we know the distance to, we primarily carried out \texttt{PROSPECTOR} SED fitting in the first two scenarios described in Section  \ref{sec:SEDfitting}, i.e., (i) dust as a free parameter and redshift fixed, (ii) dust fixed to zero and redshift fixed.

The results for all 26 UDGs with spectroscopic redshifts are summarised in Tables \ref{tab:prospector_results} (with dust) and \ref{tab:prospector_results_nodust} (without dust), divided according to the environment that the galaxies reside in. The results for the remaining three field UDGs with no spectroscopic redshifts are discussed in Section \ref{sec:redshift}.
The maximum likelihood SED models comparing scenarios (i) and (ii) for DF44 are shown in Fig. \ref{fig:sed_DF44} and for all of the remaining studied galaxies are shown in Appendix \ref{sec:appendixA}. Below we analyse the results presented in Table \ref{tab:prospector_results}, together with the SED fits and corner plots shown in Fig. \ref{fig:sed_DF44} and Appendix \ref{sec:appendixA}.

\begin{figure}
    \centering
    \includegraphics[width=\columnwidth]{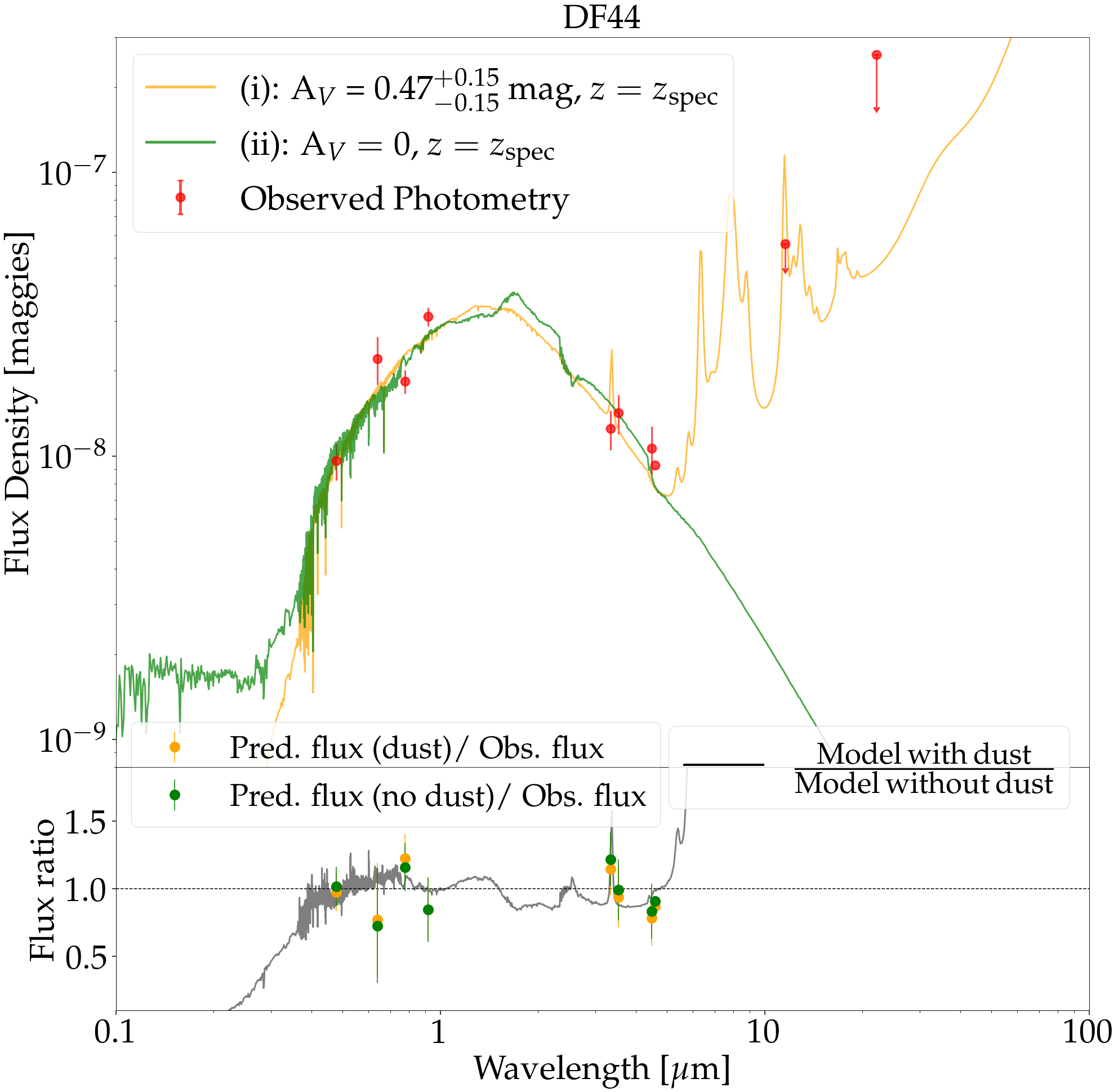}
    \caption{SED fitting results for DF44. \textit{Top:} SED fits comparing Prospector configurations with (i) dust as a free parameter and (ii) dust fixed to zero. Configuration (i) is shown with the yellow curve, the same for configuration (ii) with the green line. \textit{Bottom:} The grey line shows the SED with dust divided by the SED without dust to highlight the differences between the models. Yellow points are the predicted fluxes from the models with dust divided by the observed flux. Green points are the predicted fluxes from the models without dust divided by the observed fluxes.}
    \label{fig:sed_DF44}
\end{figure}

Looking at the SED fits (see Fig. \ref{fig:sed_DF44} and the left-hand side of Figs. \ref{fig:corner_PUDG-R24}--\ref{fig:corner_VCC1052}), a clear difference in the shape of the spectra is seen when comparing models with and without dust. This difference is most clear in the near- and mid-IR wavelength regime, e.g., models with dust have noticeably stronger infrared emission. A difference can also be seen in the optical range, with stronger signs of absorption with dust. Based on these features, the models with dust provide overall better fits to both the optical and IR data. In particular, the corrections for dust in the optical seem to improve significantly the final result (average $\chi^2_{\rm red}=1.1 \pm 0.3$ for the models with dust and $\chi^2_{\rm red}=3.2 \pm 0.3$ for the models without), hinting at the possibility of these galaxies having some amount of dust, even if small. These differences are easier to see in the lower panel of Fig. \ref{fig:sed_DF44} (and all Figs. \ref{fig:corner_PUDG-R24}--\ref{fig:corner_VCC1052}), where we show the ratio of the model with dust divided by the model without dust. We also show the ratio of the predicted fluxes from the models with dust and without dust compared to the observed fluxes for a better comparison of the fits. Models with and without dust are similar in the regions where the ratio (grey line) is close to 1. Similarly, the closer to 1 the ratios of predicted/observed fluxes are, the better the model fits the data. These ratios are proxies for the $\chi^2$ of each fit.  

When looking at the posterior distributions (corner plots found in Appendix \ref{sec:appendixA}, see right-hand side panels of Figs. \ref{fig:corner_PUDG-R24}--\ref{fig:corner_VCC1052}) and best-fit parameters (see Table \ref{tab:prospector_results}), we can see how the fitted parameters change with the inclusion of dust. The aperture stellar masses statistically increase with the inclusion of dust (average of 0.1 dex), since this addition makes the galaxies redder, implying larger mass-to-light ($M_{\star}/L$) ratios. The star formation time scales statistically increase with the inclusion of dust, resulting in slightly more extended star formation histories when dust is allowed. The posterior ages slightly change with the inclusion of dust, on average becoming 1 Gyr younger, which is expected given that fixing the reddening to zero effectively pushes the stellar populations to older ages \citep[as found for early-type galaxies with no dust,][]{Jones_11}. 
The metallicity changes significantly with the inclusion of dust, returning much more metal-poor populations in the scenarios with dust than in those without. 
    
Looking at the corner plots in Appendix \ref{sec:appendixA}, we can see that the age--metallicity degeneracy is broken, as it was shown by \cite{Pandya_18}. Nonetheless, as suggested by the same figures, the metallicity--dust degeneracy still plays a major role in these SED fits. These two parameters are completely correlated, i.e., the more dust that is added the more metal-poor the population becomes. The finding of more metal-poor stellar populations when dust is included can be interpreted as the dust providing an additional source of reddening, and thus a lower metallicity is necessary to counterbalance this extra red component.
Additionally, we show that the upper limits in the mid-IR rule out elevated dust attenuation, i.e., A$_{\rm V} < 0.88$ mag, for every galaxy in our sample.
    
From the results of the models with dust presented in Table \ref{tab:prospector_results}, we find an average age of $7.2\pm1.2$ Gyr for the three galaxies in the field with known spectroscopic redshifts \citep[if we treat PUDG-R24 as a recently accreted field UDG,][]{Gannon_22}. We find an average age of $8.5\pm0.8$ Gyr for galaxies in groups and $9.0\pm1.4$ Gyr for galaxies in clusters. Thus, a correlation between age and environmental density can be seen, i.e., galaxies in the field are on average younger than their cluster counterparts. 
As for the metallicity, in the models with dust, the galaxies in the field display an average metal-poor population with [$Z$/H] = $-1.3\pm0.2$ dex. The ones in groups have an average of $-1.09\pm0.01$ dex and the ones in clusters have an average metallicity of [$Z$/H] = $-1.3\pm0.2$ dex. Thus, we do not observe any statistical difference in the metallicity between UDGs in different environments. See Section \ref{sec:environment} for further discussion on the metallicities of our sample of UDGs.
    
Analysing the star formation time scales, we see that objects in higher density environments have shorter SFHs and are thus consistent with single burst SSPs. Field UDGs, on the other hand, show on average larger $\tau$, thus having more extended and complicated star formation histories. This finding is in agreement with literature findings of the presence of gas and even ongoing star formation in field UDGs \citep{Roman_Trujillo_17, Trujillo_17}.

The average interstellar diffuse dust attenuation coming from the SED fitting of the galaxies in our sample is A$_{\rm V} = 0.35\pm0.21$ mag. No statistically significant difference was found in the dust content between galaxies in our sample residing in different environments.

As for the models without dust (Prospector configuration (ii)), most conclusions remain the same, i.e., the higher the density the higher the age. Also, more extended SFHs were found in the field than in clusters. We see, however, an overall much more metal-rich population, with an average metallicity of $-0.7\pm0.3$ dex (as opposed to an overall average metallicity of [$Z$/H] = $-1.3\pm0.2$ dex in the models with dust). Ages are also systematically older in the models without dust, with an average of $10.6\pm1.3$ Gyr (overall average of $8.7\pm1.4$ Gyr for models with dust). 

We explore the inclusion of dust in the models further in both Sections \ref{sec:redshift} and \ref{sec:SEDuptothetask}.

\renewcommand{\arraystretch}{1.8}
\begin{table*}
\scalebox{0.8}{
\begin{threeparttable}
\caption{\texttt{PROSPECTOR} SED fitting results with dust for galaxies with spectroscopic redshifts.}
\begin{tabular}{c|c|c|c|c|c|c|c} \hline
\multirow{2}{*}{} & \multirow{2}{*}{Galaxy} & \multirow{2}{*}{Configuration} & \multirow{2}{*}{$\log$(M$_\star$/M$_\odot$)} & [$Z$/H] & $\tau$ & Age & $A_{V}$ \\ 
& & & & [dex] & [Gyr] & [Gyr] & [mag] \\ 
& (1) & (2) & (3) & (4) & (5) & (6) & (7) \\ \hline

\multirow[t]{6}{*}{\textbf{Field}} & 

\textbf{PUDG-R24 \textcolor{blue}{(GC-poor)}} & (i): A$_V\neq0$; z = z$_{\rm spec}$ & $8.72^{+0.13}_{-0.15}$ & $-1.12^{+0.38}_{-0.27}$ & $1.33^{+1.90}_{-0.93}$ & $6.32^{+1.86}_{-1.196}$ & $0.32^{+0.19}_{-0.17}$  \\[0.1cm]  
 \cline{2-8}

& \textbf{DGSAT I (No GC info)}  & (i): A$_V\neq0$; z = z$_{\rm spec}$ & $8.65^{+0.11}_{-0.13}$ & $-1.55^{+0.34}_{-0.30}$ & $1.47^{+1.98}_{-1.00}$ & $7.80^{+1.99}_{-2.13}$ & $0.26^{+0.10}_{-0.09}$  \\[0.1cm]  

\cline{2-8}

& \textbf{M-161-1 (No GC info)} & (i): A$_V\neq0$; z = z$_{\rm spec}$ & $8.13^{+0.15}_{-0.19}$ & $-1.17^{+0.19}_{-0.20}$ & $4.38^{+3.33}_{-2.83}$ & $8.95^{+3.48}_{-3.81}$ & $0.09^{+0.10}_{-0.07}$   \\ \hline

\multirow[t]{8}{*}{\textbf{Group}}  &  \textbf{NGC 1052-DF4 \textcolor{blue}{(GC-poor)}}  & (i): A$_V\neq0$; z = z$_{\rm spec}$ & $8.48^{+0.04}_{-0.06}$ & $-1.08^{+0.35}_{-0.22}$ & $0.65^{+0.53}_{-0.38}$ & $8.76^{+2.91}_{-1.51}$ & $0.16^{+0.07}_{-0.06}$  \\[0.1cm]

 \cline{2-8}

&  \textbf{NGC 1052-DF2 \textcolor{blue}{(GC-poor)}} & (i): A$_V\neq0$; z = z$_{\rm spec}$ & $8.48^{+0.15}_{-0.16}$ & $-1.11^{+0.35}_{-0.28}$ & $2.67^{+2.77}_{-1.86}$ & $7.97^{+1.83}_{-2.73}$ & $0.27^{+0.13}_{-0.17}$  \\[0.1cm]  
 \cline{2-8}

& \textbf{DF03 \textcolor{blue}{(GC-poor)}} & (i): A$_V\neq0$; z = z$_{\rm spec}$ & $8.12^{+0.15}_{-0.17}$ & $-1.08^{+0.17}_{-0.18}$ & $4.09^{+3.40}_{-2.59}$ & $9.23^{+2.48}_{-2.73}$ & $0.10^{+0.12}_{-0.07}$  \\  

 \hline

\multirow[t]{44}{*}{\textbf{Cluster}}  &
 
\textbf{DFX1 \textcolor{red}{(GC-rich)}} & (i): A$_V\neq0$; z = z$_{\rm spec}$ & $8.46^{+0.15}_{-0.16}$ & $-1.19^{+0.31}_{-0.22}$ & $3.65^{+3.43}_{-2.30}$ & $8.13^{+2.56}_{-2.50}$ & $0.10^{+0.11}_{-0.07}$ \\[0.1cm]  

 \cline{2-8} 
 
& \textbf{DF26 \textcolor{red}{(GC-rich)}} & (i): A$_V\neq0$; z = z$_{\rm spec}$ & $8.61^{+0.15}_{-0.18}$ & $-1.06^{+0.36}_{-0.30}$ & $2.61^{+3.46}_{-1.82}$ & $8.67^{+2.36}_{-2.47}$ & $0.31^{+0.22}_{-0.19}$ \\[0.1cm]  

 \cline{2-8} 
 
& \textbf{DF02 \textcolor{blue}{(GC-poor)}} & (i): A$_V\neq0$; z = z$_{\rm spec}$ &   $8.31^{+0.13}_{-0.13}$ & $-0.97^{+0.20}_{-0.20}$ & $5.88^{+2.78}_{-3.05}$ & $9.12^{+2.04}_{-2.09}$ & $0.80^{+0.14}_{-0.25}$ \\[0.1cm]  

 \cline{2-8} 
 
& \textbf{DF07 \textcolor{red}{(GC-rich)}} & (i): A$_V\neq0$; z = z$_{\rm spec}$ & $8.60^{+0.15}_{-0.18}$ & $-1.34^{+0.66}_{-0.47}$ & $3.57^{+2.99}_{-2.44}$ & $10.34^{+2.81}_{-3.54}$ & $0.39^{+0.16}_{-0.21}$  \\[0.1cm]  

 \cline{2-8} 

& \textbf{Y358 \textcolor{red}{(GC-rich)}} & (i): A$_V\neq0$; z = z$_{\rm spec}$ & $8.60^{+0.16}_{-0.18}$ & $-1.20^{+0.69}_{-0.54}$ & $4.03^{+3.50}_{-2.70}$ & $10.06^{+2.72}_{-3.75}$ & $0.43^{+0.25}_{-0.25}$ \\[0.1cm] 

 \cline{2-8} 

& \textbf{DF44 \textcolor{red}{(GC-rich)}} & (i): A$_V\neq0$; z = z$_{\rm spec}$ & $8.63^{+0.11}_{-0.12}$ & $-1.53^{+0.36}_{-0.32}$ & $2.62^{+2.52}_{-1.71}$ & $10.98^{+2.21}_{-3.23}$ & $0.47^{+0.15}_{-0.15}$ \\[0.1cm]  

 \cline{2-8} 
 
& \textbf{DFX2 (No GC info)} & (i): A$_V\neq0$; z = z$_{\rm spec}$ & $8.40^{+0.05}_{-0.04}$ & $-1.10^{+0.23}_{-0.21}$ & $8.58^{+1.00}_{-1.61}$ & $6.64^{+1.05}_{-0.47}$ & $0.33^{+0.07}_{-0.07}$  \\[0.1cm]  
 
 \cline{2-8} 

& \textbf{PUDG-R16 \textcolor{blue}{(GC-poor)}}  & (i): A$_V\neq0$; z = z$_{\rm spec}$ & $8.69^{+0.07}_{-0.10}$ & $-1.12^{+0.34}_{-0.25}$ & $0.79^{+0.85}_{-0.47}$ & $9.28^{+1.25}_{-2.10}$ & $0.42^{+0.16}_{-0.18}$ \\[0.1cm]

\cline{2-8} 

& \textbf{DF40 \textcolor{blue}{(GC-poor)}} & (i): A$_V\neq0$; z = z$_{\rm spec}$ & $8.38^{+0.18}_{-0.25}$ & $-1.17^{+0.77}_{-0.59}$ & $5.08^{+3.17}_{-3.07}$ & $8.33^{+3.97}_{-4.16}$ & $0.56^{+0.31}_{-0.30}$ \\[0.1cm]  
\cline{2-8} 

  &  \textbf{DF23 \textcolor{red}{(GC-rich)}} & (i): A$_V\neq0$; z = z$_{\rm spec}$ & $8.10^{+0.17}_{-0.20}$ & $-1.59^{+0.45}_{-0.29}$ & $3.65^{+3.74}_{-2.51}$ & $9.65^{+3.17}_{-3.87}$ & $0.10^{+0.12}_{-0.07}$  \\[0.1cm]  
 
 \cline{2-8}

& \textbf{Y436 \textcolor{red}{(GC-rich)}} & (i): A$_V\neq0$; z = z$_{\rm spec}$ & $8.05^{+0.17}_{-0.17}$ & $-1.34^{+0.25}_{-0.38}$ & $3.08^{+3.42}_{-2.22}$ & $7.60^{+1.72}_{-2.46}$ & $0.40^{+0.08}_{-0.13}$ \\[0.1cm]   
 \cline{2-8} 
 
& \textbf{Y534 \textcolor{red}{(GC-rich)}}  & (i): A$_V\neq0$; z = z$_{\rm spec}$ & $8.17^{+0.16}_{-0.19}$ & $-1.28^{+0.63}_{-0.49}$ & $3.67^{+3.53}_{-2.48}$ & $10.16^{+2.70}_{-3.82}$ & $0.33^{+0.24}_{-0.21}$ \\[0.1cm] 

 \cline{2-8}  
 
 & \textbf{DF17 \textcolor{red}{(GC-rich)}} & (i): A$_V\neq0$; z = z$_{\rm spec}$ & $8.57^{+0.14}_{-0.18}$ & $-1.62^{+0.43}_{-0.26}$ & $3.53^{+3.33}_{-2.45}$ & $9.92^{+2.80}_{-3.67}$ & $0.08^{+0.10}_{-0.06}$ \\[0.1cm]  
 \cline{2-8} 
 
& \textbf{DF25 \textcolor{blue}{(GC-poor)}} & (i): A$_V\neq0$; z = z$_{\rm spec}$ & $8.24^{+0.18}_{-0.23}$ & $-1.14^{+0.66}_{-0.62}$ & $5.01^{+3.20}_{-3.10}$ & $8.36^{+4.02}_{-4.55}$ & $0.62^{+0.43}_{-0.38}$  \\[0.1cm]  
 
 \cline{2-8} 
 
& \textbf{DF08 \textcolor{red}{(GC-rich)}} & (i): A$_V\neq0$; z = z$_{\rm spec}$ & $8.33^{+0.14}_{-0.17}$ & $-1.55^{+0.45}_{-0.32}$ & $2.57^{+3.14}_{-3.56}$ & $10.46^{+2.53}_{-3.56}$ & $0.14^{+0.15}_{-0.10}$  \\[0.1cm]  

 \cline{2-8} 

& \textbf{DF46 \textcolor{blue}{(GC-poor)}} & (i): A$_V\neq0$; z = z$_{\rm spec}$ &  $8.04^{+0.17}_{-0.16}$ & $-1.22^{+0.33}_{-0.20}$ & $3.86^{+3.52}_{-3.60}$ & $8.92^{+2.63}_{-2.67}$ & $0.09^{+0.12}_{-0.06}$  \\[0.1cm]
 
 \cline{2-8}

& \textbf{VCC1287 \textcolor{red}{(GC-rich)}} & (i): A$_V\neq0$; z = z$_{\rm spec}$ & $8.43^{+0.15}_{-0.16}$ & $-1.61^{+0.45}_{-0.28}$ & $3.16^{+3.54}_{-2.27}$ & $10.31^{+2.64}_{-3.54}$ & $0.16^{+0.11}_{-0.09}$ \\[0.1cm] 
   
 \cline{2-8}  

& \textbf{DF06 \textcolor{red}{(GC-rich)}} & (i): A$_V\neq0$; z = z$_{\rm spec}$ & $8.20^{+0.19}_{-0.22}$ & $-1.29^{+0.90}_{-0.52}$ & $5.78^{+2.81}_{-3.30}$ & $5.12^{+5.76}_{-3.83}$ & $0.88^{+0.09}_{-0.18}$  \\[0.1cm]  
 
 \cline{2-8} 

& \textbf{VCC1884 \textcolor{blue}{(GC-poor)}} & (i): A$_V\neq0$; z = z$_{\rm spec}$ & $8.04^{+0.16}_{-0.22}$ & $-1.12^{+0.76}_{-0.62}$ & $4.90^{+3.14}_{-2.91}$ & $9.10^{+3.37}_{-4.15}$ & $0.65^{+0.21}_{-0.28}$ \\[0.1cm]

 \cline{2-8} 
 
& \textbf{VCC1052 \textcolor{blue}{(GC-poor)}} & (i): A$_V\neq0$; z = z$_{\rm spec}$ & $8.13^{+0.17}_{-0.17}$ & $-1.13^{+0.76}_{-0.63}$ & $4.58^{+3.39}_{-2.88}$ & $8.51^{+2.80}_{-2.75}$ & $0.55^{+0.23}_{-0.28}$ \\

 \hline 
 
\end{tabular}
\begin{tablenotes}
      \small
      \item \textbf{Note.} UDGs are separated by the environment that they reside in (see Table \ref{tab:properties}). Columns are: (1) Galaxy ID with GC-richness in parentheses (i.e., rich $geq$ 20 GCs, poor < 20 GCs); (2) \texttt{PROSPECTOR} configuration; (3) Total stellar mass; (4) Metallicity; (5) Star formation time scale; (6) Mass-weighted age; (7) Dust reddening; `--' stands for fixed parameters.
\end{tablenotes}
\label{tab:prospector_results}
\end{threeparttable}}
\end{table*}

\subsection{Comparing \texttt{Prospector} configurations (iii) and (iv): photometric redshifts and SED fitting results for field UDGs}
\label{sec:redshift}

\begin{table*}
\scalebox{0.8}{
\begin{threeparttable}
\caption{\texttt{PROSPECTOR} SED fitting results with dust for galaxies without spectroscopic redshift.}
\begin{tabular}{c|c|c|c|c|c|c|c|c} \hline
\multirow{2}{*}{} & \multirow{2}{*}{Galaxy} & \multirow{2}{*}{Configuration} & \multirow{2}{*}{$\log$(M$_\star$/M$_\odot$)} & [$Z$/H] & $\tau$ & Age & $A_{V}$ & \multirow{2}{*}{z} \\ 
& & & & [dex] & [Gyr] & [Gyr] & [mag] & \\ 
& (1) & (2) & (3) & (4) & (5) & (6) & (7) & (8) \\ \hline

\multirow[t]{10}{*}{\textbf{Field}} & \textbf{LSBG-490 (No GC info)}  & (iii): A$_V\neq0$; $z \neq z_{\rm spec}$ & $8.16^{+0.36}_{-0.47}$ & $-1.19^{+0.72}_{-0.57}$ & $3.09^{+4.28}_{-2.12}$ & $6.04^{+5.12}_{-3.05}$ & $0.54^{+0.22}_{-0.25}$ & $0.018^{+0.013}_{-0.009}$ \\   
  \cline{2-9} 
 
& \textbf{LSBG-378 (No GC info)} & (iii): A$_V\neq0$; $z \neq z_{\rm spec}$ & $8.36^{+0.45}_{-0.51}$ & $-0.91^{+0.63}_{-0.73}$ & $3.12^{+3.92}_{-2.28}$ & $5.72^{+5.19}_{-3.47}$ & $0.26^{+0.20}_{-0.17}$ & $0.016^{+0.014}_{-0.007}$ \\  
  
 \cline{2-9}  
 
& \textbf{LSBG-044 (No GC info)}  & (iii): A$_V\neq0$; $z \neq z_{\rm spec}$ & $8.38^{+0.34}_{-0.45}$ & $-1.04^{+0.57}_{-0.62}$ & $1.22^{+2.83}_{-0.88}$ & $6.59^{+5.21}_{-4.29}$ & $0.29^{+0.22}_{-0.19}$ & $0.032^{+0.013}_{-0.013}$ \\  
  \hline 

\end{tabular}
\begin{tablenotes}
      \small
      \item \textbf{Note.} UDGs are separated by the environment that they reside in (see Table \ref{tab:properties}). Columns are: (1) Galaxy ID with GC-richness in parentheses; (2) \texttt{PROSPECTOR} configuration; (3) Total stellar mass; (4) Metallicity; (5) Star formation time scale; (6) Mass-weighted age; (7) Dust reddening; (8) Redshift. `--' stands for fixed parameters.
\end{tablenotes}
\label{tab:prospector_results_redshift}
\end{threeparttable}}
\end{table*}

Due to the faint nature of UDGs it is impractical to pursue a large campaign of spectroscopic redshifts to establish their true (physical) sizes. One way to overcome this is to estimate the photometric redshift of the galaxies \citep[see][for a different approach on estimating photometric redshifts]{Greene_22}.

To fit the redshifts, we use two \texttt{PROSPECTOR} configurations, as described in Section \ref{sec:analysis}: (iii) dust and redshift as free parameters, (iv) $A_{V}$ fixed to zero and free redshift.

\begin{figure}
    \centering
    \includegraphics[width=\columnwidth]{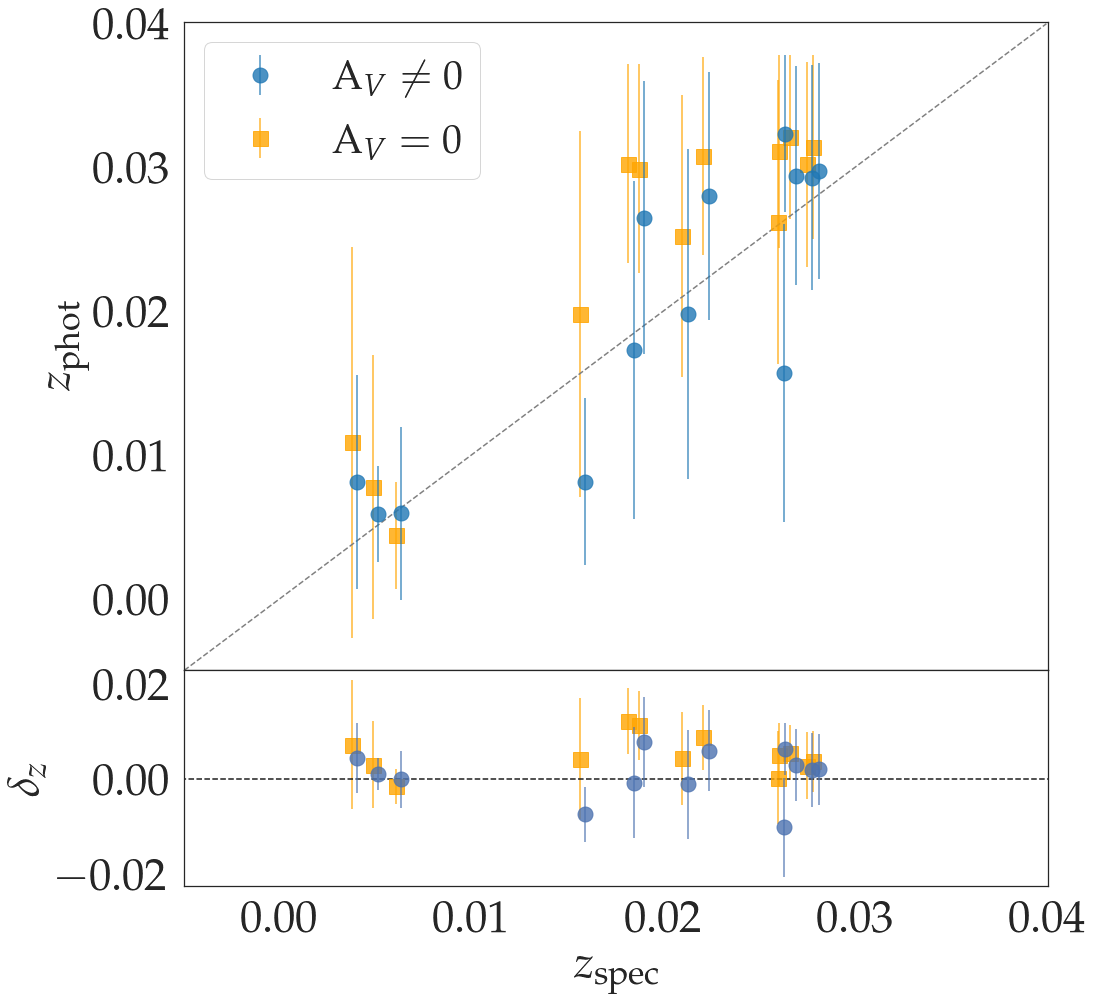}
    \caption{Comparison of redshifts of UDGs obtained via SED fitting and the spectroscopic values known in the literature. Blue points are the results from the SED models with dust as a free parameter. Orange points are the result of the models without dust ($A_{V}$ = 0). The lower panel shows the difference between the redshifts obtained with SED fitting and the values known in the literature ($\delta_z = z_{\rm phot} - z_{\rm spec}$). Small shifts between the orange and blue points were applied for visibility. The models with dust provide better results, having an overall higher accuracy and a smaller offset than the models without dust.}
    \label{fig:comp_distances}
 \end{figure}

In Fig. \ref{fig:comp_distances} we show the comparison between the recovered redshifts in both configurations and the spectroscopic redshifts listed in Table \ref{tab:properties}. 
To measure how well we are recovering the photometric redshifts we use two metrics commonly employed for this purpose \citep{Molino_20, ViniciusLima_22}. These are: 
\begin{enumerate}
    \item Precision:  
    \begin{equation} \sigma_{\text{NMAD}} = 1.48 \times \text{median} \left( \frac{\delta_z - \text{median}(\delta_z)}{1 + z_{\rm spec}} \right).\end{equation}
    \item Mean redshift bias:
     \begin{equation} \mu_{\rm Bias} = \Bar{\delta_z},\end{equation}
\end{enumerate}
where $\delta_z = z_{\rm phot} - z_{\rm spec}$.

We note that although the outlier fraction is another metric used to evaluate the accuracy of the photometric redshift estimates, we do not have a large enough sample to analyse such a metric. 

For the models without dust, we find $\sigma_{\rm NMAD} = 0.05$ and $\mu_{\rm Bias} = -0.03$, while for the models with dust, we find significantly better results with $\sigma_{\rm NMAD} = 0.02$ and $\mu_{\rm Bias} = -0.01$. The $\delta_z$ uncertainty equates to a $\delta(cz)$ recessional velocity uncertainty, which using our assumed Hubble constant, equates to a distance error of the order of 30--50 Mpc.
In fact, we see that when the dust is fixed to zero, there is a systematic behaviour where all galaxies are pushed further away ($z_{\rm phot} > z_{\rm spec}$, with an average distance push of 40 Mpc). We see that the results when the dust is a free parameter are consistently closer to the expected values and do not show such systematic errors. This seems to imply that the code is trying to compensate for the lack of dust by pushing objects further away. We interpret this as an intrinsic redness in the galaxies that can only be explained by the presence of dust or by the galaxy being further distant. We note that the fact that all error bars are large and cross the expected line may indicate that the distance uncertainties are overestimated. 

These results, combined with the discussion in Section \ref{sec:results_1_2}, all seem to indicate that the models with dust are a better representation of the SED of these galaxies. Whether this dust component is simply an artificial addition of the code to deal with systematic errors introduced by fitting old stellar populations \citep{Leja_19} or a real physical property of the galaxies, cannot be disentangled with this dataset. Only further investigation of the mid- and far-infrared emission of UDGs can provide clues to the presence or not of dust. We reiterate that the amount of dust introduced is always small ($A_{V} < 0.9$ mag for every galaxy) and always consistent with zero within 3$\sigma$. See Section \ref{sec:dust} for further discussion on this topic.

\subsubsection{Stellar populations of field UDGs with previously unknown redshifts}

If the redshift is a free parameter, the estimate of the stellar population properties of field UDGs with no spectroscopic redshift becomes possible. In Tables \ref{tab:prospector_results_redshift} (with dust) and \ref{tab:prospector_results_redshift_nodust} (without dust), we provide the recovered stellar populations for the three field UDGs where no spectroscopy is available: LSBG-490, LSBG-378 and LSBG-044.

We note that we classify these galaxies as field ones \citep{Greco_18}, but some of the galaxies found by \cite{Greco_18} turned out to be associated with host galaxies and could alternatively be classified as group ones (Hayes et al. in prep.). We do not have the information of whether the three galaxies included in our sample are associated with any host galaxies, thus we keep the field classification. However, we bear in mind the caveat that it is difficult to classify the environments of field UDGs without a spectroscopic redshift and careful checking against potential host galaxies. See \cite{Polzin_21} for the kind of careful analysis that is required to classify the environment of these galaxies.

In agreement with what was found in Section \ref{sec:results_1_2}, these field galaxies consistently have younger ages (with an average of $6.1\pm0.4$ Gyr in the models with dust) than their cluster counterparts. They are moderately metal-poor in the models with dust with an average metalicity of [$Z$/H] = $-1.0\pm 0.1$ dex. 

With the recovered redshifts, all of the galaxies in our sample meet the size criteria ($R_{\rm e} > 1.5$ kpc) to be classified as UDGs.
This method of recovering photometric redshifts and stellar populations of galaxies with unknown distances sets a pathway to building up a statistically significant sample of population properties of UDG candidates across the sky.

\section{Discussion}
\label{sec:discussion}
\subsection{Is SED fitting recovering reliable stellar population properties of UDGs?}
\label{sec:SEDuptothetask}

\begin{figure*}
    \centering
    \includegraphics[width=\textwidth]{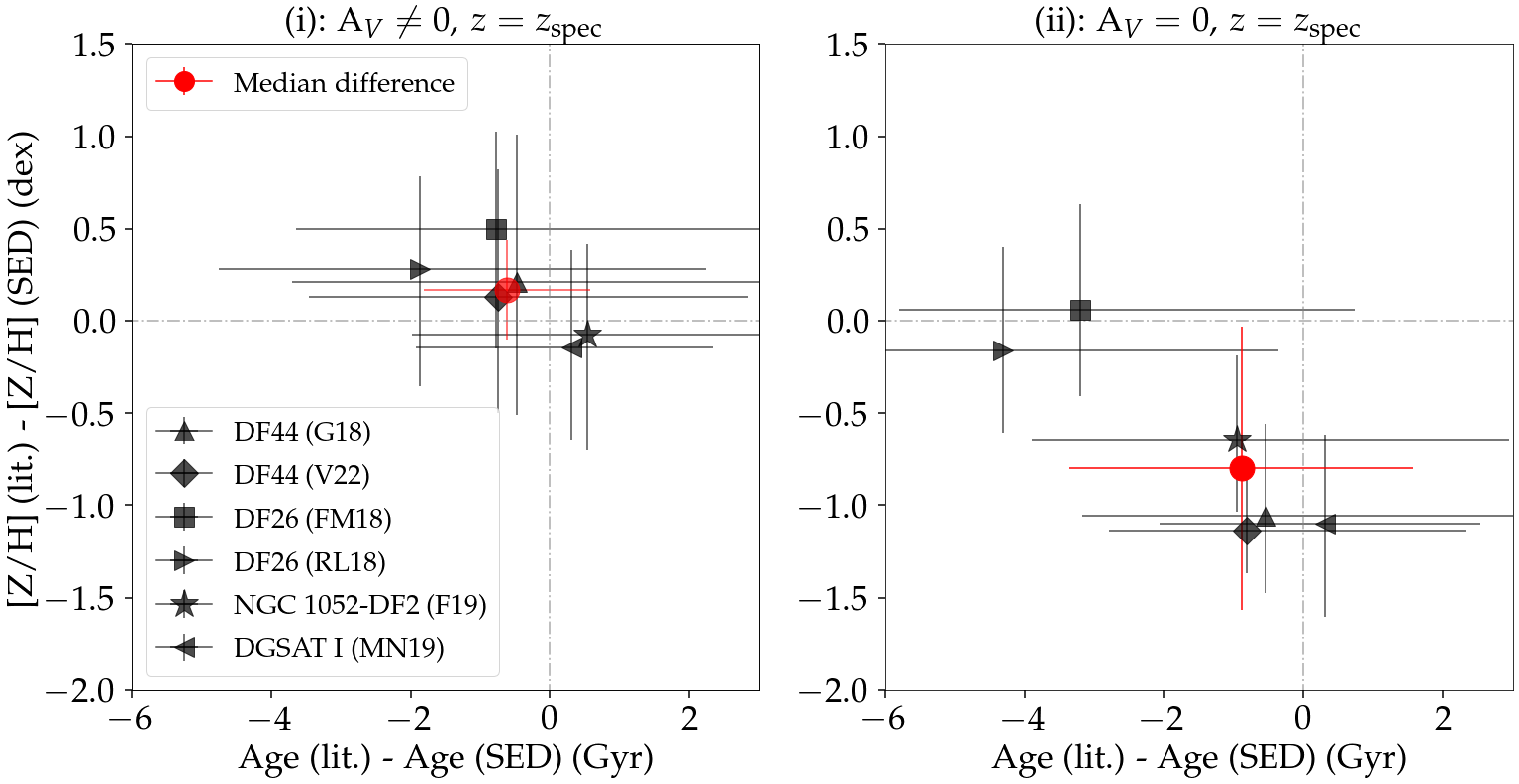}
    \caption{ Comparison of stellar population properties obtained in the current study with those in the literature. Left: models with dust as a free parameter. Right: models with no dust. Both plots compare the age (x-axis) and metallicity (y-axis) between the results obtained with SED fitting in our current study and those obtained with spectroscopy in the literature. The different markers stand for the different studies in the literature being compared: triangle \protect\citep[DF44 from ][]{Gu_18}, diamond \protect\citep[DF44 from ][]{Villaume_22}, square \protect\citep[DF26 from ][]{Ferre-Mateu_18}, left-facing triangle \protect\citep[DF26 from ][]{Ruiz-Lara_18}, star \protect\citep[NGC 1052-DF2 from ][]{Fensch_19}, and right-facing triangle \protect\citep[DGSAT I from ][]{MartinNavarro_19}. Red filled circles are the median difference between all the studies being compared. We conclude that the models with dust provide better results for the recovered stellar population properties of UDGs.}
    \label{fig:comparison_literature}
\end{figure*}

Among our sample galaxies, four were previously studied with spectroscopy in the literature and thus provide a great test to see if our results can reproduce the galaxies' properties or not. These galaxies include the well studied Coma cluster UDG DF44, as well as another Coma cluster UDG, DF26. The third one is NGC 1052-DF2, a galaxy located in or near the NGC 1052 group and hence a great comparison to the other two located in high-density environments. As a final comparison, we use the field UDG DGSAT I to test even further the stellar population dependence on the environment. We note that DF17 and DF07 were also studied in the literature \citep{Gu_18} and could be included in our comparisons. However, the metallicities of these galaxies are quoted in [Fe/H] by \cite{Gu_18}, and differently from DF44, we do not have the alpha abundance information \citep{Villaume_22} to correctly convert their [Fe/H] values into [$Z$/H] to properly compare them to our results. Thus we do not include these galaxies in our comparisons. We do not plot or compare our results with those of \cite{Kadowaki_17} for the same reason.

In Fig. \ref{fig:comparison_literature}, we show the comparison of the main stellar population properties (e.g., age and metallicity) of these four galaxies with the literature values, where we have implemented models with and without the inclusion of dust. In the next subsections we briefly discuss our findings for each of these UDGs, comparing the spectroscopic results to those reported both in Fig. \ref{fig:comparison_literature} and in Tables \ref{tab:prospector_results} and \ref{tab:prospector_results_nodust}.

\subsubsection{DF44}

We provide a comparison of the stellar population properties recovered for DF44 with two spectroscopic studies in the literature, \citealt{Gu_18} (hereafter, G18) and \citealt{Villaume_22} (hereafter, V22).
The results of G18 and V22 are very similar, although obtained using very different datasets. 
G18, using MaNGA/SDSS data, found that DF44 has an old stellar population, with an age of 10.47$^{+1.29}_{-1.74}$ Gyr and an iron content of [Fe/H] = --1.25$^{+0.33}_{-0.39}$ dex (which we convert to [$Z$/H] using the alpha abundance provided by V22: [$Z$/H]$=-1.33^{+0.33}_{-0.39}$ dex).
V22, using much deeper spectra from Keck/KCWI, derived a slightly younger stellar population, with an age of 10.23$^{+0.73}_{-0.90}$ Gyr, an iron abundance of [Fe/H] = --1.33$^{+0.05}_{-0.04}$ dex and an alpha abundance of [Mg/Fe] = --0.10$^{+0.06}_{-0.06}$, resulting in a total stellar metallicity of [$Z$/H] = $-1.41 \pm 0.08$ dex. 
Both studies are consistent within the uncertainties, leading to the conclusion that DF44 is primarily comprised of old and metal-poor stellar populations.

In our SED fitting setup, we have recovered for the models without dust a slightly older and much more metal-rich stellar population than the studies in the literature as it can be seen in Table \ref{tab:prospector_results_nodust}.
On the other hand, in the models with dust (Table  \ref{tab:prospector_results}), we see that our results are much closer to those found in the literature for both the age and metallicity, being strongly consistent with both G18 and V22.

These results show that the inclusion of a dust reddening component as small as $A_{V} \sim 0.5 \pm 0.2$ mag brings the stellar populations much closer to those found in the literature with spectroscopy, especially when taking the metallicity into consideration.

As a final comment, we note the recent work of \cite{Webb_22}. They fitted DF44 with \texttt{Prospector} using KCWI spectra together with ultraviolet to near-IR photometry (including some of the same photometry as we used in the current paper). Using two different star formation histories, they found that if an exponentially declining SFH is assumed, DF44 is consistent with an age of $10.65 \pm 0.2$ Gyr, a metallicity of [$Z$/H] = $-1.20 \pm 0.01$ dex, and a dust content of $A_V = 0.51^{+0.38}_{-0.62}$ mag (and thus consistent with our findings in this paper). If, on the other hand, a bursty SFH is adopted, DF44 may be much older and even consistent with an ancient age of $13.04 \pm 0.05$ Gyr. This emphasizes the notion that the choice of the SFH can have a significant impact on the SED fitting results and that our assumption of an exponentially declining SFH may be accompanied by many caveats, as previously discussed in Section \ref{sec:analysis}. 

\subsubsection{DF26}
DF26 was studied with spectroscopy by \citealt{Ferre-Mateu_18} (hereafter, FM18) and by \citealt{Ruiz-Lara_18} (hereafter, RL18).
FM18 and RL18 found that DF26 hosts a slightly younger and more metal-rich stellar population when compared to DF44. 
FM18 found, using Keck-DEIMOS spectra, that DF26 has a luminosity-weighted age of $7.9 \pm +1.8$ Gyr, and an overall metallicity of [$Z$/H] = $-0.56 \pm 0.16$ dex. 
RL18, using the OSIRIS spectrograph on the Gran Telescopio de Canarias, derived a luminosity-weighted age of 6.8$^{+1.20}_{-1.23}$ Gyr and a luminosity-weighted metallicity of [$Z$/H] = --0.78$^{+0.08}_{-0.08}$ dex.
\texttt{PROSPECTOR} only delivers mass-weighted ages and metallicities, making the comparison with these works harder. We note though that FM18 stated that the mass-weighted ages for the objects studied by them would be expected to be 1-2 Gyr older than the luminosity-weighted ones.

We can see that for the models without dust we find a large difference in age, but the metallicity is similar to that reported by RL18 and FM18.
When looking at the models with dust, on the other hand, the ages are much closer to the expected, but the metallicities have a larger discrepancy. However, both results with and without dust are consistent with spectroscopy within the uncertainties.

\subsubsection{NGC 1052-DF2}
\label{sec:ngc1052-df2}
NGC 1052-DF2 was studied using VLT/MUSE data by \citealt{Fensch_19} (hereafter, F19). They found an age of $8.9 \pm 1.5$ Gyr, and a metallicity of [$Z$/H] = $-1.07 \pm 0.12$ dex.

For this galaxy, when looking at its recovered stellar populations with SED fitting in Tables \ref{tab:prospector_results} and \ref{tab:prospector_results_nodust}, we see that the age difference for both models with and without dust is small and consistent with the results reported in the literature. However, the metallicity output for the models without dust is strongly different from the one obtained with spectroscopy (although consistent within 1$\sigma$), while the model with dust attenuation delivers much closer results.

\subsubsection{DGSAT I}
DGSAT I stellar populations were recovered using both spectroscopy \citep[][hereafter MN19]{MartinNavarro_19} and SED fitting with \texttt{PROSPECTOR} \citep{Pandya_18}.
\cite{Pandya_18} also ran models with \texttt{PROSPECTOR} with the addition or not of interstellar diffuse dust, which we discuss below, where their results without dust are included inside parentheses in the following. They found an age of $6.81^{+4.08}_{-3.02}$ ($7.12^{+3.79}_{-2.79}$) Gyr, a metallicity of [$Z$/H] = $-0.63^{+0.35}_{-0.62}$ ($-0.27^{+0.25}_{-0.22}$) dex, and a dust reddening of $A_{V} < 0.26$ mag.

MN19 have studied DGSAT I with spectroscopy, finding a mass-weighted age of $8.1 \pm 0.4$ Gyr and a metallicity of [$Z$/H] = $-1.7 \pm 0.4$, with an incredibly high alpha enhancement ([Mg/Fe] $\sim +1.5$ dex).

Similarly to what was found by \cite{Pandya_18}, we can see that the addition of dust in our models brings the metallicities to a lower level, indicating again that the included dust is ``absorbing'' part of the redness of the images and only more metal-poor stellar populations can explain the observed colours of these UDGs.

DGSAT I has quite an unusual chemical abundance, as suggested by MN19, with a [Mg/Fe] enhancement 10 times higher than the most chemically enriched systems studied to date. This unexpected chemical abundance can be connected to the recently detected blue and irregular low surface brightness clump on top of DGSAT I's disk. This clump was associated, using \textit{Hubble Space Telescope} data, with a recent ($\sim$500 Myr) episode of star formation in the galaxy \citep{Janssens_22}. This is in agreement with the finding of an extended star formation history for DGSAT I by \cite{MartinNavarro_19}. This alpha enhancement in DGSAT I is difficult to detect with broad-band SED fitting techniques, as these are not focused on specific absorption lines and thus only the overall shape of the spectrum can be recovered. This limitation might explain why \cite{Pandya_18} found a much more metal-rich population than MN19. However, in this work, with the inclusion of the mid-IR \textit{WISE} bands and using a more recent version of \texttt{Prospector}, we find metallicities much closer to those found with spectroscopy, meaning that the inclusion of these bands and a broader coverage of the spectrum may be key to separating the metallicity and alpha enhancement of the galaxies. Another explanation for the differences may be that \cite{Pandya_18} found an optical colour for DGSAT I inconsistent with the one used in our study coming from \cite{Janssens_22} and the one found by \cite{MartinezDelgado_16}.

In fact, as discussed in Appendix \ref{sec:appendixB}, we see that the exclusion of the \textit{WISE} bands has a strong effect on the estimate of the dust extinction and metallicity of the galaxies. With the \textit{WISE} bands, the dust posterior peaks at smaller values because the 12 and 22 $\mu$m upper limits help to constrain the amount of dust found. Because of the dust--metallicity degeneracy, we also find a much more constrained estimate of the metallicity.

\subsubsection{VCC1287}

Although VCC1287 was not previously studied with spectroscopy, in this Section we provide a comparison of the stellar population properties recovered for it in this study and in the one of \cite{Pandya_18}, both using \texttt{Prospector} and models with and without dust attenuation.
We discuss their results below, with their recovered parameters for the model without dust included inside parentheses.
\cite{Pandya_18} found an age for VCC1287 of $> 8.66$ ($> 7.74$) Gyr, a metallicity of [$Z$/H] $<-1.55$ ($-1.56^{+0.52}_{-0.19}$) dex, and a dust reddening of $A_{V} < 0.16$ mag.

Since \cite{Pandya_18} quoted only lower limits for their recovered ages and metallicities, it is hard to fully compare our results. Bearing in mind this caveat, we show (as per Tables \ref{tab:prospector_results} and \ref{tab:prospector_results_nodust}) that our results are consistent with those found by them, with a higher agreement in the models with dust than in those without.

\subsubsection{Median difference between SED fitting results and spectroscopy}

As discussed above, the parameters recovered with the setup where the dust content is free are closer to the ones obtained with spectroscopy. To provide a final comparison and assess which results better reproduce the stellar populations of the UDGs, we provide the median age and metallicity difference for the models with and without dust in Fig. \ref{fig:comparison_literature}.
For the SED results with dust, we find a median difference of 0.8 Gyr in age and 0.1 dex in metallicity.
For the case of the SED models with the dust fixed to zero, we see a higher discrepancy between our results and all of the ones used for comparison in literature, reaching a median difference of 1.3 Gyr in age and 0.9 dex in metallicity.

These tests effectively demonstrate that the results with dust are better if we take the spectroscopic results as the baseline. Although all recovered stellar populations (with and without dust) are consistent with the literature within uncertainties, the inclusion of dust as a free parameter seems to better constrain the stellar population parameters, delivering results on average closer to those expected.

\subsubsection{Dust in UDGs?}
\label{sec:dust}

While the studies of \cite{Pandya_18} and \cite{Barbosa_20} (mean $A_V = 0.1$ mag) both hint at the possibility of the presence of some dust in UDGs, neither included bands in the mid and far-IR to test its presence. 
When we look into the values of $A_V$ obtained for all of our sample of galaxies (as shown in Tables \ref{tab:prospector_results}, \ref{tab:prospector_results_redshift}, \ref{tab:prospector_results_nodust} and \ref{tab:prospector_results_redshift_nodust}), they are small (${A}_{V} < 0.88$ mag), but not particularly reliable, given that they are coming from upper limits in the \textit{WISE} photometry, rather than proper detections. The true nature of dust in these galaxies, or the actual amount of dust present in each one of them, is beyond our capabilities with this dataset. However, it is interesting to note that the models with dust provide on average smaller reduced $\chi^2$ and the presence of dust, as discussed in Sections \ref{sec:results} and \ref{sec:SEDuptothetask}, brings the age and metallicity (and the redshift, as discussed in Section \ref{sec:redshift}) closer to what was found spectroscopically. This implies that regardless of whether the presence of dust is real or not, the code needs to include this small amount of dust to properly recover the properties of the galaxies. 

Furthermore, we note the study of \cite{Pandya_18}, which besides fitting the two previously mentioned UDGs, fitted a dwarf elliptical galaxy, VCC1122. This is a brighter dwarf where the \textit{Spitzer}-IRAC 8.0 $\mu$m data were available and with enough signal-to-noise for a dust detection if it were present in the galaxy, and the dust recovered was nearly zero. Without the \textit{Spitzer}-IRAC 4.5 $\mu$m band, there was dust inferred, similar to what we find in this study. Their findings show that with the inclusion of the \textit{Spitzer}-IRAC 4.5 $\mu$m band has two effects: 1) the recovered dust content goes to nearly zero and 2) the galaxy gets more metal-rich. This raises the question whether the lack of the \textit{Spitzer}-IRAC 4.5 $\mu$m band is driving higher amounts of dust and thus decreasing the recovered metallicities. For the sake of comparison, we fit the same dwarf elliptical galaxy in this study. This is further discussed in Section \ref{sec:MZR}.

Cluster UDGs are known to have little-to-no ongoing star formation activity \citep{Ferre-Mateu_18}. Due to the relatively short timescale of dust survival we therefore do not expect UDGs to harbour a significant dust content. This is similar to what is observed for early-type galaxies and/or dwarf ellipticals \citep{Jones_04,Jones_11}. However, there are many reasons why a small amount of dust could be part of the galaxies. The first one, of course, is that some galaxies indeed have dust.
On the other hand, \cite{Leja_19} have hinted at the possibility that there may be systematic errors introduced by fitting old stellar populations using SED fitting techniques while assuming a specific SFH shape for the galaxies or in the underlying SSP models. These systematics may be incorporated in the dust component, and thus its addition to the models delivers truer stellar population properties.
Additionally, this finding could be related to incorrect Galactic dust corrections rather than dust internal to the galaxies, but we note that we find no correlation between the Galactic $A_{V}$ of the galaxies and their intrinsic dust reddening from \texttt{PROSPECTOR} fits. 

Also, we note that when fitting Milky Way globular clusters without correcting for Galactic reddening, \cite{Johnson_21} found dust posteriors consistent with the literature. However, the dust values that \cite{Johnson_21} found are systematically higher than the literature by $\sim$0.2 mag on average. This demonstrates that \texttt{PROSPECTOR}'s $A_{V}$ posteriors may not be reflecting real properties of the sources, but rather again could be an artificial addition of the code to better fit the data. Lastly, we note the recent work of \cite{Janssens_22}. They found from the small spread in the GC colours, that DGSAT I is consistent with having a very small amount of dust, much smaller than the one inferred in this work. The same conclusion can be drawn for NGC 1052-DF2 and NGC 1052-DF4 based on the monochromatic GC population found by \cite{vanDokkum_22b}. This again suggests that this amount of dust added may be not a sign of physical dust but rather an artificial addition of the code.

To further test this hypothesis, we have fitted one galaxy in our sample, NGC 1052-DF4, with two other SED fitting codes, \texttt{Bagpipes} \citep{Carnall_18} and \texttt{Cigale}. We use again two different setups, one with dust and one without. Independently of the code, we find that the resulting stellar populations are more metal-poor once dust is allowed, similarly to what is found with \texttt{PROSPECTOR}. The reddening found with \texttt{Cigale} for NGC 1052-DF4 was $A_{V} = 0.12$ mag, and 0.08 mag with \texttt{Bagpipes}, both within the uncertainties of the $A_{V} = 0.16 \pm 0.07$ mag result with \texttt{PROSPECTOR}. 
This leads to the conclusion that either SED fitting techniques in general may need this dust addition in order to properly recover the stellar populations of old, metal-poor galaxies (especially such faint ones as UDGs), or the dust inferred is real. The addition of data in the far-IR or radio regimes would be able to robustly test for the presence of dust in such galaxies, or even deeper IR data would be useful to put more stringent upper limits.

Bearing in mind the caveats for dust in UDGs discussed thus far, we conclude that the models with dust provide better results and thus, from this point on, we use and discuss our \texttt{Prospector} results only for the models with dust attenuation.

\subsection{Stellar population dependence on environment and GC-richness}
\label{sec:environment}

\begin{figure*}
    \centering
    \includegraphics[width=\columnwidth]{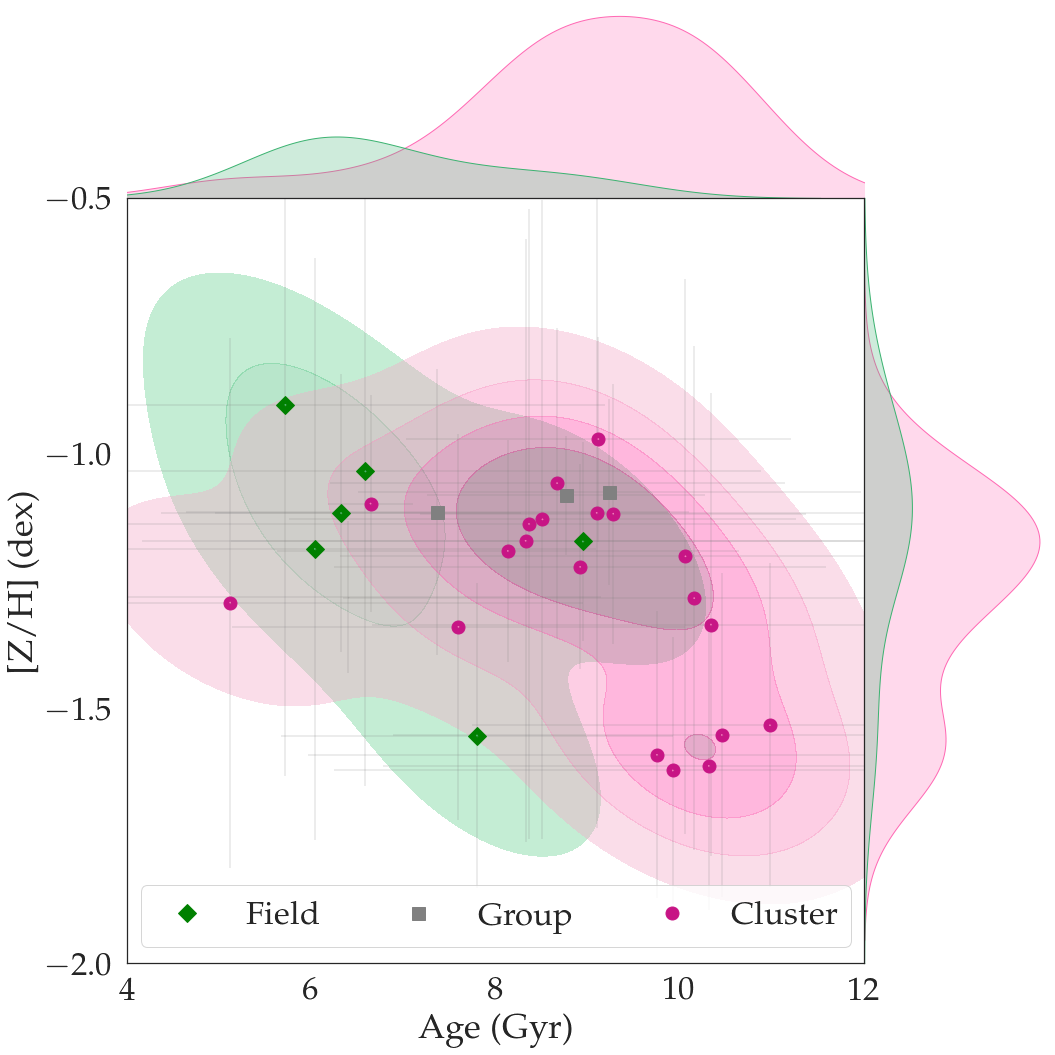}
    \includegraphics[width=\columnwidth]{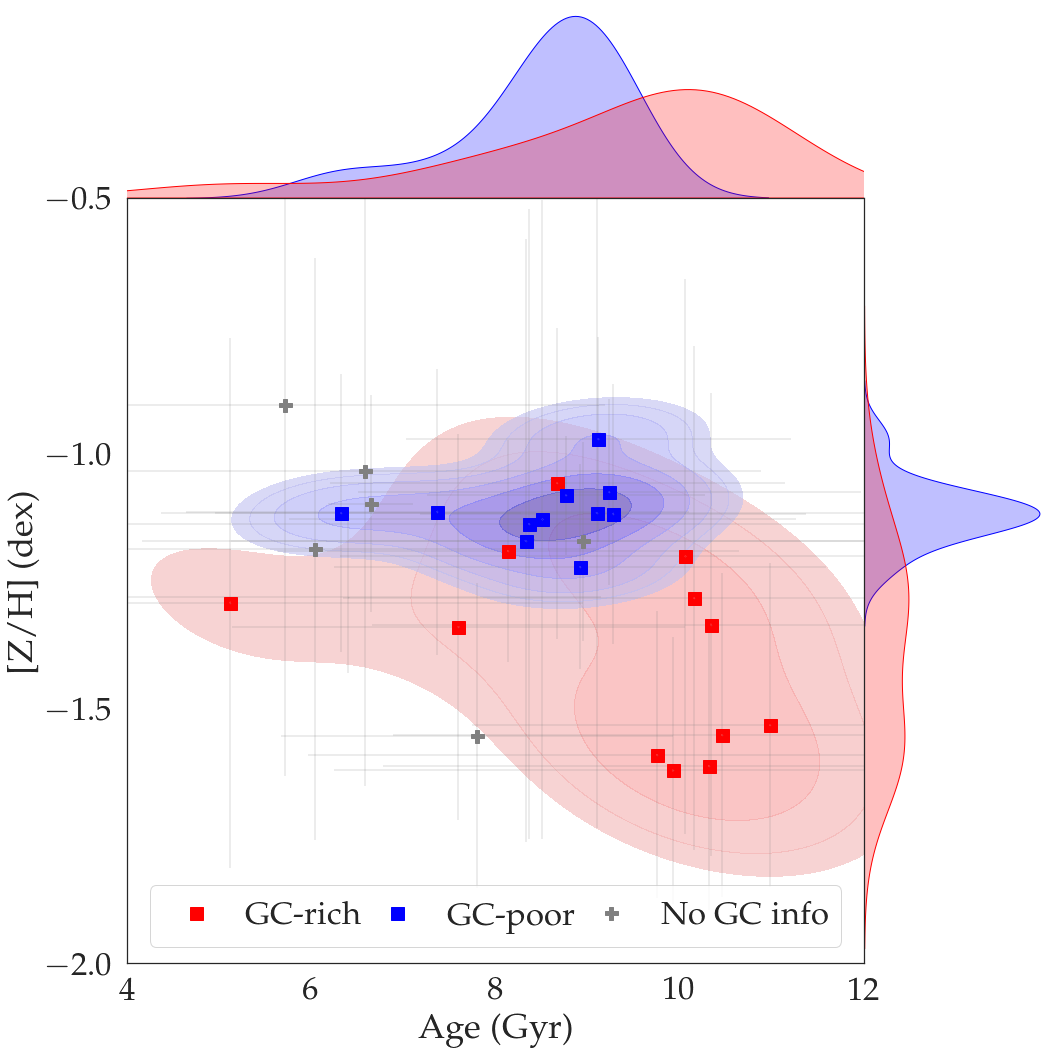}
    \caption{Comparison of stellar population properties against the environment and GC-richness of the UDGs for the models with dust included. Left: Magenta contours show the density of the magenta points and stand for the stellar populations of the cluster UDGs in our sample. Green contours are derived from the density of green points and show the age and metallicity of field UDGs. Grey squares are group UDGs in the sample. Right: Blue contours show the stellar populations of the GC-poor UDGs in our sample, while red contours are the GC-rich UDGs. Marginal smoothed histograms show the distribution of age (above) and metallicity (right) for the different UDG populations (i.e., cluster vs. field and GC-poor vs. GC-rich). We find evidence of a stellar population dependence with the environment and GC-richness. Field UDGs are younger than cluster ones, while GC-poor UDGs are systematically more metal-rich than the GC-rich ones.}
    \label{fig:comp_environment}
\end{figure*}

Cluster UDGs have been shown to be old ($\sim$10 Gyr) and metal-poor objects \citep{Ferre-Mateu_18, Gu_18, Chilingarian_19, Ruiz-Lara_18, Kadowaki_17}. UDGs in low-density environments, on the other hand, have been shown to host younger ($\sim$7 Gyr) stellar populations \citep{Rong_17,Roman_Trujillo_17,Pandya_18,Barbosa_20}. Many of them have current star formation, with divergent findings with respect to whether they are more or less metal-rich than their cluster counterparts \citep[e.g., ][]{Barbosa_20}. 
Additionally, \cite{Ferre-Mateu_18} showed that there exists an age dependence with projected clustocentric distance, i.e., UDGs are younger at larger projected clustocentric radii. Similarly, \cite{Alabi_18} and \cite{Kadowaki_21} found a colour dependence with the environment, with bluer UDGs residing in lower-density environments than the redder ones. This trend also applies for normal dwarf and giant galaxies and has been studied for decades now \citep{Dressler_80, Thomas_10,Tiwari_20}.

One of our goals in this work is to probe the stellar population properties of the UDGs in our sample and to test if by using SED fitting alone we can distinguish between UDGs that live in different environments. 

On the left panel of Fig. \ref{fig:comp_environment}, we show the 2D distribution of stellar population properties (age/metallicity) of the UDGs we studied colour-coded by the environments that they reside in. The density contours shown in the plots were derived using a kernel density estimate. They represent the data by using a continuous probability density curve in the two dimensional plane. We can see that, in agreement with the findings of \cite{Pandya_18} and \cite{Ferre-Mateu_18}, quenched UDGs in field environments are systematically younger (mean age$_{\rm Field}$ = $6.7 \pm 1.1$ Gyr) than the cluster ones (mean age$_{\rm Cluster}$ = $9.0 \pm 1.4$ Gyr), although these populations are still consistent within the uncertainties. We do not find any clear metallicity dependence on the environment. We note that we do not have a large enough number of group UDGs in our sample to comment on trends in this ``transitional'' density environment.

Additionally, on the right hand panel of Fig. \ref{fig:comp_environment}, we show the UDGs in the metallicity--age plane colour-coded by their GC-richness (see Table \ref{tab:properties}). This is the first time that the stellar populations of UDGs have been investigated according to their GC-richness, and it is interesting to see that the GC-poor UDGs are consistently more metal-rich (average [$Z$/H]$_{\rm GC-poor} = -1.1 \pm 0.1$ dex) than their GC-rich counterparts (average [$Z$/H]$_{\rm GC-rich} = -1.4 \pm 0.2$ dex).
Although some of the GC richness classifications are uncertain (as discussed in Section \ref{sec:data}), a clear distinction in the metallicity of the two populations is observed.
We note the study of \cite{Ferre-Mateu_18} and, although they did not look directly at the GC-richness of their sample of UDGs, they stated that none of the UDGs had GC-like stellar populations. In particular, they found that none of the UDGs was older than 10 Gyr, which seems to disagree with some of our results. However, as discussed in Section \ref{sec:introduction}, spectroscopic studies are biased to the brightest galaxies and thus there may be a selection effect in the UDGs studied by \cite{Ferre-Mateu_18}, which may explain why they did not find any UDGs with old, GC-rich populations.

This separation in metallicity between the GC-poor and the GC-rich UDG populations is further explored in Section \ref{sec:MZR}.

\subsection{Scaling relations: Clues to the origins of UDGs}
\label{sec:MZR}

\begin{figure*}
    \centering
    \includegraphics[width=\textwidth]{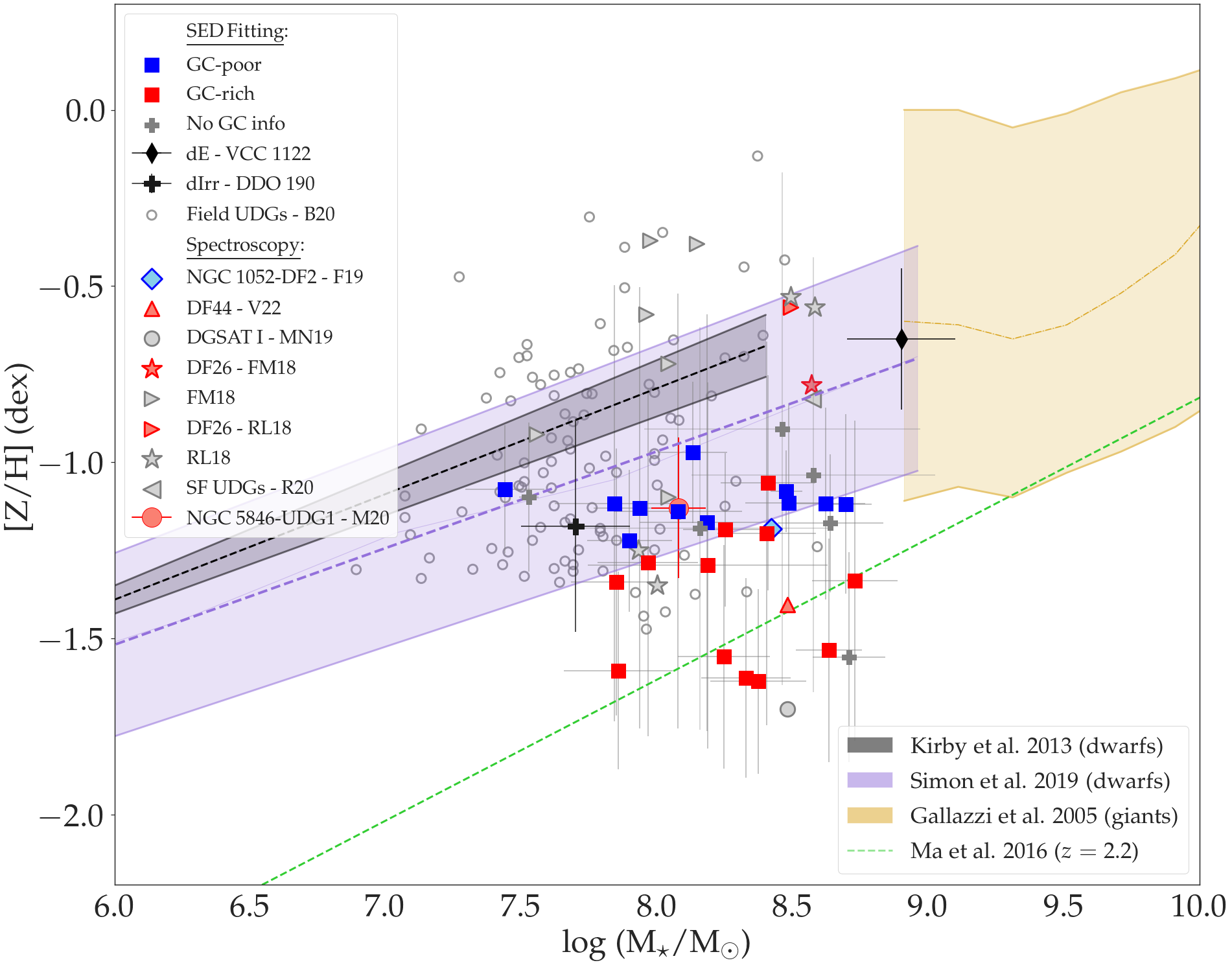}
    \caption{Stellar mass--metallicity distribution of UDGs. Squares show the results obtained from SED fitting with \texttt{PROSPECTOR} for the 29 UDGs in our sample. UDGs are colour-coded by their GC-richness. Blue colours stand for GC-poor UDGs, red are GC--rich UDGs and grey points are the UDGs for which we do not have any GC information. The diamond is the metallicity and stellar mass of NGC 1052-DF2 as obtained by \protect\cite{Fensch_19}. The circle shows the results for DGSAT I obtained by \protect\cite{MartinNavarro_19}. The triangle stand for the results obtained for DF44 by \protect\cite{Villaume_22}. Right-facing triangles are the results for the UDGs analysed by \protect\cite{Ferre-Mateu_18}. The red circle shows the results obtained for NGC 5846-UDG1 (assuming [$\alpha$/Fe]=0.3) by \protect\cite{Muller_20}. Stars are results from \protect\cite{Ruiz-Lara_18}. Left-facing triangle are the results for the star forming UDGs analysed by \protect\cite{Rong_20}. These UDGs from the literature are all from spectroscopic studies and there are some duplicated galaxies in the plot (DGSAT I, DF26, NGC1052-DF2 and DF44).
     Small circles are field UDGs analysed by \protect\cite{Barbosa_20}. The black thin diamond is our fit with \texttt{Prospector} of the dwarf elliptical VCC1122. The black plus sign is the local group dIrr DDO 190.  The \protect\cite{Kirby_13} MZR for dwarf galaxies is shown with the black dashed line. The \protect\cite{Gallazzi_05} MZR for giant galaxies is shown with the dash-dotted golden line. The \protect\cite{Simon_19} MZR for dwarf galaxies is shown with the purple dashed line. The dashed green line is the evolving MZR at redshift $z=2.2$ from \protect\cite{Ma_15}. We conclude that GC-poor UDGs have on average higher metallicities, falling within the MZR for dwarf galaxies. This indicates that a puffed-up dwarf scenario may be appropriate for them. On the other hand, some GC-rich UDGs exhibit extremely metal-poor stellar populations and thus may be better explained by a failed galaxy scenario.}
    \label{fig:MZR}
\end{figure*}

We explore in this section the positioning of our UDG sample on the stellar mass -- metallicity relation \citep[MZR, ][]{Gallazzi_05,Kirby_13,Simon_19} as compared to non-UDGs.

Since the output metallicities from \texttt{Prospector} are in total stellar metallicities, i.e., [$Z$/H], we applied a correction to the \cite{Kirby_13} and \cite{Simon_19} relations, originally in [Fe/H], of +0.3 dex. These relations were derived by measuring the metallicities of individual stars in nearby dwarf galaxies. To find this correction, we used the conversion between [$Z$/H] and [Fe/H] from \citep{Vazdekis_15}:

\begin{equation}
[Z/{\rm H}] = [{\rm Fe/H}] + 0.75[\alpha{\rm /Fe}].
\label{eq:kirby}
\end{equation}

We use the published values of [$\alpha$/Fe] and [Fe/H] in \cite{Kirby_20} for five dwarf spheroidal galaxies around M31 (which were part of the initial sample used to derive the \cite{Kirby_13} relation) to fit the MZR in both [Fe/H] and [$Z$/H]. We found that the slope of the two curves is the same, but there is a shift of $0.3 \pm 0.03$ dex between them, culminating in the conversion we applied. Similarly, \cite{Simon_19}, using data on several local group dwarfs have found that these galaxies have on average alpha abundances of 0.3 dex. Again, applying Eq. \ref{eq:kirby} translates to an average shift of 0.23 dex when plotting [$Z$/H] instead of [Fe/H]. To plot the MZR from \cite{Simon_19}, we had to convert their values from log(L$_{\rm V}$/L$_{\odot}$) to log(M$_{\star}$/M$_{\odot}$) and refit the relation. To do this, we assumed an average mass-to-light (M$_{\star}$/L) ratio of 2 (i.e., suitable for old stellar populations and for dwarf ellipticals and spheroidals, \citealt{Kirby_13}) and fitted a curve to the newly converted values. An average (M$_{\star}$/L) of 1.8 was found for the UDG in our sample using \texttt{Prospector}, reinforcing that this choice of (M$_{\star}$/L) is appropriate. The linear relation is best parameterised by:

\begin{equation}
    [Z/{\rm H}] = (0.27 \pm 0.02) \times \log({\rm M}_{\star}/{\rm M}_{\odot}) - (3.16 \pm 0.14). 
\end{equation}

The plotted relation reflects what was mentioned in \cite{Simon_19} that they found a scatter in metallicity that was 0.25 dex larger than that found by \cite{Kirby_13}. 
After applying this conversion, we note that the \cite{Kirby_13}, \cite{Simon_19} and \cite{Gallazzi_05} relations agree well with each other (within the uncertainties).

In the case of UDGs, this conversion is extremely important because it may determine if they lie above or below the MZR, which can be directly connected to their formation history.
Most studies of the stellar populations of UDGs done so far \citep{Pandya_18,Ferre-Mateu_18,Ruiz-Lara_18} plotted [$Z$/H] values on top of the \cite{Kirby_13} relation in [Fe/H], which may have affected their conclusions. 
It is important to keep in mind the caveat that MZRs, both for dwarfs and giants, have a strong dependence on age \citep{Gallazzi_05,Hidalgo_17} and environment \citep[see][and references therein]{Peng_Maiolino_14}, and thus any interpretations must take these factors into consideration.

In Fig. \ref{fig:MZR} we show the MZR for our sample of UDGs, compared to the relation found for local universe dwarfs \citep{Kirby_13,Simon_19} and giant galaxies \citep{Gallazzi_05}. We also plot the results found for UDGs in other studies using spectroscopy \citep{Ruiz-Lara_18, Ferre-Mateu_18, Gu_18,Fensch_19,Rong_20,Villaume_22} in Fig. \ref{fig:MZR}. For NGC 5846-UDG1, we plot the results from \cite{Muller_20} assuming [$\alpha$/Fe]=0.3. 

We also plot the UDG stellar population results from SED fitting from \cite{Barbosa_20}. We note that in their paper, they plot their metallicities as [Fe/H] values. However, they state in the text that they used BASE models to perform the fits. These models assume solar scaled abundances, i.e., [$Z$/H] = [Fe/H] ([$\alpha$/Fe]=0). Thus, even though they plot these as [Fe/H], we plot the exact same values as [$Z$/H] in our paper. 

Additionally, we fit with \texttt{Prospector} two dwarf galaxies that bracket our UDG stellar mass range. One normal dwarf elliptical galaxy and one local group dwarf irregular galaxy, VCC1122 and DDO 190, respectively. We did this to further check if we can reproduce the dwarf MZR or if the routine is artificially applying a systematic offset in the metallicity of all sources. We find that VCC1122 is consistent with both dwarf MZRs and with the results obtained previously for this galaxy by \cite{Pandya_18}. We also show that DDO 190 follows the dwarf MZR proposed by \cite{Simon_19} and lies slightly below the \cite{Kirby_13} MZR, still being consistent with it within errors. The recovered stellar mass (log(M$_{\star}$/M$_\odot) = 7.69 \pm 0.2$) and metallicity ([Z/H] $= - 1.35^{+0.26}_{-0.21}$ dex) for DDO 190 are consistent with those reported with spectroscopy by \cite{DDO_06}. 
With these two tests we show that we can reproduce the local MZR, giving confidence that the unusual UDG results are not a product of systematics in the methods.

Our SED fitting results shown in Fig. \ref{fig:MZR} are from the models with dust (\texttt{Prospector} configuration (i) for galaxies with spectroscopic redshifts and configuration (iii) for galaxies with unknown distances). These are the ones found to be more consistent with spectroscopic measurements in the literature (see Section \ref{sec:SEDuptothetask}). Bearing in mind the caveats in Section \ref{sec:results} about dust in UDGs, our sample lies systematically below the MZR for dwarf galaxies. Conversely, if we were to plot the results without dust in the models, the UDGs would be distributed around much more metal-rich populations (mean [$Z$/H] $= -0.7 \pm 0.5$ dex) compared to mean [$Z$/H] $= -1.2 \pm 0.2$ dex with dust.
However, we do note that the GC-poor UDGs have higher amounts of dust (average $A_{V} = 0.4 \pm 0.2$  mag) than the GC-rich ones (average $A_{V} = 0.3 \pm 0.2$ mag). We therefore discount the possibility that the dust is the main driver of the low metallicities found for our sample of UDGs, given that the galaxies with lower metallicities do not have the largest amounts of dust.
Therefore, the remainder of this Section we comment on our results with ``dust''.

Our results shown in Fig. \ref{fig:MZR} are consistent with some of the literature studies using spectroscopy \citep[see results in Section \ref{sec:SEDuptothetask},][]{MartinNavarro_19,Villaume_22,Fensch_19,Ruiz-Lara_18,Ferre-Mateu_18}. However, a fair share of the UDGs studied by \cite{Ruiz-Lara_18} and \cite{Ferre-Mateu_18} (not present in this study) are inconsistent with our results and lie above the relation. It is interesting to notice that the star forming UDGs \citep{Rong_20} lie well above all of the quiescent UDGs present in the current study.
Our results are also consistent with those from \cite{Barbosa_20}, where they show that their field UDGs scatter around the \cite{Kirby_13} MZR for dwarfs, but heavily weighted to lower metallicities. 

It is interesting to notice that the stellar masses obtained by \cite{Barbosa_20} are systematically smaller than the ones obtained from all other studies in the literature, indicating that their stellar masses may be underestimated. If their stellar masses are underestimated by a factor of 2 (0.3 dex), for example, they would lie in the same stellar mass range as all of the other UDGs. If this shift is applied, then most of their UDGs would also lie below the MZR, similar to our results.

One key prediction of the ``puffed-up dwarf'' scenario is that if UDGs were formed from dwarfs, they should have stellar population properties similar to those of other dwarfs. Therefore, they should be fairly consistent with the scaling relations of dwarfs, such as the MZRs proposed by \cite{Kirby_13} and \cite{Simon_19}. If, on the other hand, they are consistent with a failed galaxy scenario, they would be more metal-poor than the relation at a fixed stellar mass, since failed galaxies are expected to be poor in metals due to having quenched before forming metal-rich stars and/or made up of disrupted stars from metal-poor globular clusters \citep{Peng_Lim_16,Danieli_22,Naidu_22}.

By linking the positioning of the UDGs in the mass--metallicity plane to possible formation scenarios, a few things can explain why our UDGs lie below the MZR for dwarfs. If we look into puffed-up dwarf scenarios, for example, \cite{Collins_Read_22} suggested that strong feedback can eject many metals from the galaxies, diluting their metal abundance and thus making the galaxies lie below the expected MZR. Conversely, \cite{Collins_Read_22} also suggested that dwarfs that undergo tidal stripping would in turn lie above the MZR. This is because the interaction would drive star formation episodes that would increase the overall metallicity of the galaxies. We thus disfavour tidal stripping as the main formation mechanism of the UDGs studied in this work. 

Alternatively, if we look into the failed galaxy scenarios \citep{vanDokkum_15}, where a combination of early-quenching and early massive star formation occurs, we can expect that UDGs would have only or primarily ancient cluster stars with low-metallicity. This is because the early quenching would have prevented them from having further ``rounds'' of star formation. This model was further developed by \cite{Danieli_22} in order to explain the overly rich globular cluster (GC) population in the group UDG NGC 5846\_UDG1. In this case, they proposed that UDGs may be the result of early star formation in massive clumps of gas, forming a lot of GCs. With time, strong feedback coming from these clumps of gas (together with other quenching mechanisms) would quench the galaxy. In this scenario, most of the stellar light from UDGs would come from their high number of GCs, and thus a low metallicity, consistent with that of in-situ formed GCs, would be expected. A main expectation of this scenario, thus, is that UDGs would be both GC-rich and extremely metal-poor. This is exactly in line with the locus of GC-rich UDGs on Fig. \ref{fig:MZR}.

Additionally, a third formation scenario, proposed to explain quiescent galaxies in the field, dubbed the ``backsplash orbit'' \citep{Benavides_21} scenario, could be invoked to explain DGSAT I and M-161-1. This scenario would require the galaxies to be close in proximity to a massive galaxy, group or cluster and to have been completely stripped of gas. 
M-161-1 has no massive neighbours within 1.5 Mpc in projected distance \citep{Papastergis_17} and so a backsplash orbit scenario is unlikely to explain its formation. The galaxy is indeed old, even in the absence of an obvious quenching mechanism, suggesting that it belongs to the failed galaxy subpopulation. As for DGSAT I, \cite{Janssens_22} suggested, studying its GC system, that this UDG is also more likely to be a failed galaxy than a backsplash orbit one, although it may lie close in proximity to a massive group. This is because of the presence of a young overdensity ($\sim 500$ Myr) in DGSAT I's disk, which directly challenges the gas stripping requirement of the backsplash scenario. In addition, its massive GC system \citep{Janssens_22} and incredibly metal-poor stellar population \citep{MartinNavarro_19} seem to fully agree with the failed galaxy scenario proposed by \cite{Danieli_22}.

All of this evidence seem to indicate two different formation scenarios for the galaxies in our sample. GC-poor UDGs, lying slightly below the dwarf MZR, seem to be more consistent with a puffed-up dwarf scenario. These galaxies have low numbers of GCs, consistent with what is observed in regular dwarfs, and so this formation scenario seems to fit them better. Their metallicities, lower than what is observed for dwarfs, might be explained by strong stellar feedback, as suggested by \cite{Collins_Read_22} and \cite{diCintio_17}. This is also consistent with their star formation time-scales ($\tau$). These galaxies, as discussed in Section \ref{sec:results}, have more extended SFHs and are on average younger than their GC-rich counterparts.
On the other hand, some GC-rich UDGs, lying well below the dwarf MZR, have extremely metal-poor stellar populations and thus are not consistent with the puffed-up dwarf formation scenario. These UDGs are much better explained by a failed galaxy scenario. Particularly, since these galaxies are all GC-rich and have extremely low metallicities, we believe they are consistent with the formation scenario proposed by \cite{Danieli_22}, and the stellar content of these galaxies may have originally formed as (now disrupted) GCs. These GC-rich UDGs, additionally, have shorter SFHs, consistent with early quenching, as expected for failed galaxies.
In fact, these extremely metal-poor GC-rich UDGs are consistent with the evolving MZR at redshift $z= 2.2$, as derived by \cite{Ma_15} using cosmological simulations and shown in Fig. \ref{fig:MZR}. This may be an indication that these galaxies have quenched at this redshift. Interestingly, this redshift corresponds to an age of 10.5 Gyr, i.e, consistent with our mean age for the GC-rich UDGs of $9.8 \pm 1.6$ Gyr.

\subsubsection{The co-formation of NGC 1052-DF2 and NGC 1052-DF4}

NGC 1052-DF2 and NGC 1052-DF4 have been a source of debate for a few years now, since both were found to be UDGs (i.e., low-surface brightness and large effective radii) lacking dark matter \citep{vanDokkum_18,vanDokkum_19}. They also were found to host massive and extremely bright GC systems (which could indeed be classified as ultra compact dwarfs rather than GCs) \citep{vanDokkum_18, Shen_21}. All of these unique properties are accompanied by the fact that the galaxies are very close in proximity \citep[$\sim$2 Mpc in three-dimensional distance, ][]{Shen_21}. Assuming that it is not a coincidence that the galaxies share such unusual properties while being so close, a common formation scenario for them must simultaneously explain: 1) their lack of dark matter, 2) their large sizes and 3) the presence of luminous and massive GCs. 

Although there have been many attempts, no proposed formation scenario \citep{Martin_18, Ojiya_18,Trujillo_19,Monelli_19,Shin_20, Montes_20,Montes_21,Maccio_21,Jackson_21,Trujillo_Gomez_21,Lee_21,Moreno_22,Trujillo-Gomez_22} has been able to simultaneously explain all of these rather unique properties that NGC 1052-DF2 and NGC 1052-DF4 share.

Recently, \cite{vanDokkum_22} proposed that the formation of NGC 1052-DF2 and NGC 1052-DF4 (and seven other galaxies in the NGC 1052 group), resulted from a head-on interaction of two gas-rich galaxies approximately 8 Gyr ago. This scenario was proposed to be similar to that of the bullet cluster and it was dubbed the ``bullet dwarf'' scenario. In this hypothesis, the interaction would have separated baryonic and dark matter of the progenitor galaxies. With this, the remnants of the initial two interacting galaxies would be dark matter dominated and lie at the tips of a trail of dark matter free galaxies formed from the separated baryonic matter (see Fig. 1 of \citealt{vanDokkum_22}). The interaction would also have formed many massive GCs. This scenario is capable of explaining the similar and unique properties of NGC 1052-DF2 and NGC 1052-DF4, namely, being dark matter free \citep{vanDokkum_18,vanDokkum_19} and having many massive GCs \citep{Shen_21}, all of this while being close in proximity of one another.

Based on this scenario, we expect to find that the galaxies would have similar stellar population properties, as they would have been formed by the same process and same material. Moreover, we expect to find that both of them are consistent with the proposed age of the interaction of $\sim$ 8 Gyr.
We find, using SED fitting with the models including dust, that NGC 1052-DF2 has an age of $8.0^{+1.8}_{-2.7}$ Gyr and a metallicity of [$Z$/H] = $-1.1^{+0.4}_{-0.3}$ dex. Similarly, NGC 1052-DF4 has an age of $8.8^{+2.9}_{-1.5}$ Gyr and a metallicity of [$Z$/H] = $-1.1^{+0.4}_{-0.2}$ dex.

With these findings, we show that the stellar population properties of the two galaxies are compatible within 1$\sigma$, and consistent with the expected age of the collision ($\sim 8$ Gyr). Such similarity is expected in the bullet scenario, and thus our findings agree with such a formation history. However, compatible populations are a necessary but not a sufficient condition to conclude that this was indeed the formation scenario that took place. Another possibility, for example, would be that NGC 1052-DF2 and NGC 1052-DF4 are tidal dwarf galaxies \citep{Haslbauer_19}.
Our results, nevertheless, point to the bullet scenario as one of the possible formation histories for these galaxies. 
However, more studies are required to further test this hypothesis. 

\section{Conclusions}
\label{sec:conclusions}
In the current study, we have used the fully Bayesian Monte Carlo Markov Chain inference code \texttt{PROSPECTOR} to perform spectral energy distribution fitting of twenty-nine UDGs using data from the optical to the mid-IR.
We test the efficiency of \texttt{PROSPECTOR} in recovering stellar populations using four different configurations, (i) dust fixed to zero and redshift fixed to the spectroscopic redshift; (ii) dust as a free parameter and redshift fixed to the spectroscopic redshift; (iii) dust and redshift as free parameters, and (iv) dust fixed to zero and free redshift.
For the galaxies with predetermined spectroscopic redshifts, we primarily carried out fits in the first two scenarios, while the latter two were used to test our ability to recover photometric redshifts and to estimate the stellar populations of UDGs with no distance measurement available.

Using the derived stellar populations and photometric redshifts, we conclude that the presence of dust in the models consistently improves the fits, always delivering values closer to those determined from spectroscopy. 
The amount of dust found in the galaxies is on average ${A}_{V} = 0.4 \pm 0.1$ mag, reaching a maximum value of $0.88 \pm 0.2$ mag for DF06. It is beyond the scope of this paper to understand whether this finding can be treated as a sign of physical dust in the galaxies or simply an artificial addition of the code to improve the fit. 

As for the recovered photometric redshifts, we show, using the galaxies with known distances, that we can achieve a redshift precision of $\sigma_{\rm NMAD} = 0.02$ and an offset of $\mu_{\rm bias} = -0.01$ with SED fitting. With this, we estimate the photometric redshifts of the three galaxies in the field where no distance measurement was available, finding that they all meet the size criteria to be considered UDGs. This method of recovering photometric redshifts can be expanded to all of the known UDG candidates with wide photometric coverage, making it possible to test how many actually meet the size requirement to be classified as UDGs.

We find an age dependence on the environment, with UDGs in the field being on average younger and slightly more metal-rich than their cluster counterparts. We also see a dependence with the GC-richness, with GC-poor UDGs being more metal-rich than the GC-rich ones.
We find that all the UDGs in our sample are systematically more metal-poor than what was found for regular dwarf galaxies of a comparable stellar mass. We see, however, that the GC-poor UDGs are consistent with the dwarf mass--metallicity relation (MZR), suggesting they may be puffed-up dwarfs. On the other hand, GC-rich UDGs show much lower metallicities, indicating that they may be consistent with a failed galaxy scenario.

As a byproduct, we show that NGC 1052-DF2 and NGC 1052-DF4 share similar stellar population properties, with ages consistent with 8 Gyr. This finding supports formation scenarios where the galaxies were formed together, such as the ``bullet dwarf'' \citep{vanDokkum_22} or tidal dwarf-like scenarios \citep{Haslbauer_19}. 
 
This paper provides stellar population properties of UDGs across different environments without resorting to spectroscopy. We demonstrate that SED fitting techniques, coupled with a broad wavelength coverage, may be key to statistically probing the population properties of UDGs and how these are distributed across the sky.

\section*{Acknowledgements}
We thank the referee for all comments/suggestions which improved a lot the manuscript.
We thank Pieter van Dokkum for providing GMOS data for DF44 and DFX1.
We thank Ben Johnson for all the help with dealing with \texttt{PROSPECTOR}'s priors and Viraj Pandya for insights into how to best interpret our \texttt{PROSPECTOR} results in face of such faint sources as UDGs. We thank Joel Leja and Yimeng Tang for discussion about \texttt{Prospector} systematics.
We acknowledge Song Huang for providing Subaru/ Hyper Suprime Cam data for the two Perseus cluster UDGs studied in this work.
We acknowledge David Mart\'inez-Delgado for providing Subaru optical imaging of DGSAT I (see also \citealt{MartinezDelgado_16}).
We are grateful to Stephen Gwyn for help with the astrometric and photometric calibration of the CFHT data of M-161-1.
This research was supported by the Australian
Research Council Centre of Excellence for All Sky Astrophysics in 3 Dimensions (ASTRO 3D), through project number CE170100013. 
AJR was supported as a Research Corporation for Science Advancement Cottrell Scholar. AFM acknowledges support from the Severo Ochoa Excellence scheme (CEX2019-000920-S).

This paper is based in part on observations made with and archival data obtained with the \textit{Spitzer} Space Telescope, which is operated by the Jet Propulsion Laboratory, California Institute of Technology under a contract with NASA. This paper is based in part on data from the Hyper SuprimeCam Legacy Archive (HSCLA), which is operated by the Subaru Telescope. The original data in HSCLA was collected at the Subaru Telescope and retrieved from the HSC data archive system, which is operated by Subaru Telescope and Astronomy Data Center at National Astronomical Observatory of Japan. This paper is based in part on observations from the Legacy Survey, which consists of three individual and complementary projects: the Dark Energy Camera Legacy Survey (DECaLS; Proposal ID \#2014B-0404; PIs: David Schlegel and Arjun Dey), the Beijing-Arizona Sky Survey (BASS; NOAO Prop. ID \#2015A-0801; PIs: Zhou Xu and Xiaohui Fan), and the Mayall z-band Legacy Survey (MzLS; Prop. ID \#2016A-0453; PI: Arjun Dey).
This paper is in part based on observations obtained at the Gemini Observatory, which is operated by the Association of Universities for Research in Astronomy, Inc., under a cooperative agreement
with the NSF on behalf of the Gemini partnership.
This publication makes use of data products from the Wide-field Infrared Survey Explorer, which is a joint project of the University of California, Los Angeles, and the Jet Propulsion Laboratory/California Institute of Technology, funded by the National Aeronautics and Space Administration.


\section*{Data Availability}
DECaLS data are available via the  \href{https://www.legacysurvey.org/decamls/}{Legacy survey portal}. \textit{WISE} data are available via the \href{https://wise2.ipac.caltech.edu/docs/release/allsky/}{WISE archive}. GMOS data are available via the  \href{https://archive.gemini.edu/searchform}{Gemini portal}. CFHT data are available at the \href{https://www.cadc-ccda.hia-iha.nrc-cnrc.gc.ca/en/search/?collection=CFHT&noexec=true}{CFHT database}. \textit{Spitzer}-IRAC data are available via the \href{https://irsa.ipac.caltech.edu/data/SPITZER/docs/spitzerdataarchives/}{Spitzer archive}. HSC data are proprietary and can be made available upon request.



\bibliographystyle{mnras}
\bibliography{bibli} 




\appendix
\input{Appendix_stamps.tex}




\bsp	
\label{lastpage}
\end{document}

%% file: Appendix_stamps.tex
\section{Processed postage stamps}
\label{sec:appendix_stamps}

In this appendix, we show the final processed postage stamps of the remaining UDGs studied in this work, including all the bands used. This is a continuation of Fig. \ref{fig:UDGs_stamps1}.

\begin{figure*}
    \centering
    \includegraphics[width=\textwidth]{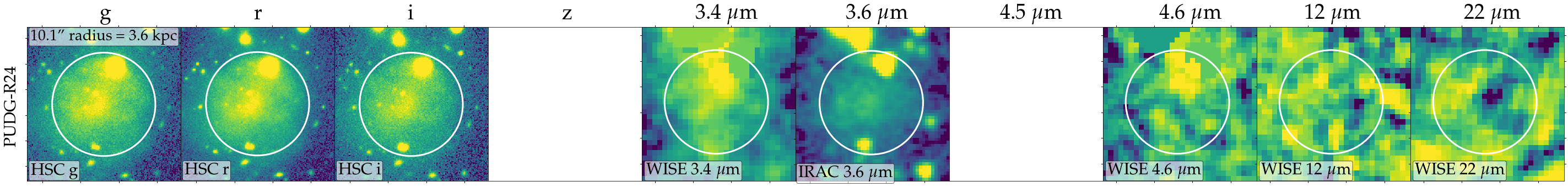}
    \includegraphics[width=\textwidth]{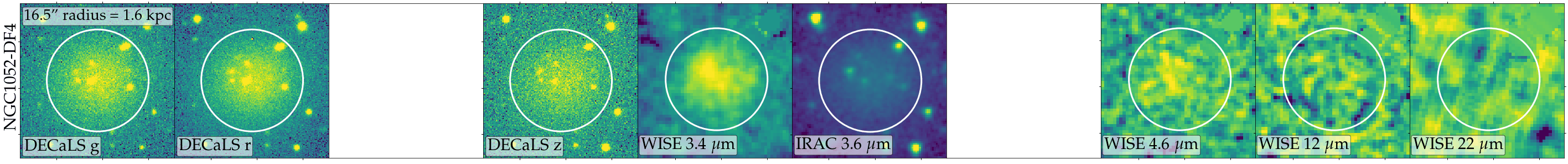}
    \includegraphics[width=\textwidth]{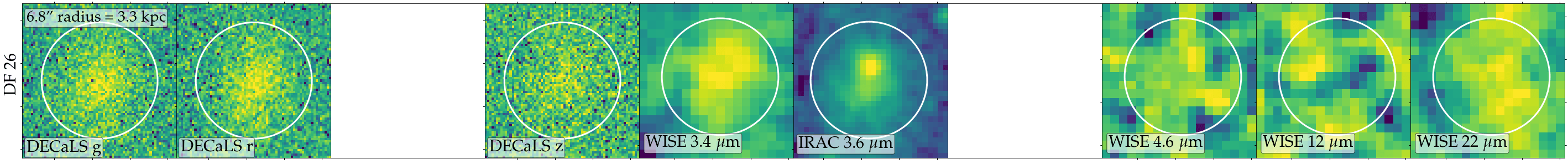}
    \includegraphics[width=\textwidth]{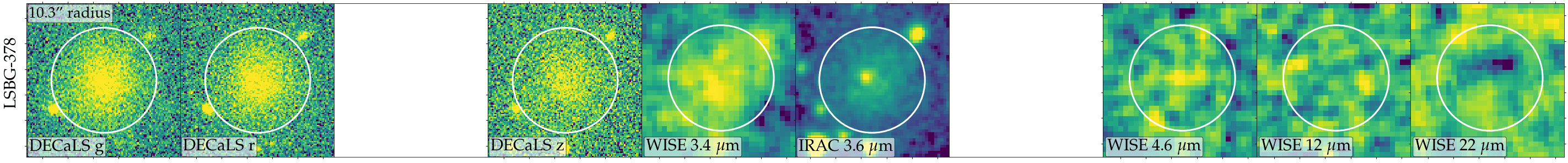}
    \includegraphics[width=\textwidth]{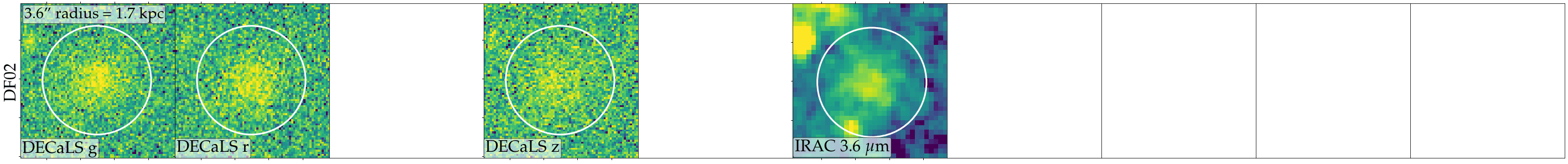}
    \includegraphics[width=\textwidth]{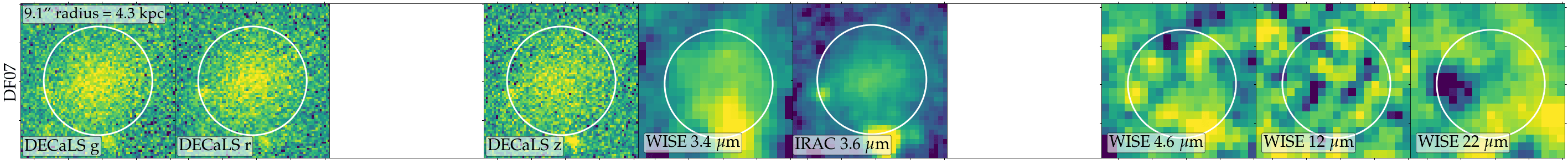}
    \includegraphics[width=\textwidth]{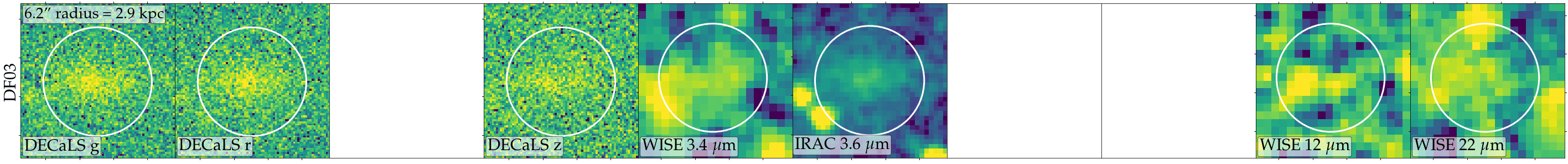}
    \includegraphics[width=\textwidth]{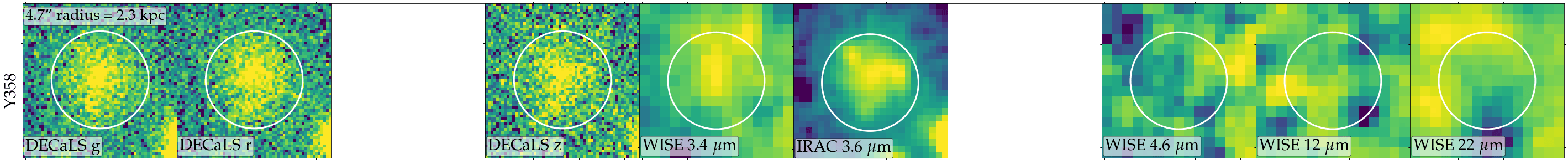}
    \includegraphics[width=\textwidth]{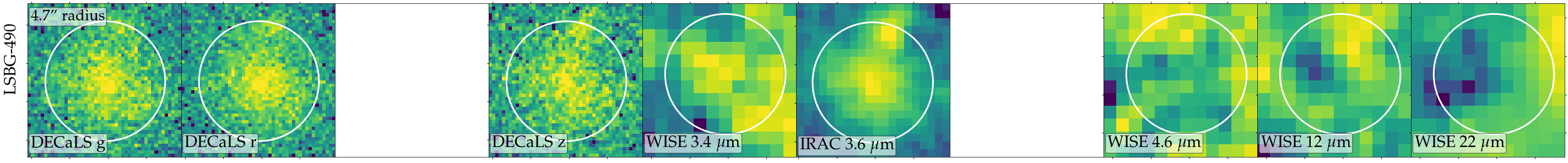}
    \includegraphics[width=\textwidth]{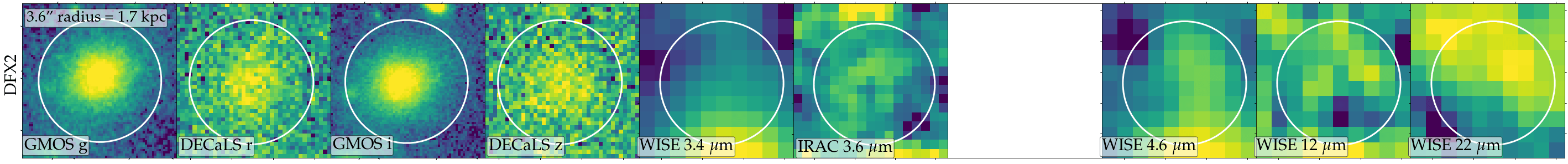}
    \includegraphics[width=\textwidth]{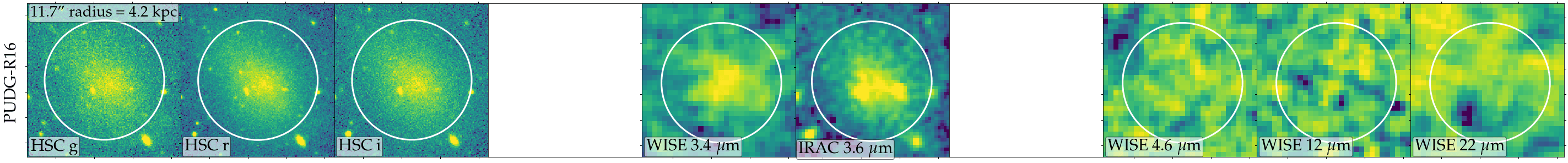}
    \includegraphics[width=\textwidth]{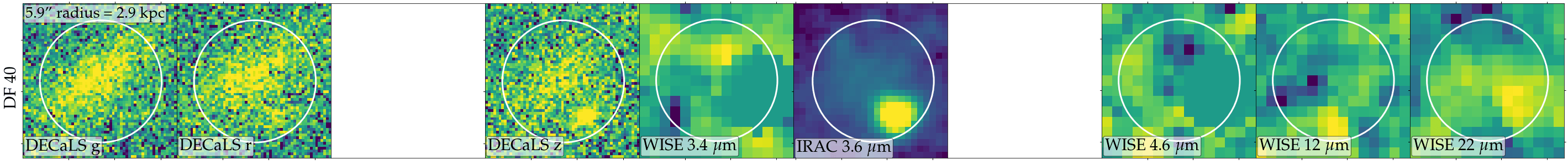}
\caption{Same as Fig. \ref{fig:UDGs_stamps1}.}
\end{figure*}

\begin{figure*}
    \centering
    \includegraphics[width=\textwidth]{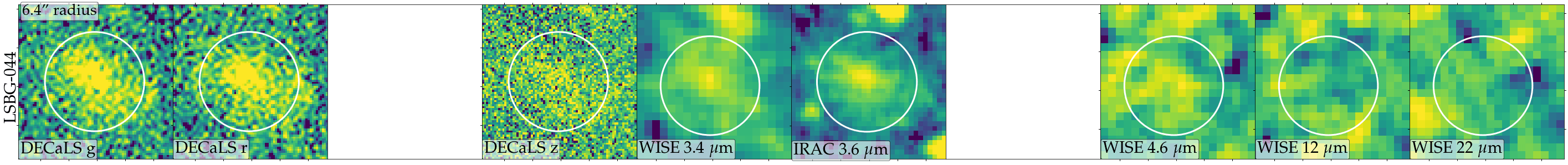}
    \includegraphics[width=\textwidth]{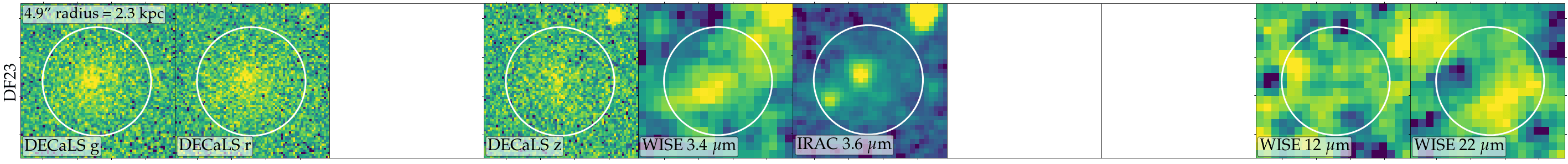}\\
    \includegraphics[width=\textwidth]{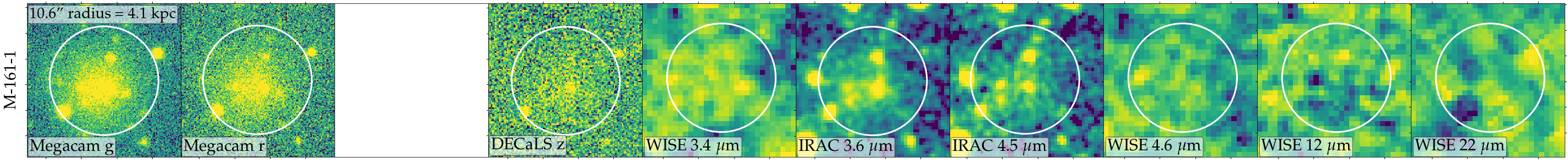}\\
    \includegraphics[width=\textwidth]{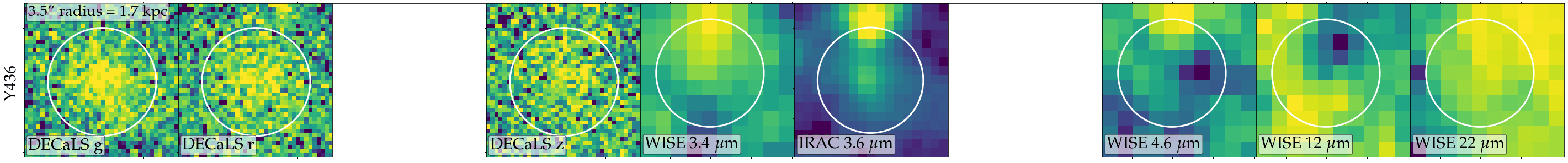}\\
    \includegraphics[width=\textwidth]{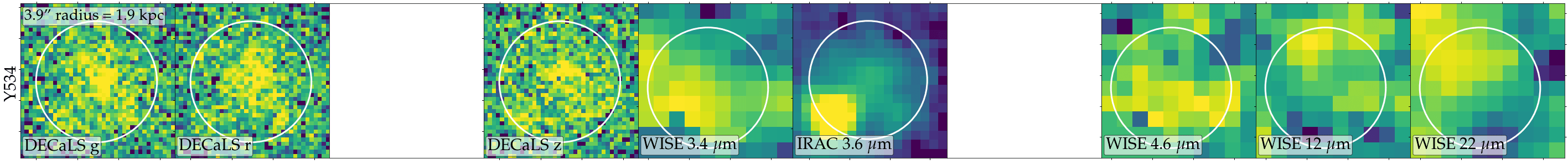}\\
    \includegraphics[width=\textwidth]{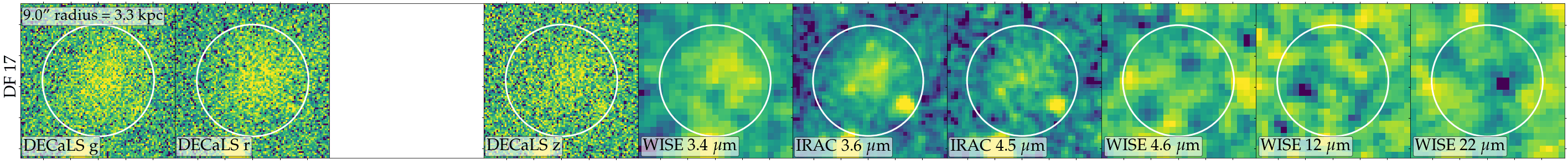}\\
    \includegraphics[width=\textwidth]{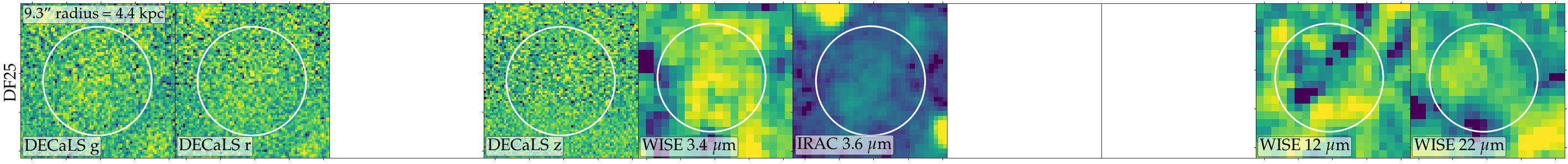}\\
    \includegraphics[width=\textwidth]{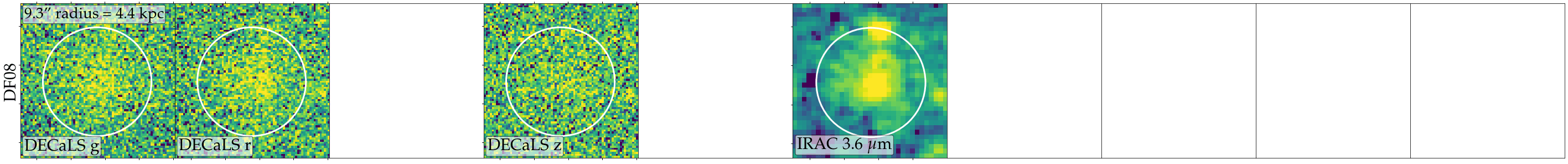}\\
    \includegraphics[width=\textwidth]{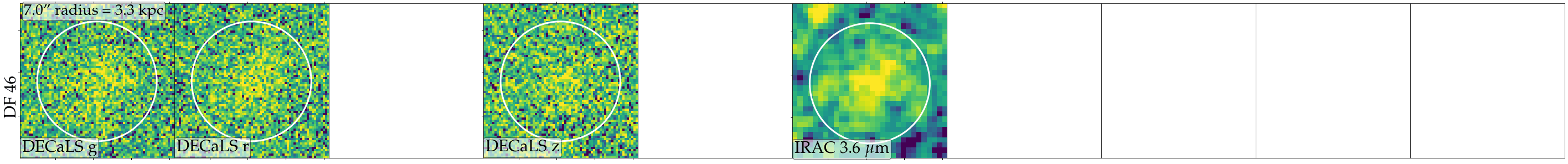}\\
    \includegraphics[width=\textwidth]{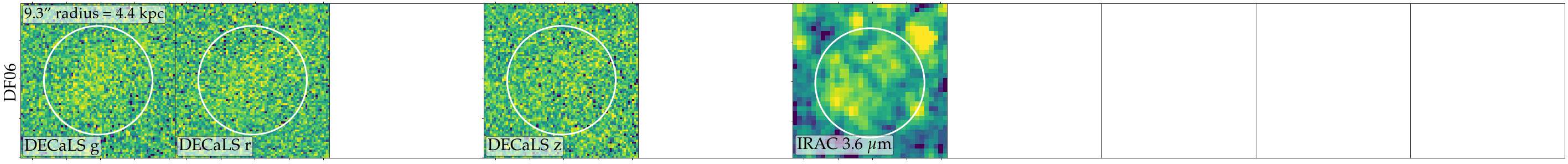}\\
    \includegraphics[width=\textwidth]{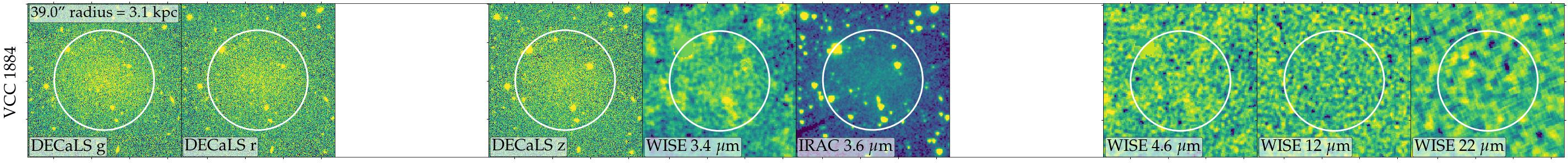}\\
    \includegraphics[width=\textwidth]{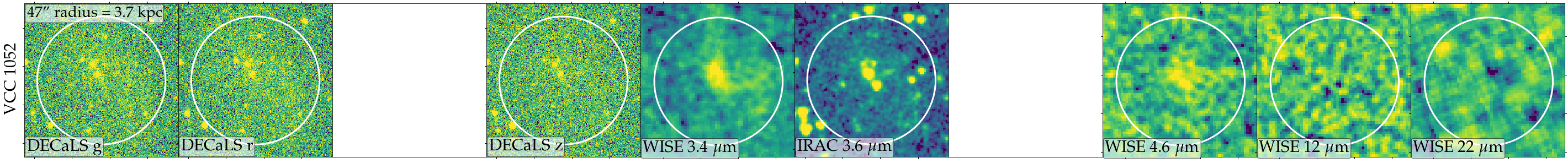}\\
    \contcaption{}
\end{figure*}

\input{AppendixA.tex}
\input{AppendixB.tex}
\input{Appendix_Table.tex}

%% file: AppendixA.tex
\section{Prospector fits}
\label{sec:appendixA}
In this appendix, we provide the full MCMC posteriors and SED fits for all the UDGs, including the models with and without dust. 

\begin{figure*}
    \centering
    \includegraphics[width=\columnwidth]{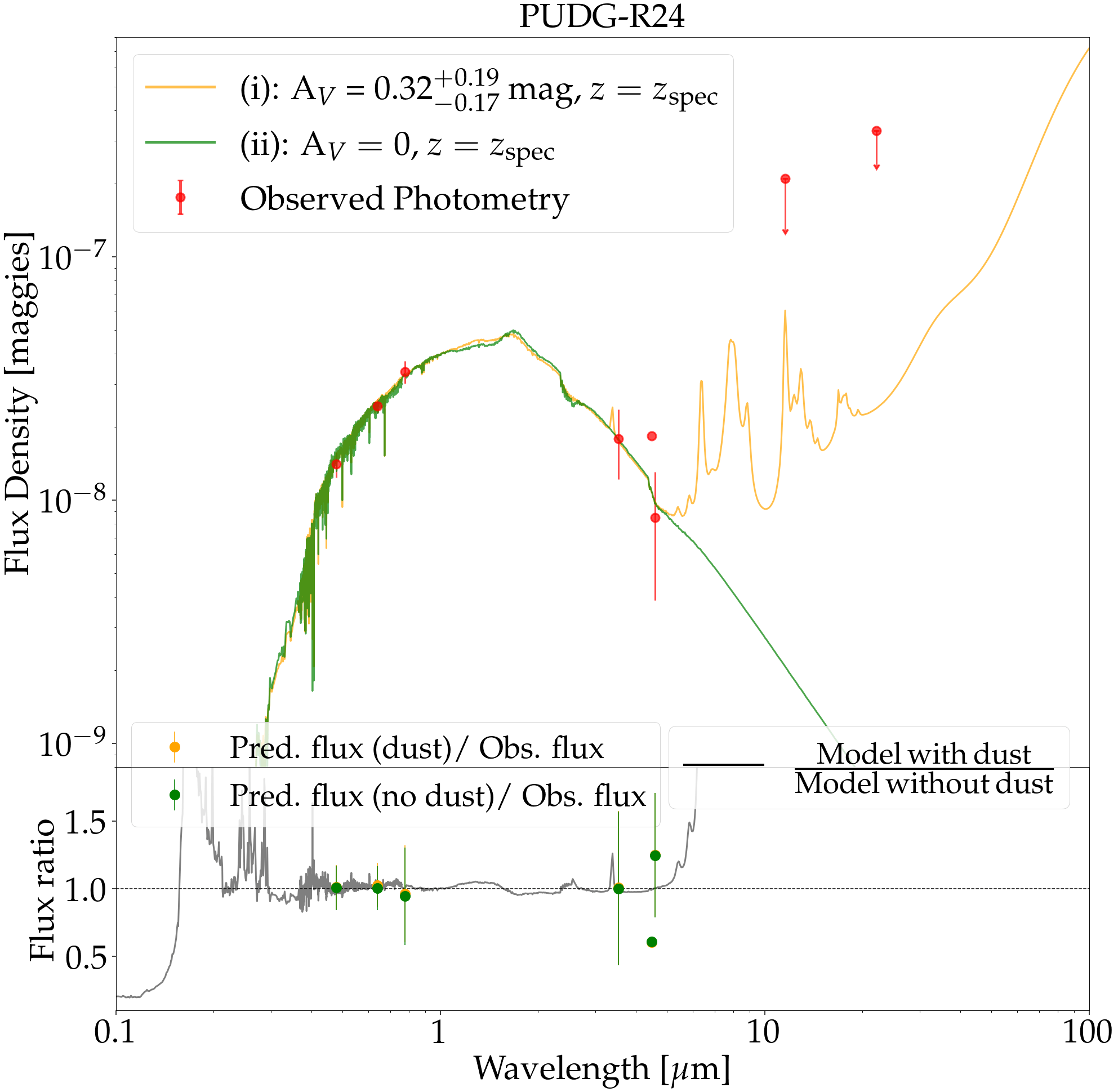}
    \includegraphics[width=\columnwidth]{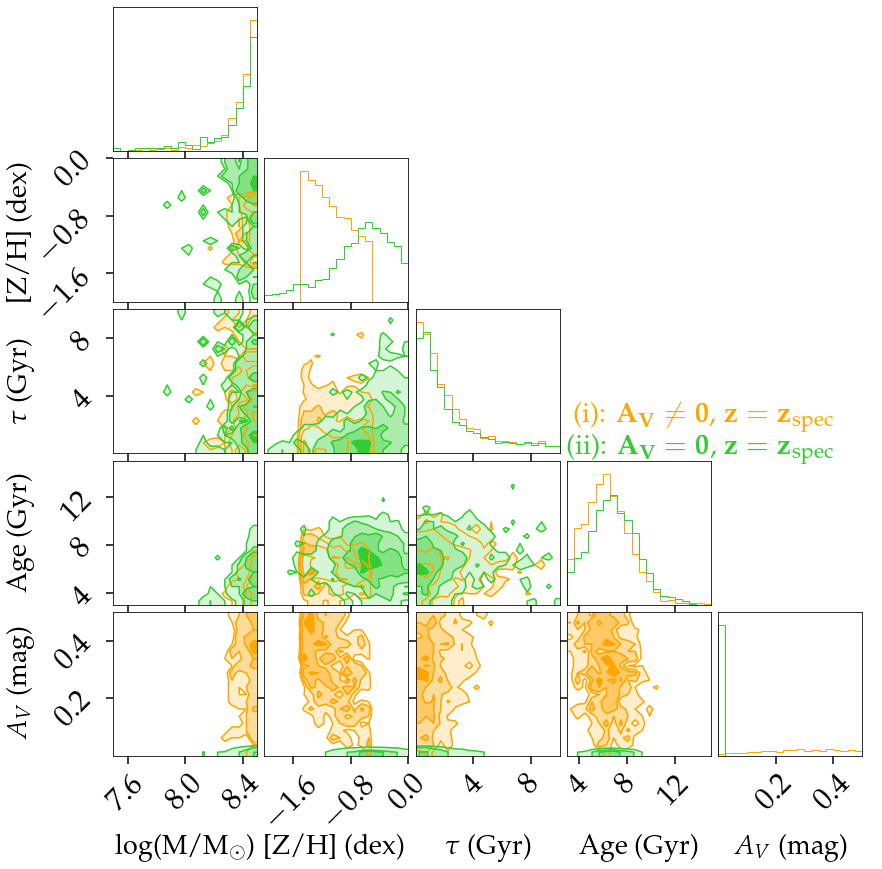}
    \caption{SED fitting results for PUDG-R24. \textit{Left:} SED fits comparing Prospector configurations with (i) dust as a free parameter and (ii) dust fixed to zero. Configuration (i) is shown with the yellow curve, the same for configuration (ii) with the green line. \textit{Right:} MCMC corner plot comparing the posterior distribution for the best fit with configuration (i) (yellow) and with configuration (ii) (green). The first panel in each column shows the 1D posterior distribution of the fitted parameter, while the remaining panels show the correlation between the parameters. This image can be read and interpreted as a covariance matrix. Columns stand for stellar mass, metallicity, star formation time scale, age and interstellar diffuse dust extinction.}
    \label{fig:corner_PUDG-R24}
\end{figure*}

\begin{figure*}
    \centering
    \includegraphics[width=\columnwidth]{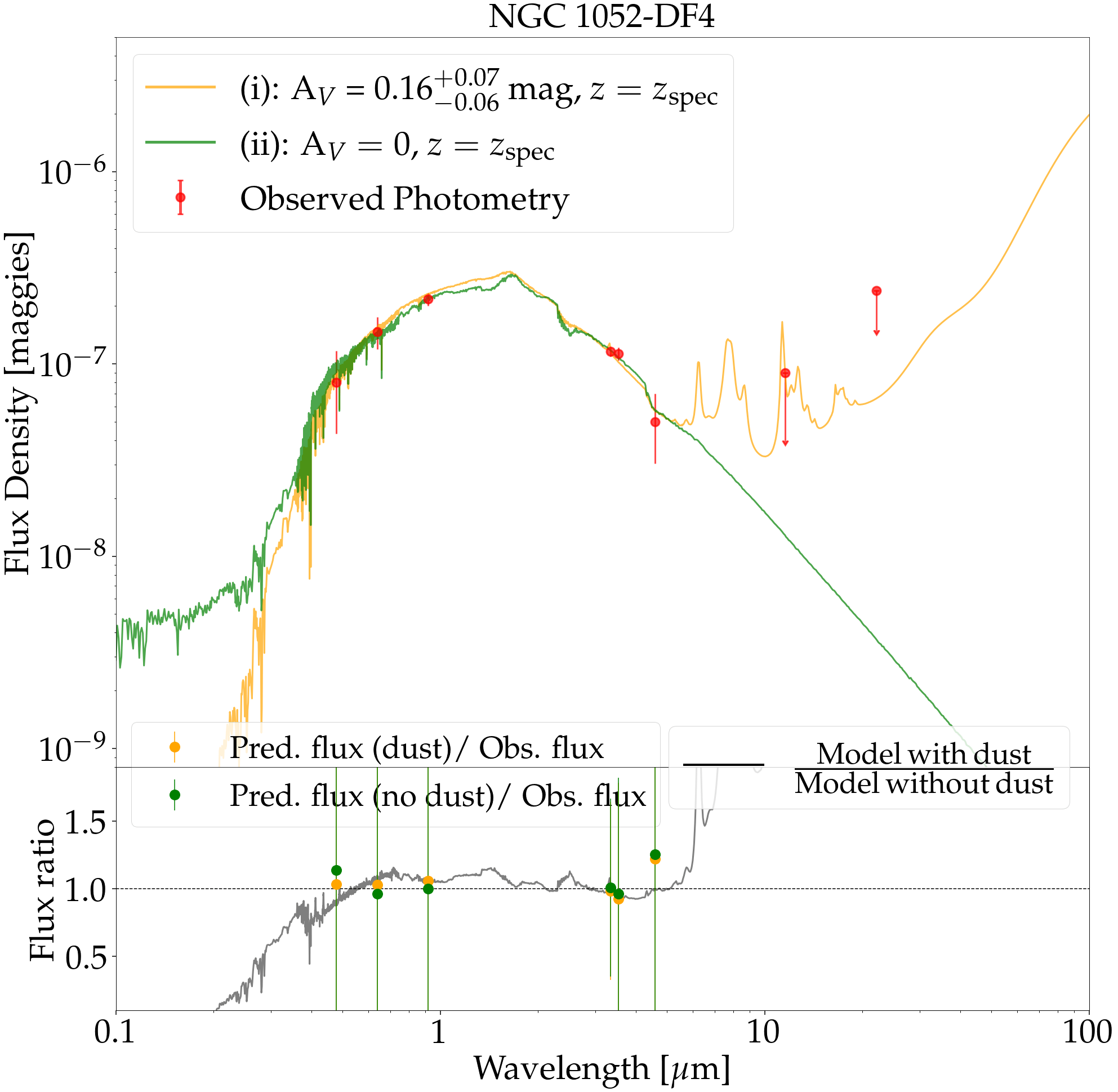}
    \includegraphics[width=\columnwidth]{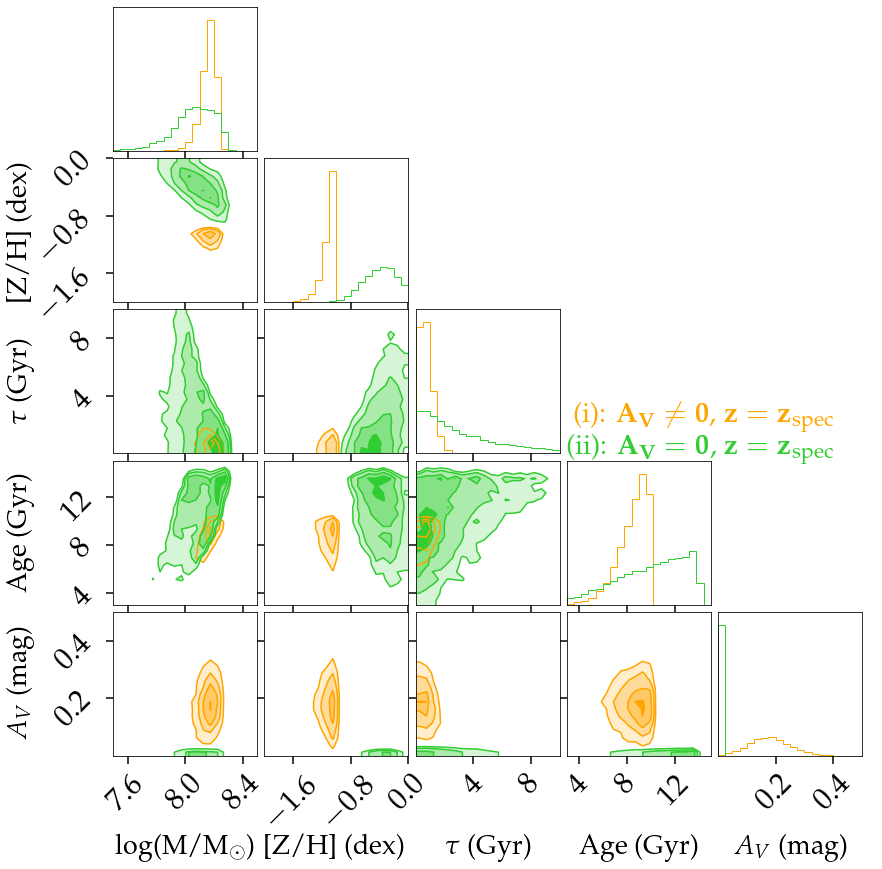}
    \caption{As Fig. \ref{fig:corner_PUDG-R24}, but for NGC 1052-DF4.}
    \label{fig:corner_NGC1052-DF4}
\end{figure*}

\begin{figure*}
    \centering
    \includegraphics[width=\columnwidth]{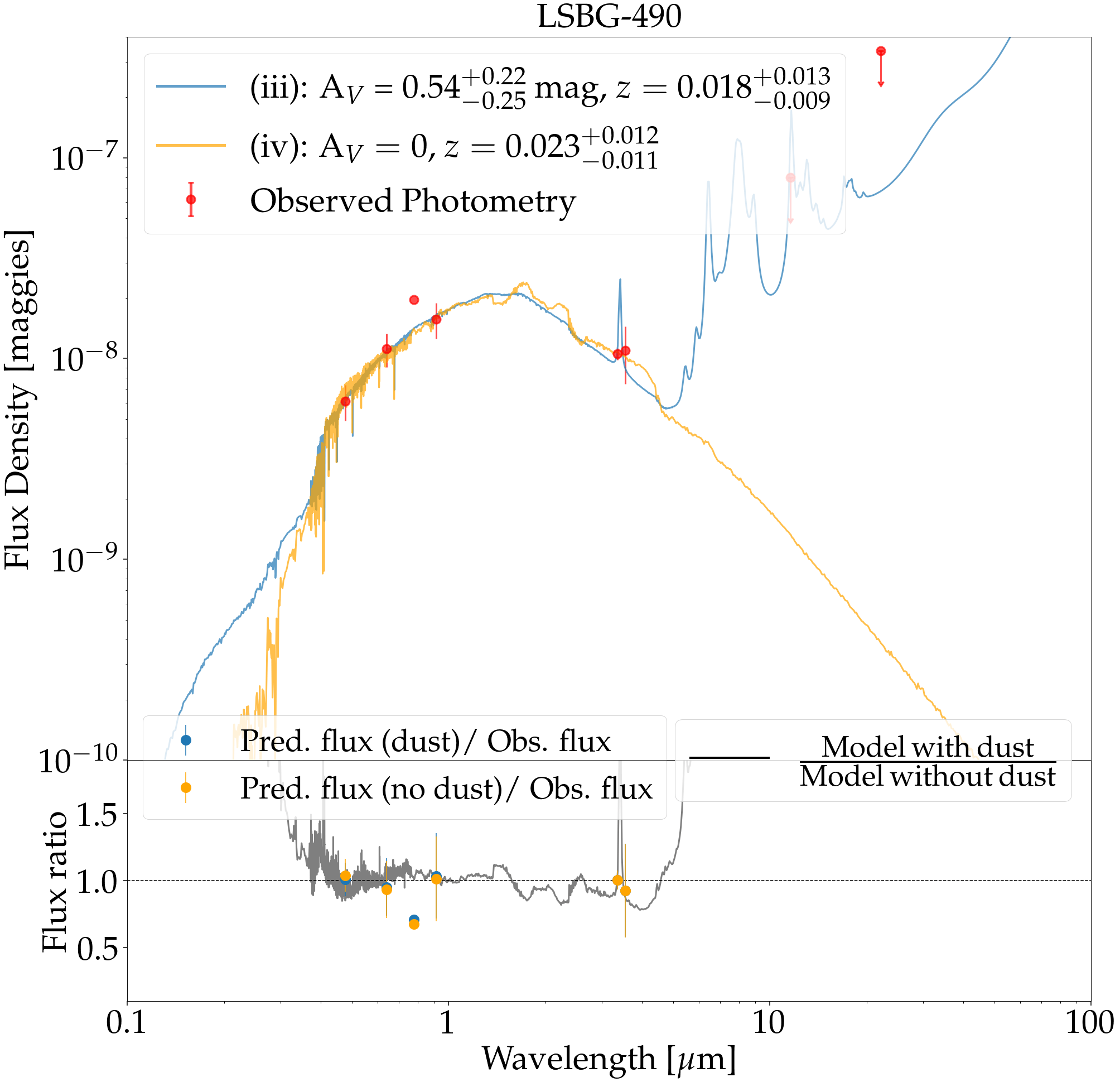}
    \includegraphics[width=\columnwidth]{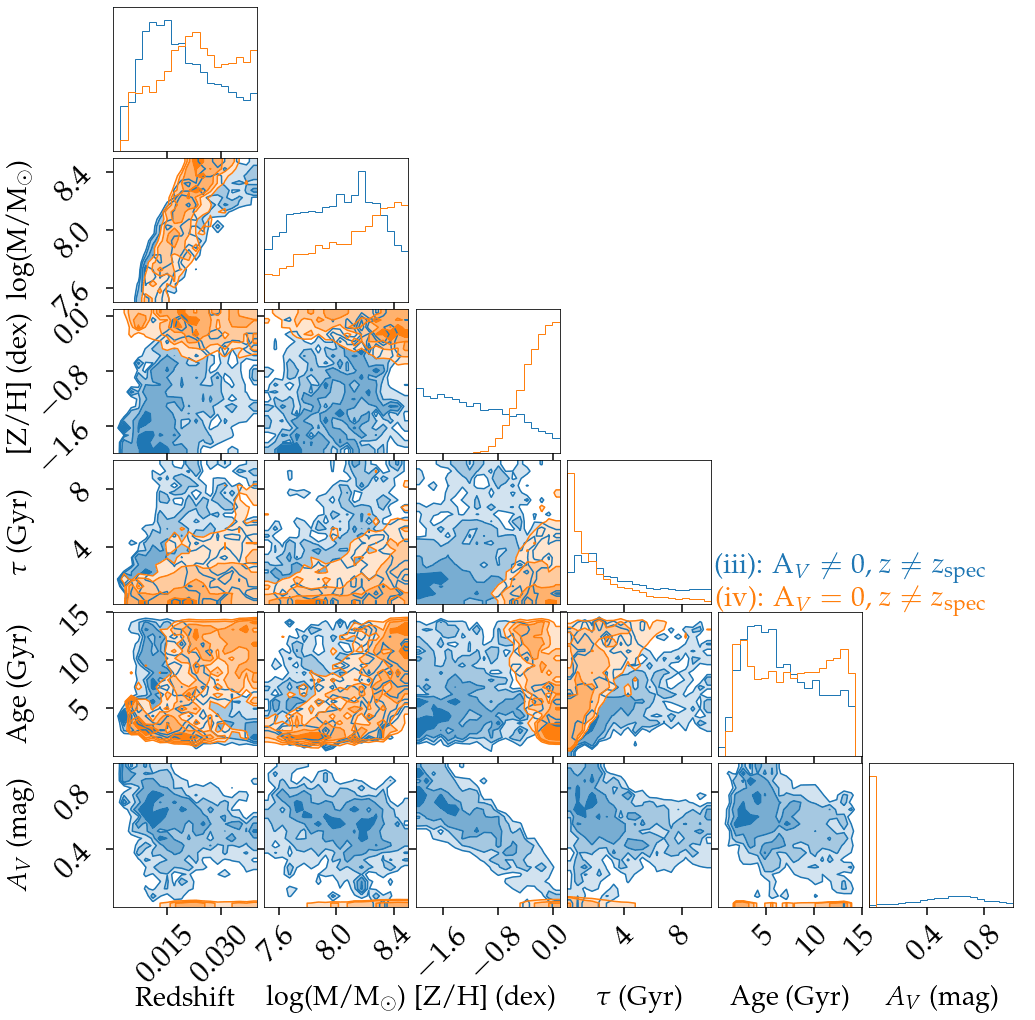}
    \caption{SED fitting results for LSBG-490. \textit{Left:} SED fits comparing Prospector configurations with (iii) dust and redshift as free parameters and (iv) dust fixed to zero and redshift as a free parameter. Configuration (iii) is shown with the blue curve, while configuration (iv) is shown with the orange line. \textit{Right:} MCMC corner plot comparing the posterior distribution for each the best fit with configuration (iii) (blue) and with configuration (iv) (orange). The first panel in each column shows the 1D posterior distribution of the fitted parameter, while the remaining panels show the correlation between the parameters. This image can be read and interpreted as a covariance matrix. Columns stand for redshift, stellar mass, metallicity, star formation time scale, age and interstellar diffuse dust extinction.}
    \label{fig:corner_LSBG-490}
\end{figure*}

\begin{figure*}
    \centering
    \includegraphics[width=\columnwidth]{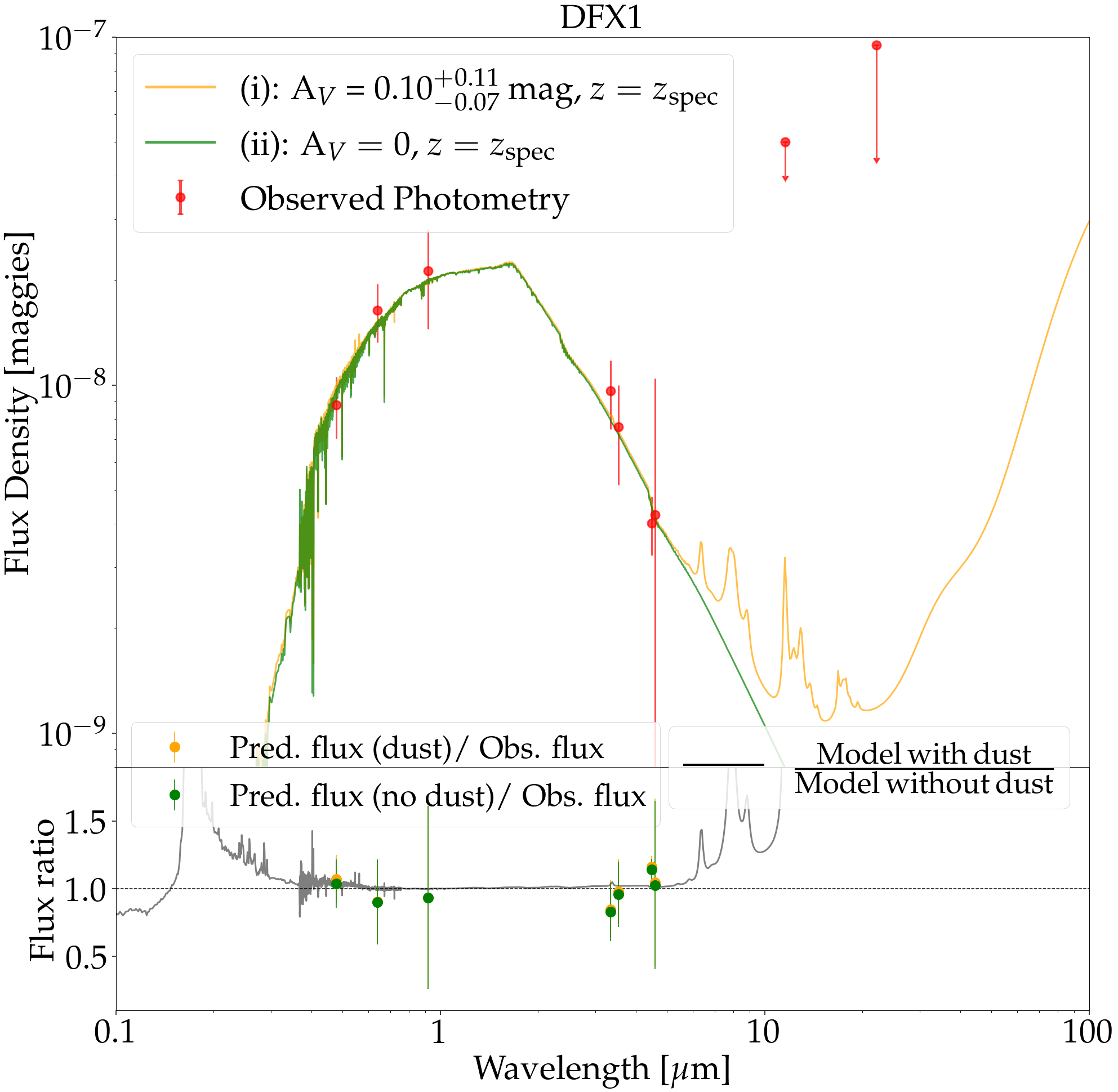}
    \includegraphics[width=\columnwidth]{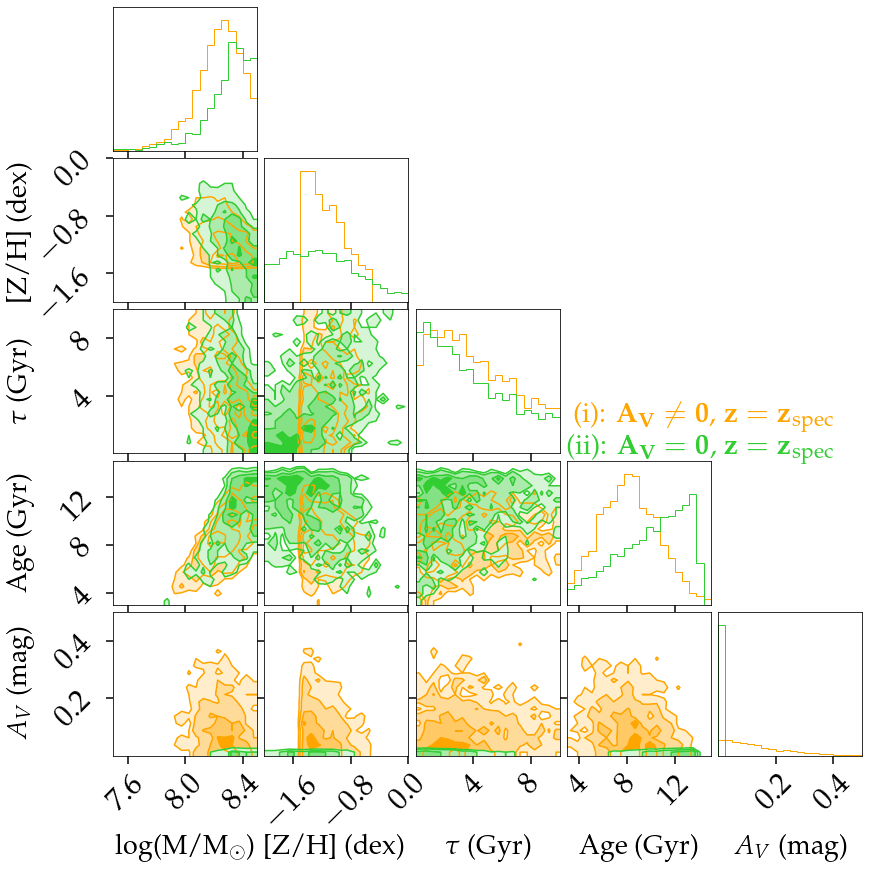}
    \caption{As Fig. \ref{fig:corner_PUDG-R24}, but for DFX1.}
    \label{fig:corner_DFX1}
\end{figure*}

\begin{figure*}
    \centering
    \includegraphics[width=\columnwidth]{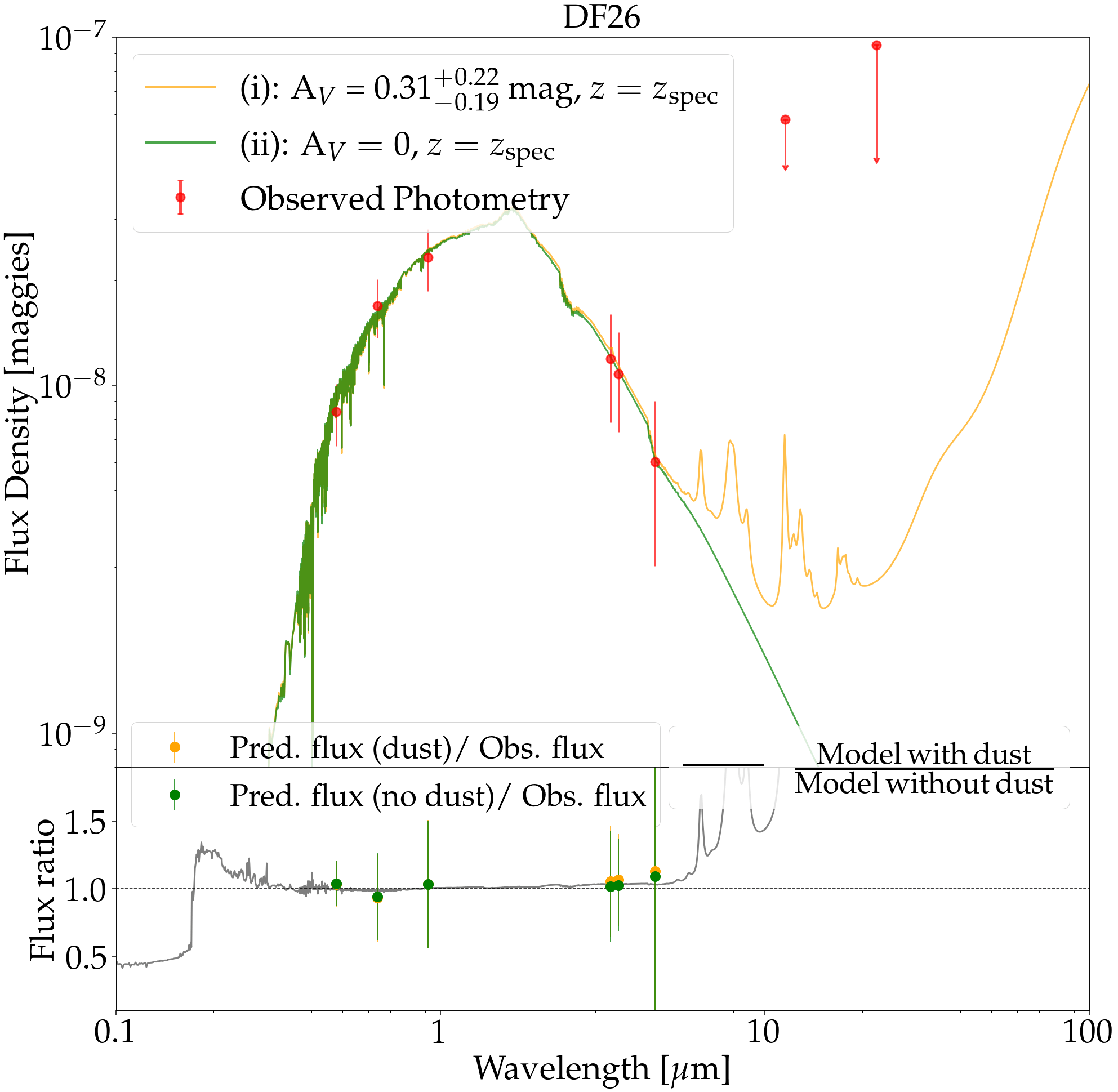}
    \includegraphics[width=\columnwidth]{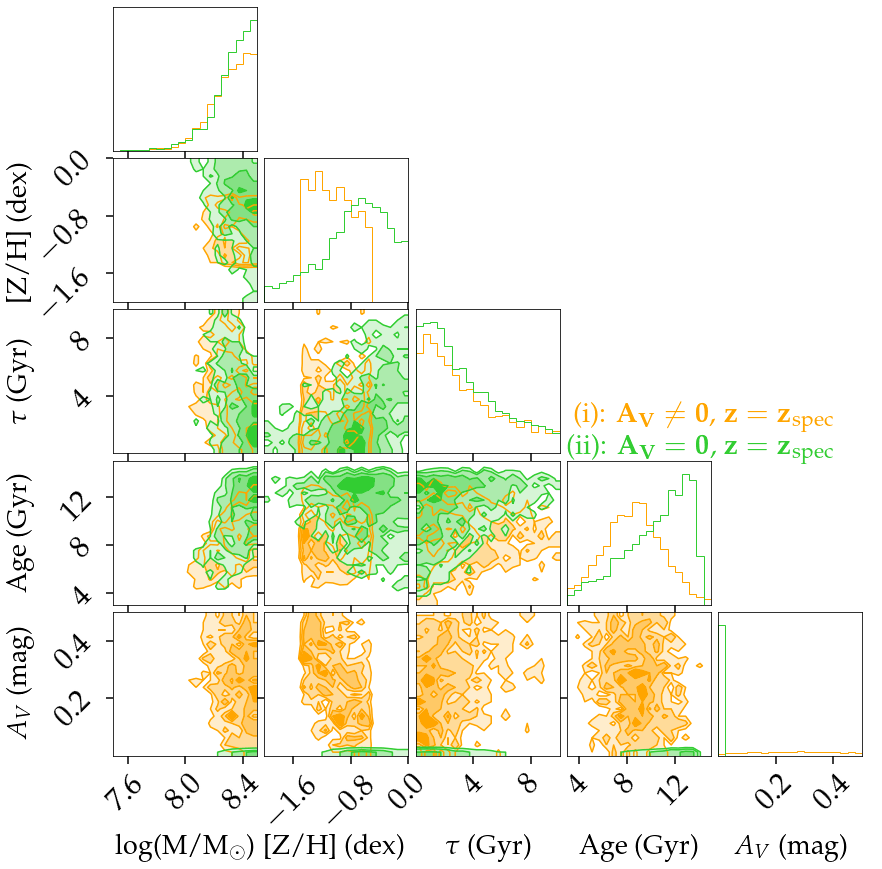}
    \caption{As Fig. \ref{fig:corner_PUDG-R24}, but for DF26.}
    \label{fig:corner_DF26}
\end{figure*}

\begin{figure*}
    \centering
    \includegraphics[width=\columnwidth]{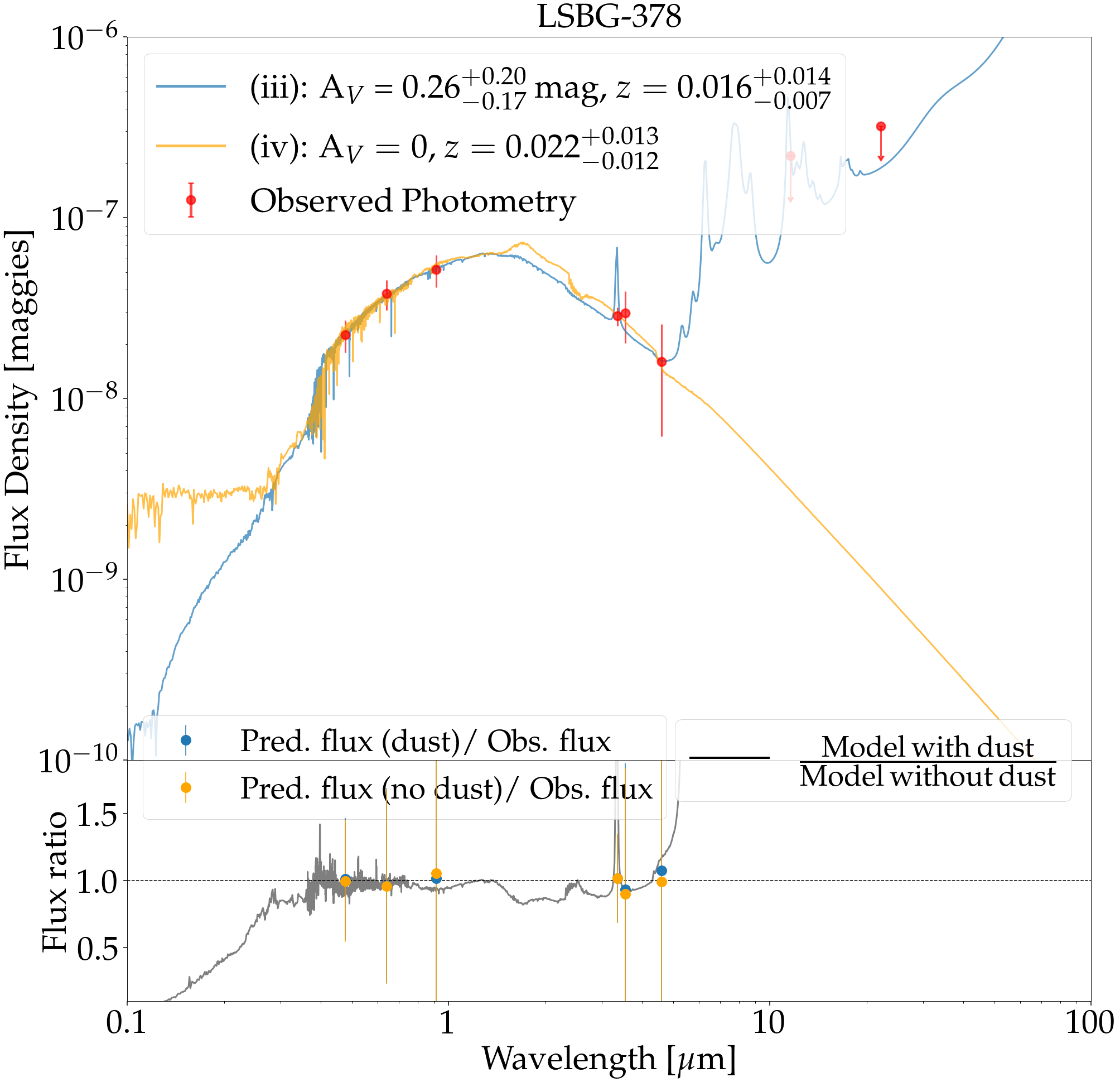}
    \includegraphics[width=\columnwidth]{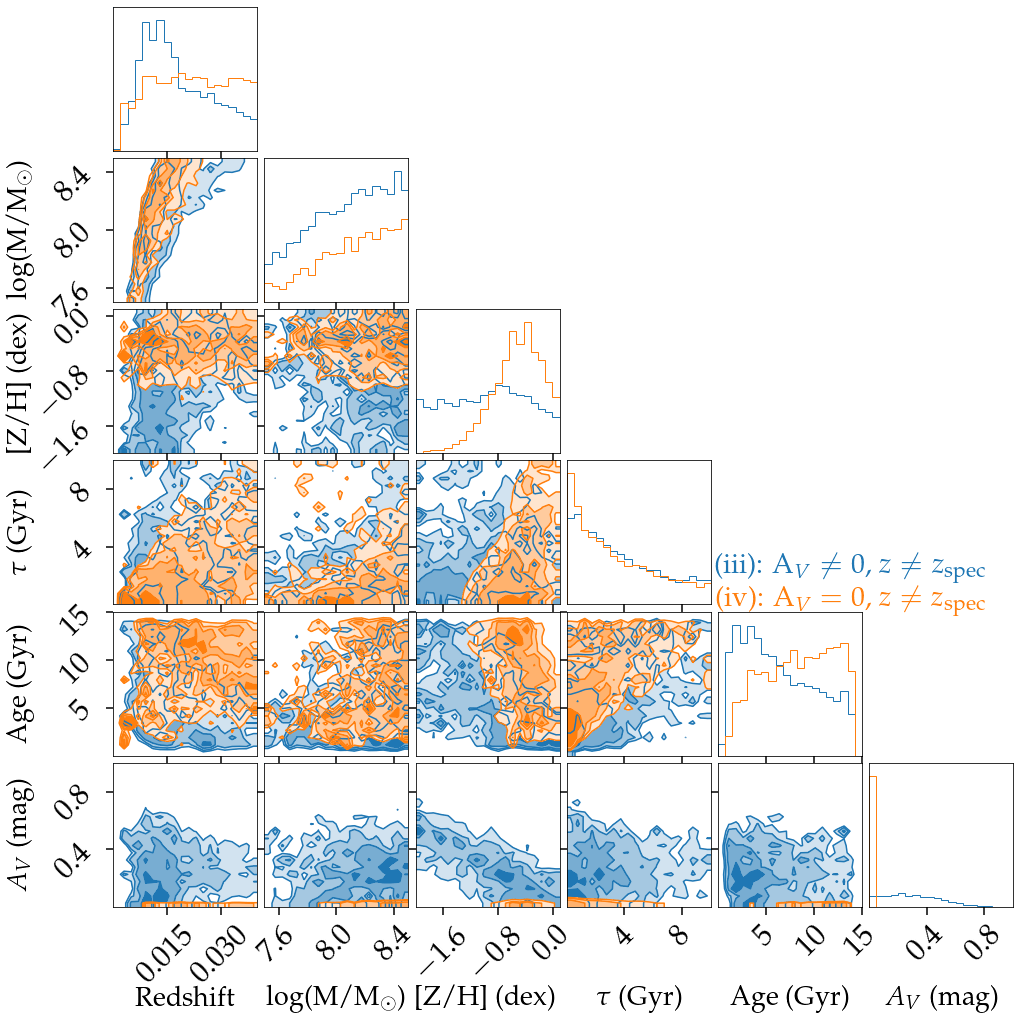}
    \caption{As Fig. \ref{fig:corner_LSBG-490}, but for LSBG-378.}
    \label{fig:corner_LSBG-378}
\end{figure*}

\begin{figure*}
    \centering
    \includegraphics[width=\columnwidth]{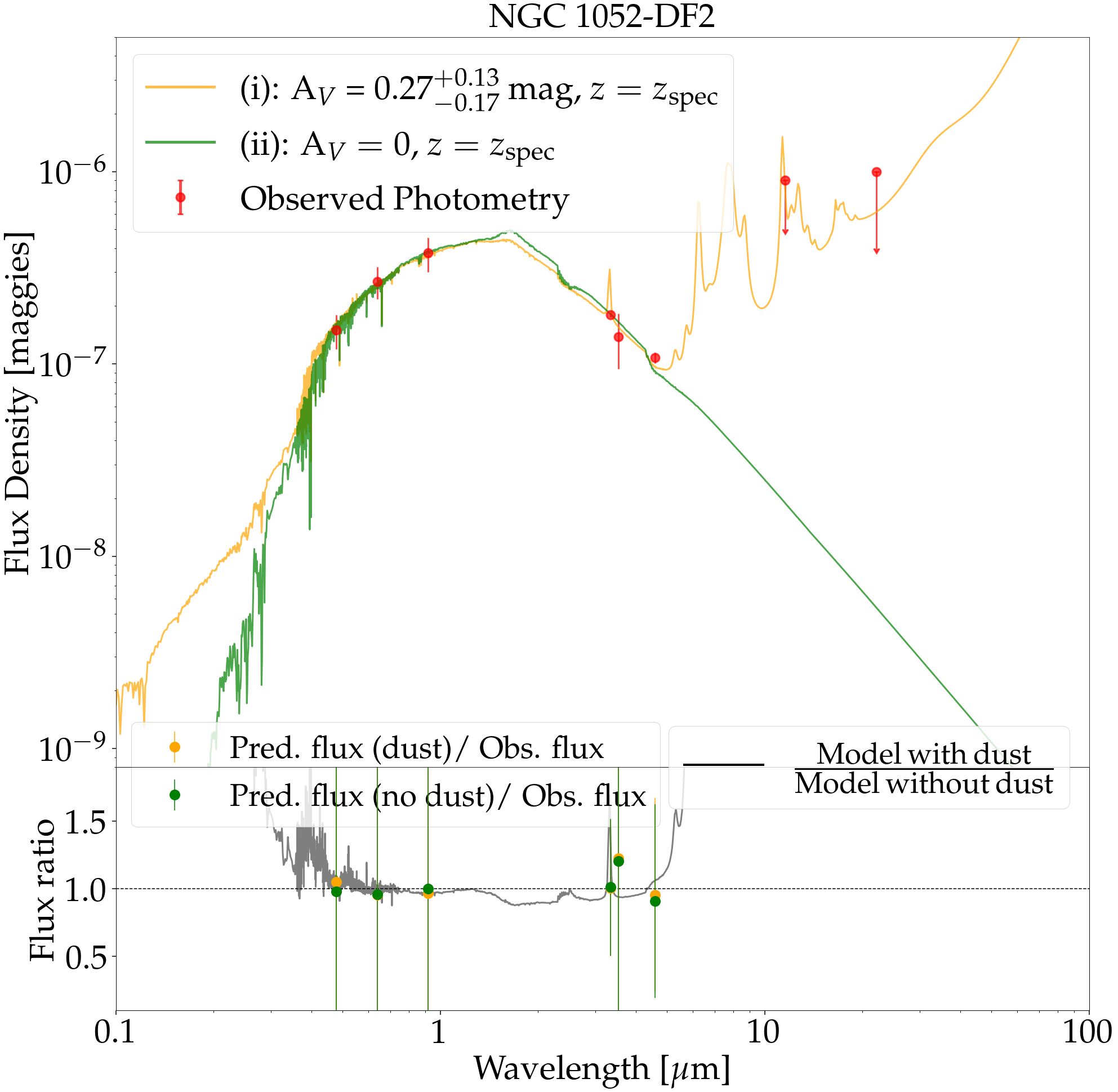}
    \includegraphics[width=\columnwidth]{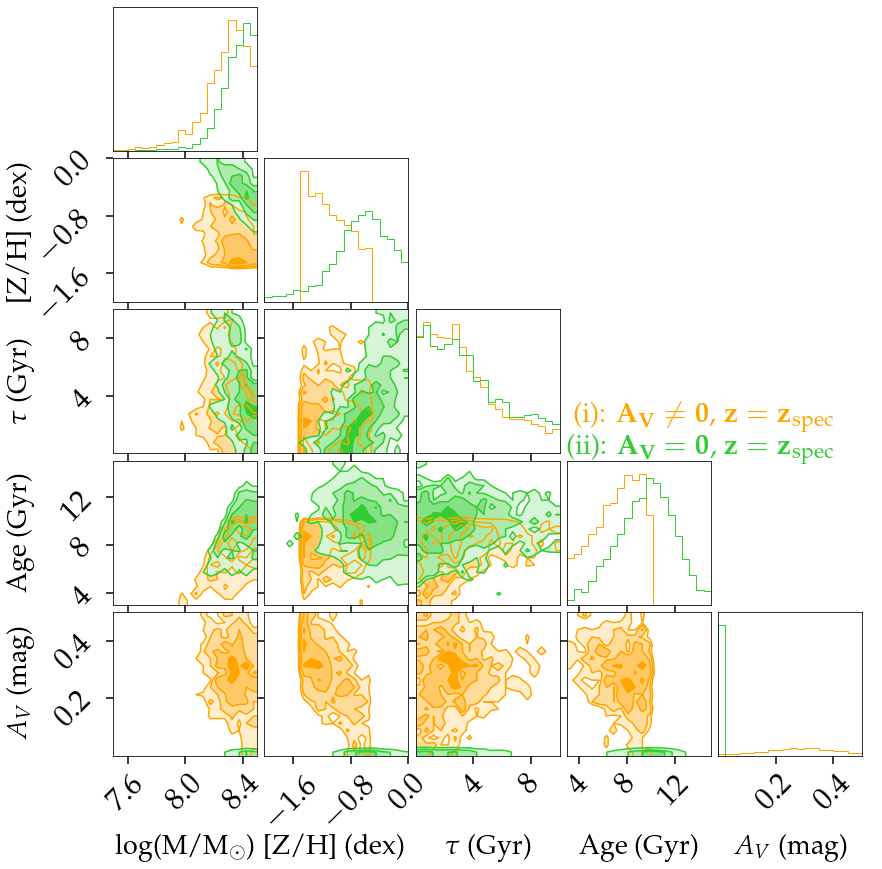}
    \caption{As Fig. \ref{fig:corner_PUDG-R24}, but for NGC 1052-DF2.}
    \label{fig:corner_NGC1052-DF2}
\end{figure*}

\begin{figure*}
    \centering
    \includegraphics[width=\columnwidth]{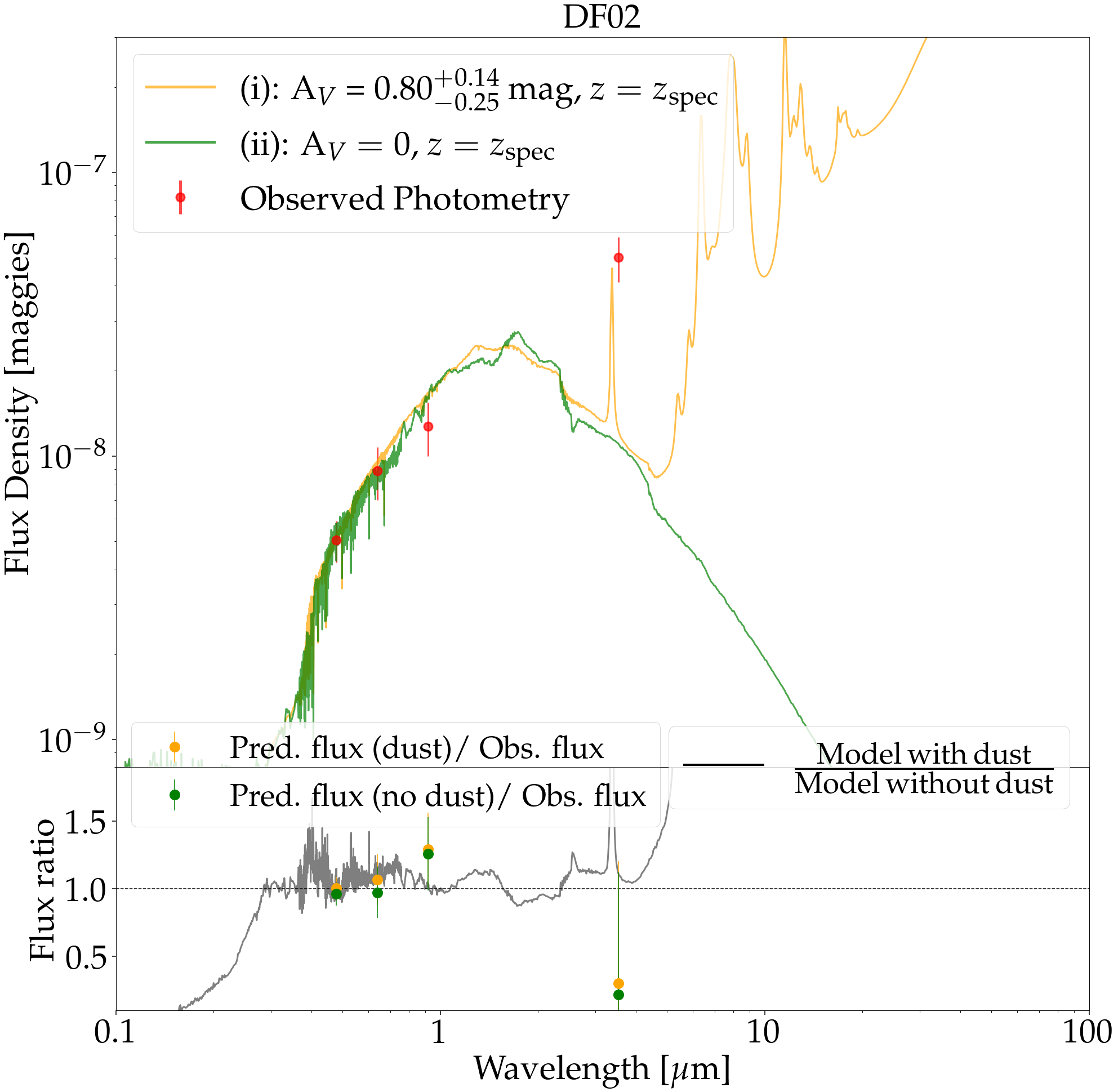}
    \includegraphics[width=\columnwidth]{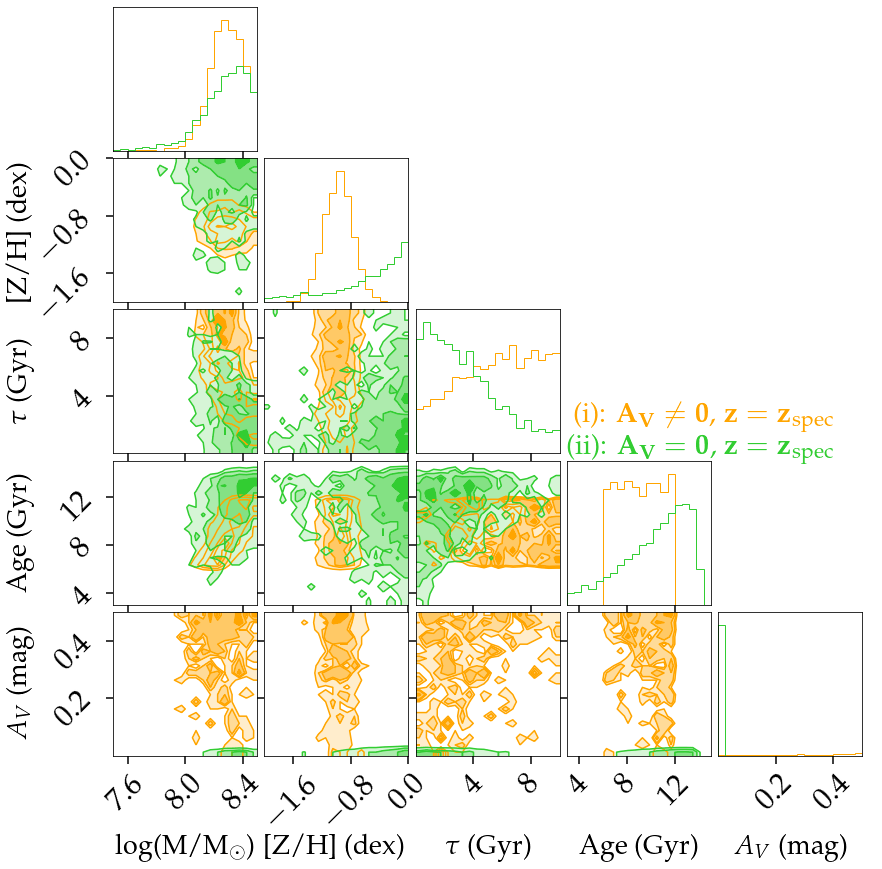}
    \caption{As Fig. \ref{fig:corner_PUDG-R24}, but for DF02.}
    \label{fig:corner_DF2}
\end{figure*}

\begin{figure*}
    \centering
    \includegraphics[width=\columnwidth]{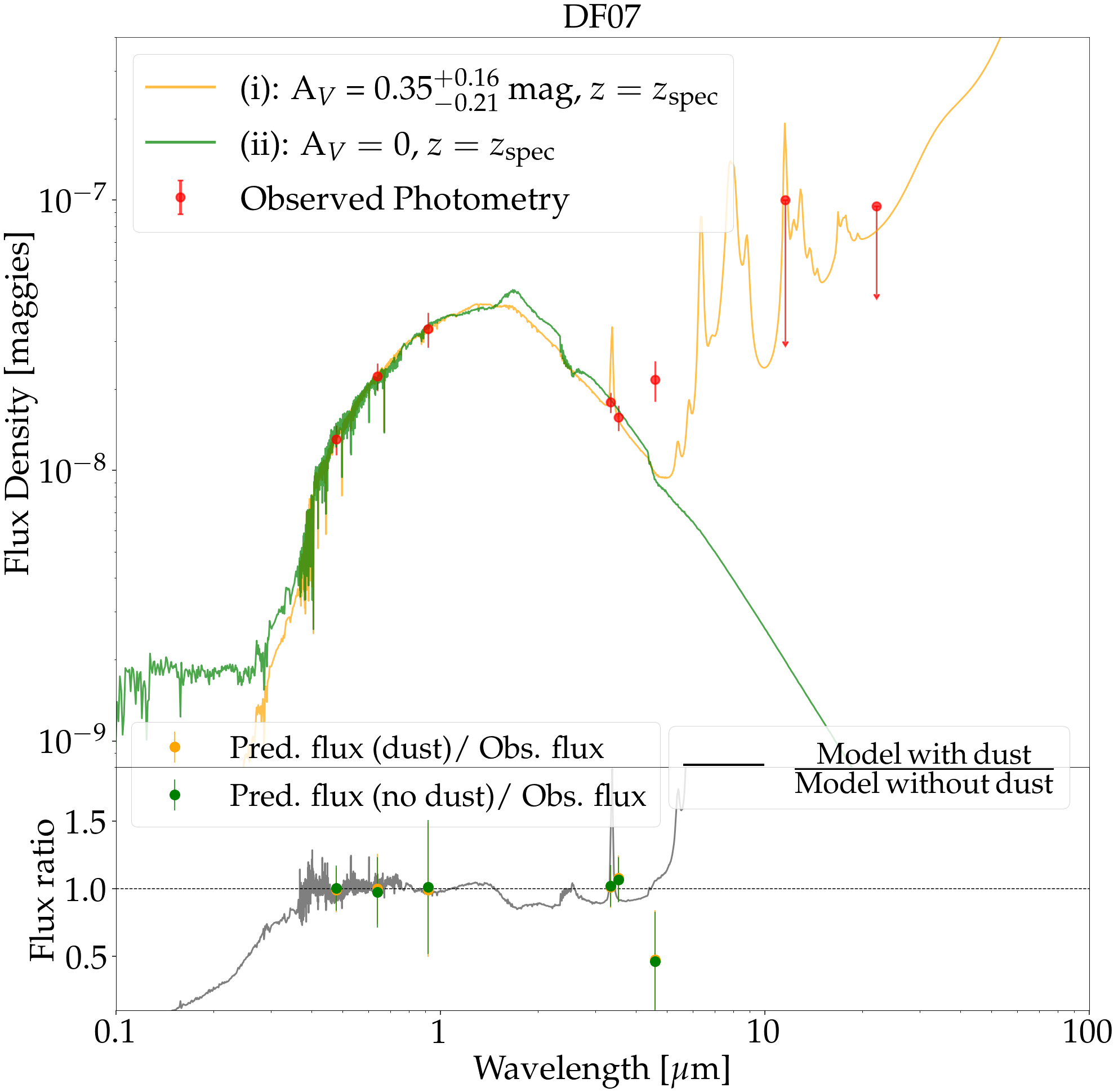}
    \includegraphics[width=\columnwidth]{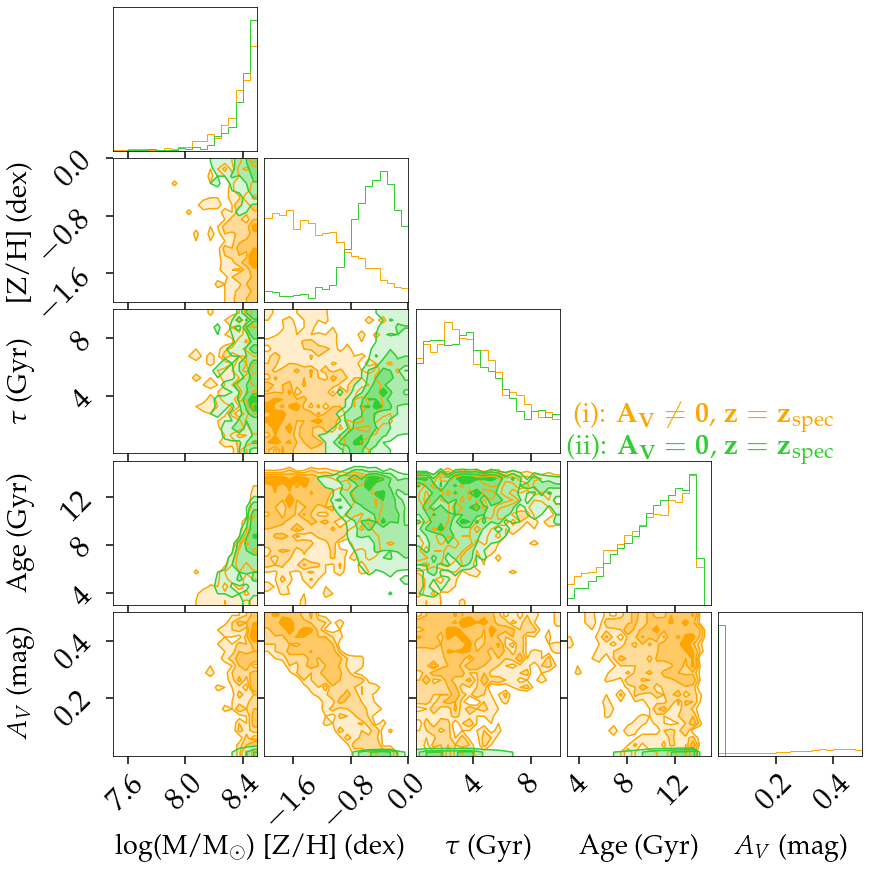}
    \caption{As Fig. \ref{fig:corner_PUDG-R24}, but for DF07.}
    \label{fig:corner_DF7}
\end{figure*}

\begin{figure*}
    \centering
    \includegraphics[width=\columnwidth]{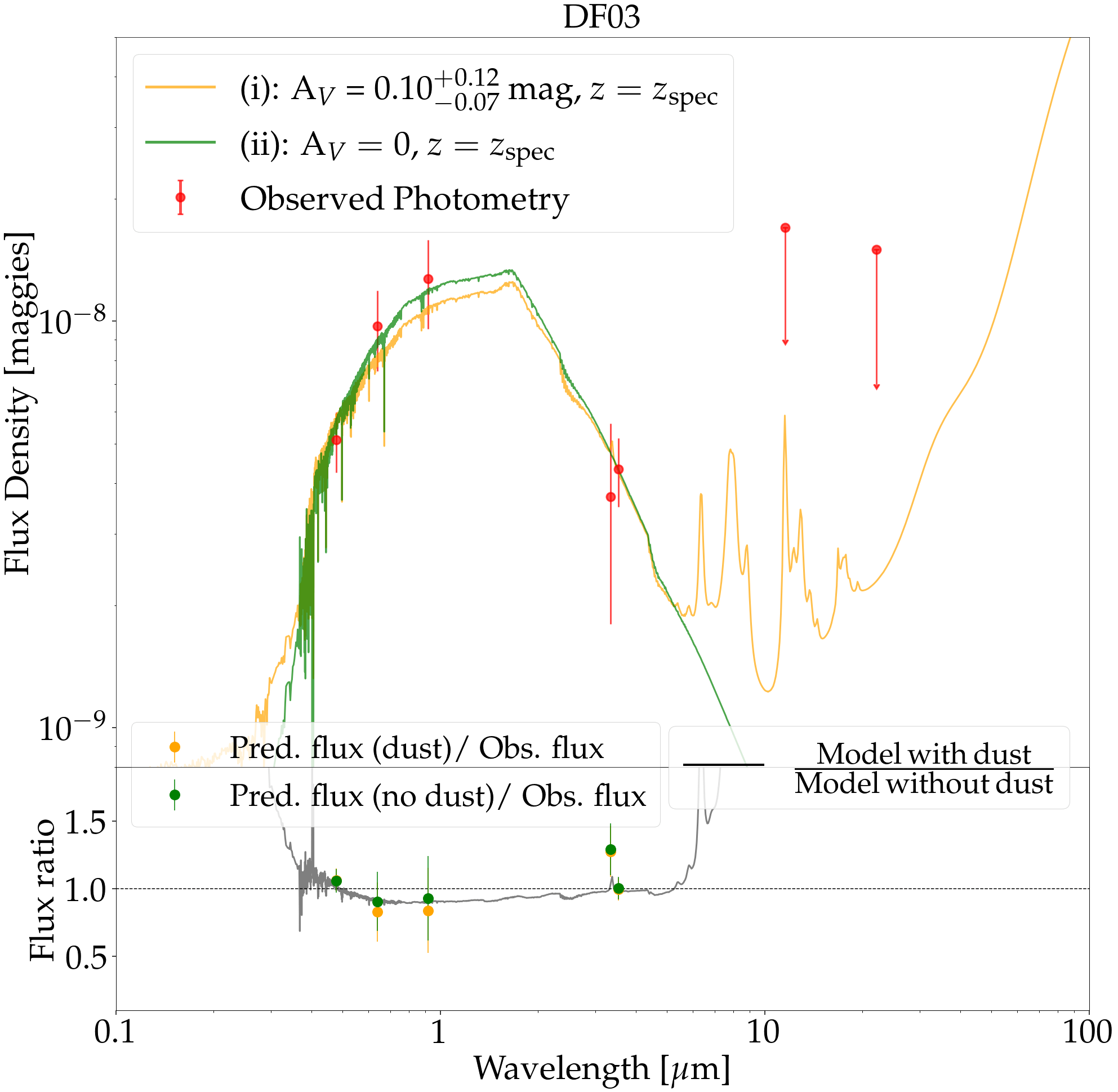}
    \includegraphics[width=\columnwidth]{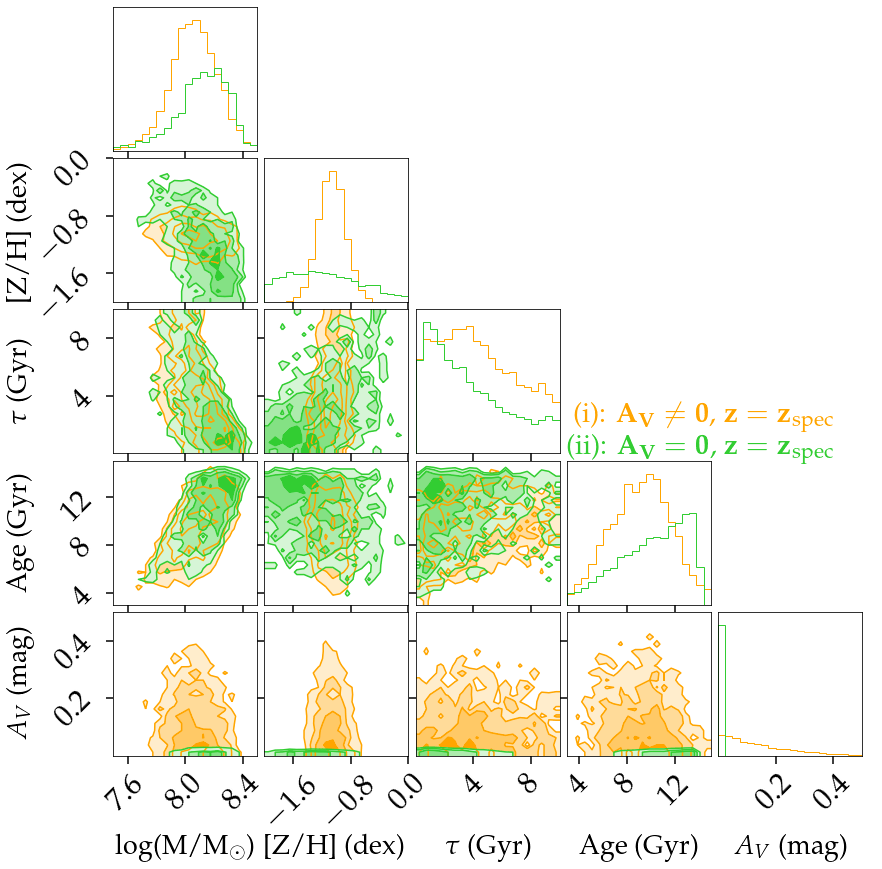}
    \caption{As Fig. \ref{fig:corner_PUDG-R24}, but for DF03.}
    \label{fig:corner_DF3}
\end{figure*}

\begin{figure*}
    \centering
    \includegraphics[width=\columnwidth]{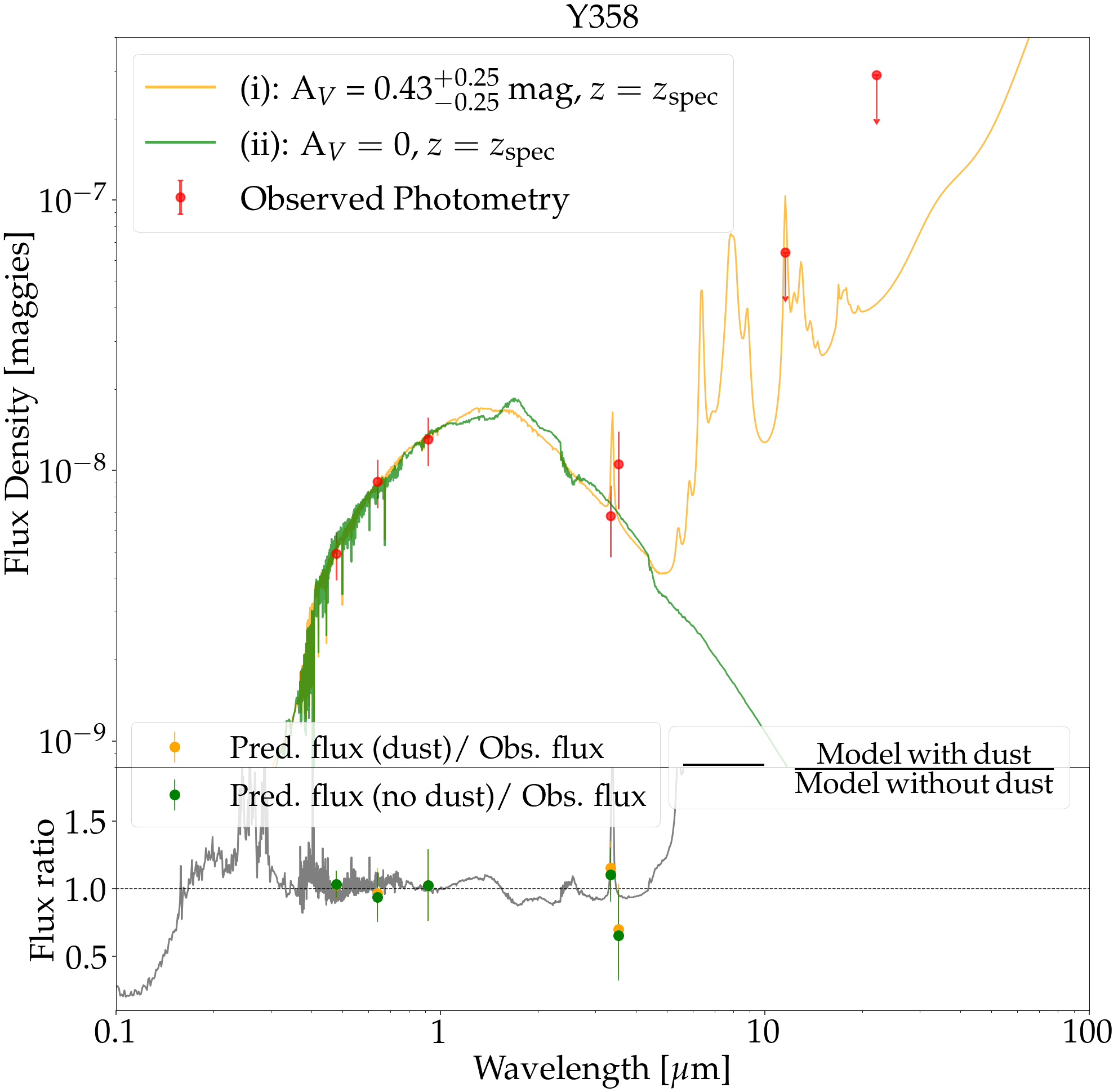}
    \includegraphics[width=\columnwidth]{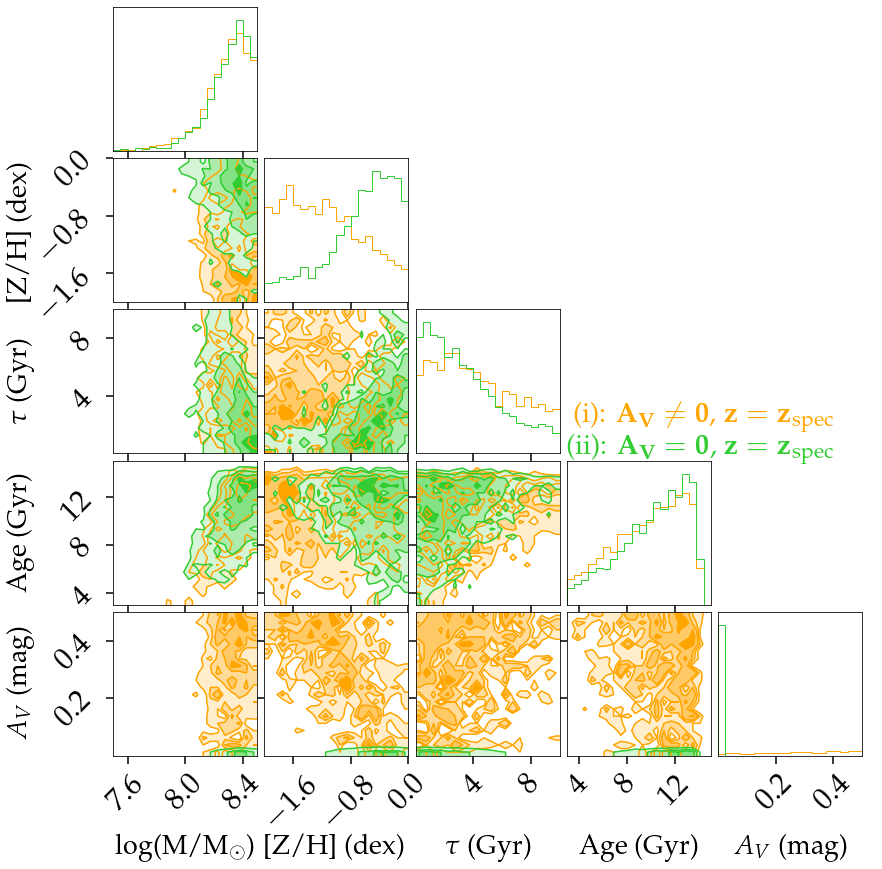}
    \caption{As Fig. \ref{fig:corner_PUDG-R24}, but for Y358.}
    \label{fig:corner_Y358}
\end{figure*}

\begin{figure*}
    \centering
    \includegraphics[width=\columnwidth]{Fig/SED/SEDfitting_DF44_fullcomp_noredshift-3-min.png}
    \includegraphics[width=\columnwidth]{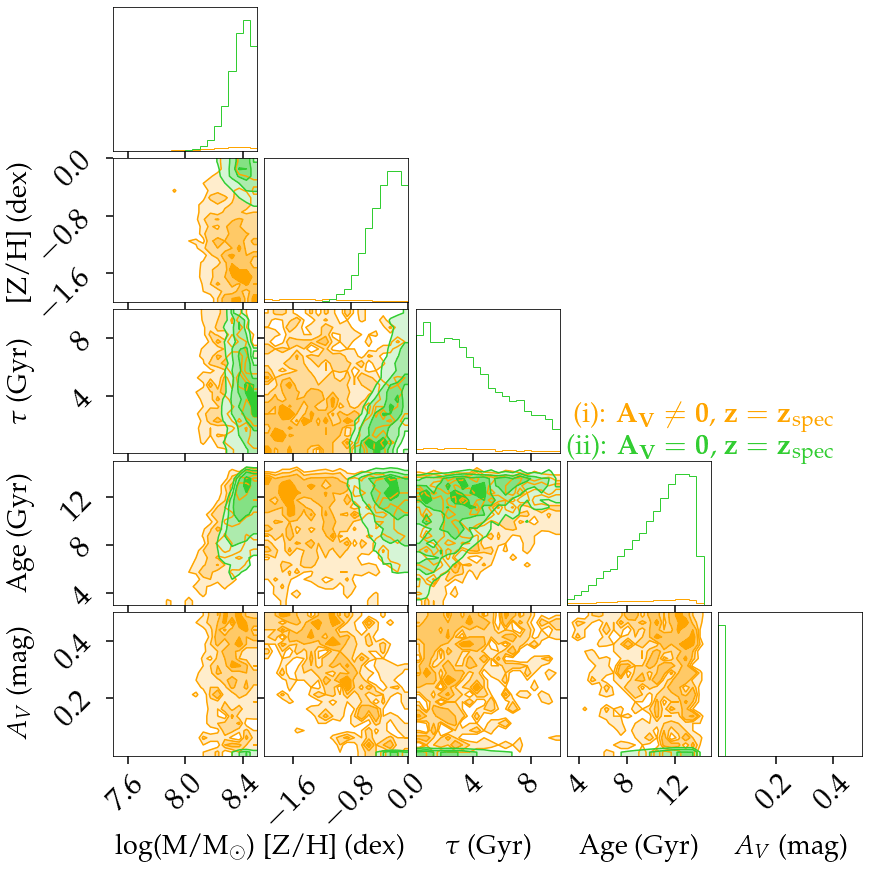}
    \caption{As Fig. \ref{fig:corner_PUDG-R24}, but for DF44.}
    \label{fig:corner_DF44}
\end{figure*}

\begin{figure*}
    \centering
    \includegraphics[width=\columnwidth]{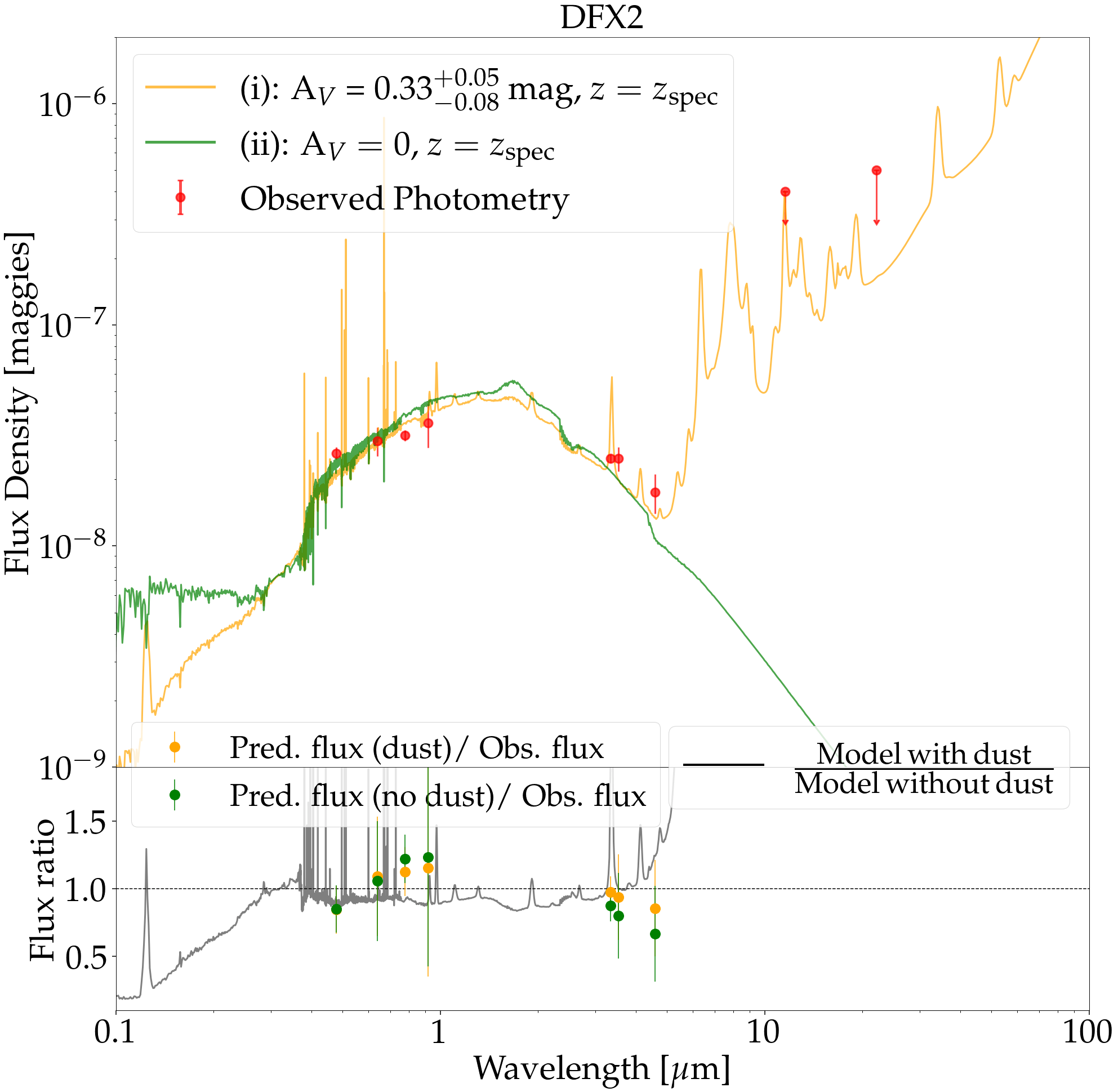}
    \includegraphics[width=\columnwidth]{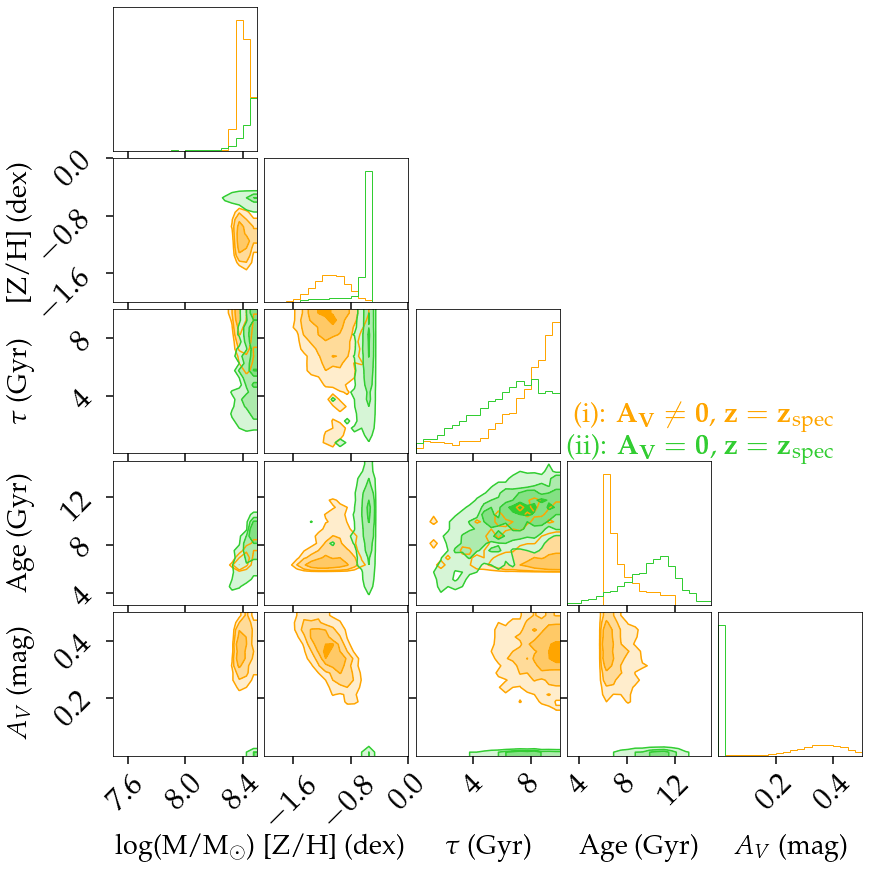}
    \caption{As Fig. \ref{fig:corner_PUDG-R24}, but for DFX2.}
    \label{fig:corner_DFX2}
\end{figure*}

\begin{figure*}
    \centering
    \includegraphics[width=\columnwidth]{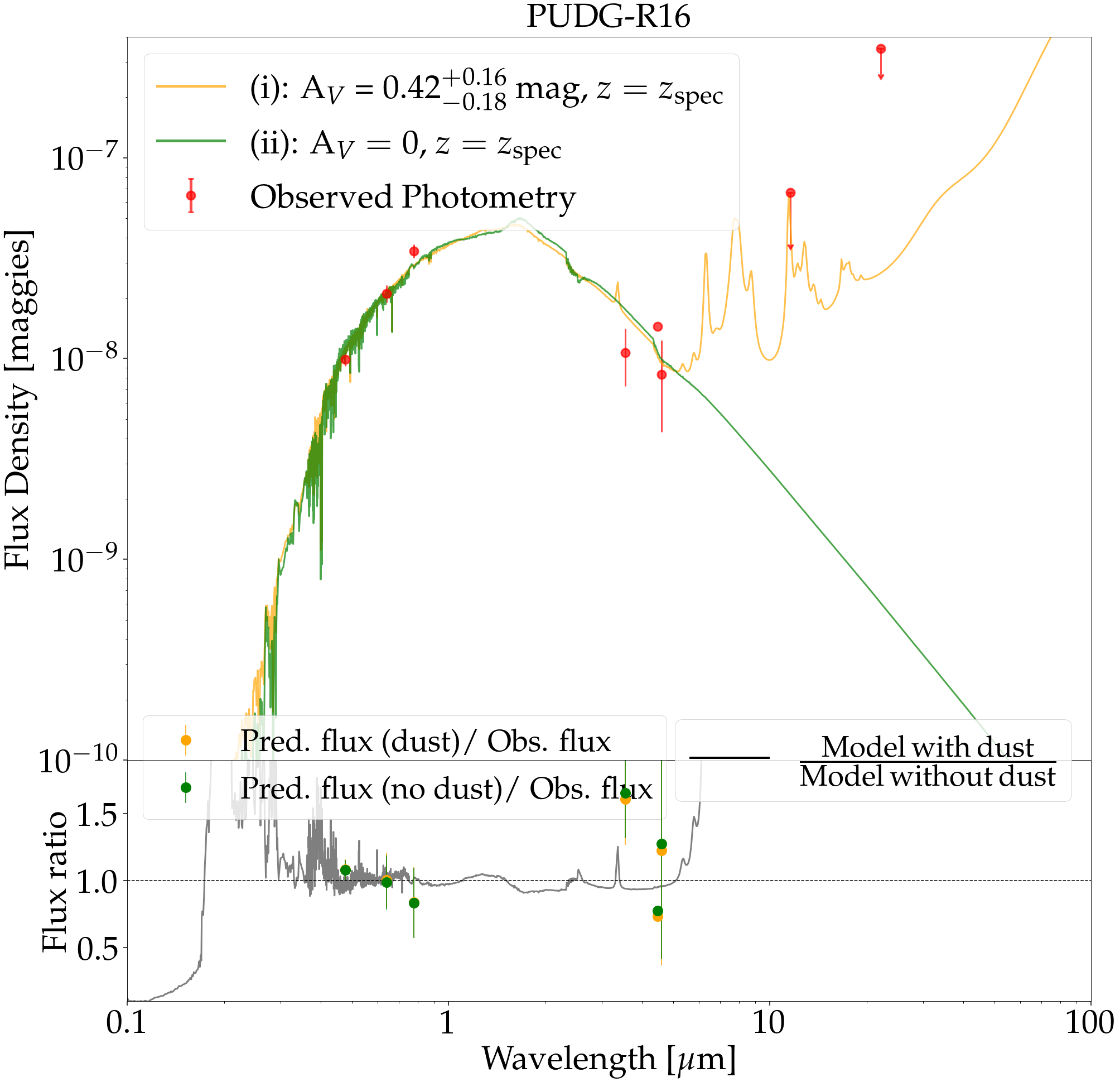}
    \includegraphics[width=\columnwidth]{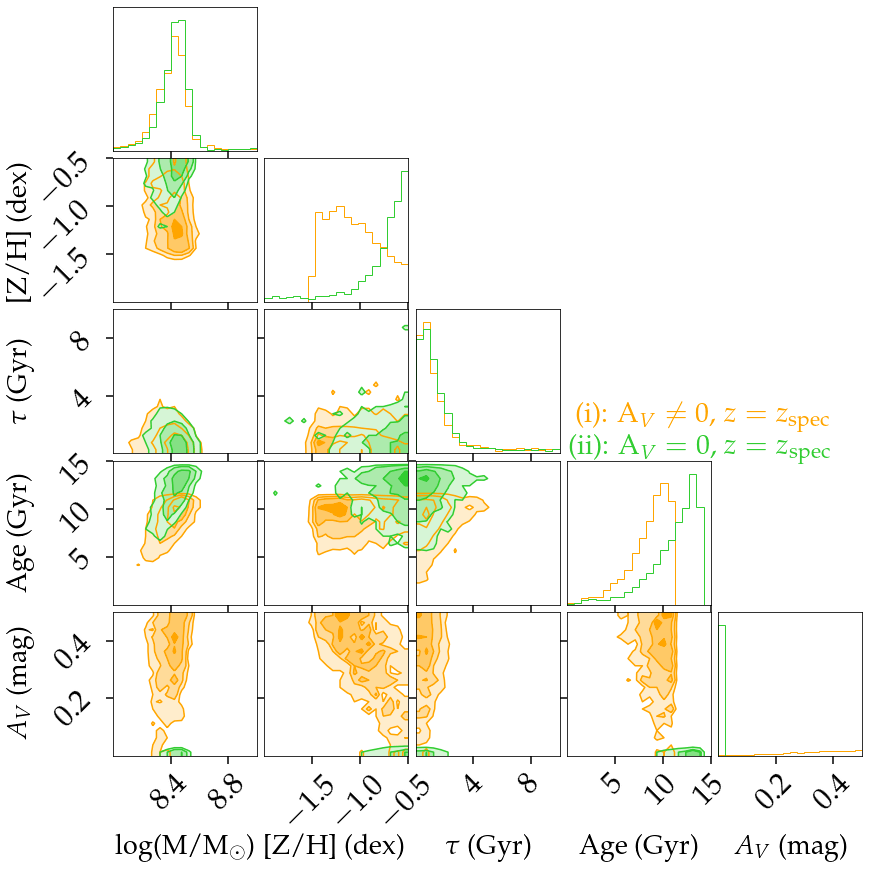}
    \caption{As Fig. \ref{fig:corner_PUDG-R24}, but for PUDG-R16.}
    \label{fig:corner_PUDG-R16}
\end{figure*}

\begin{figure*}
    \centering
    \includegraphics[width=\columnwidth]{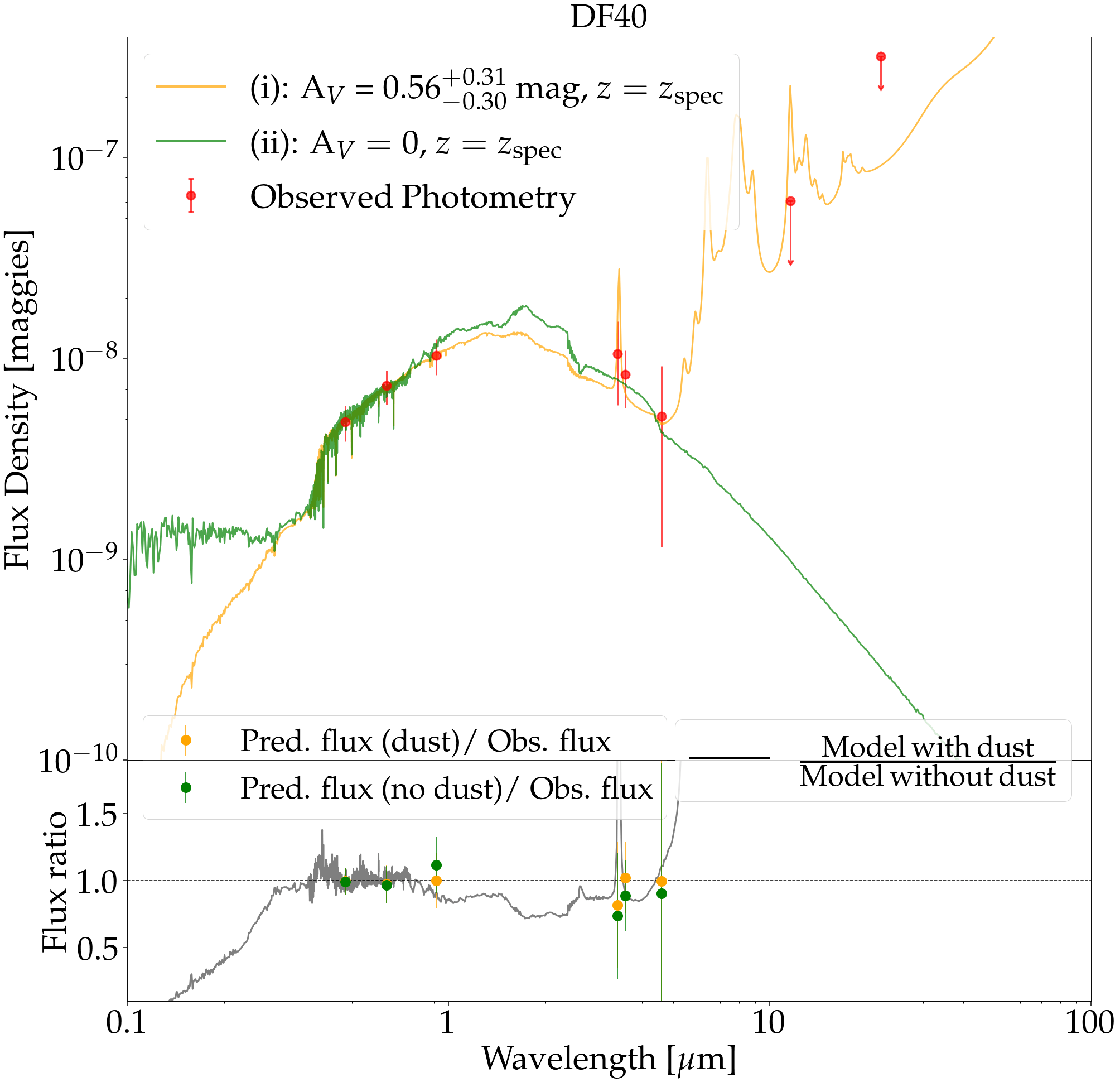}
    \includegraphics[width=\columnwidth]{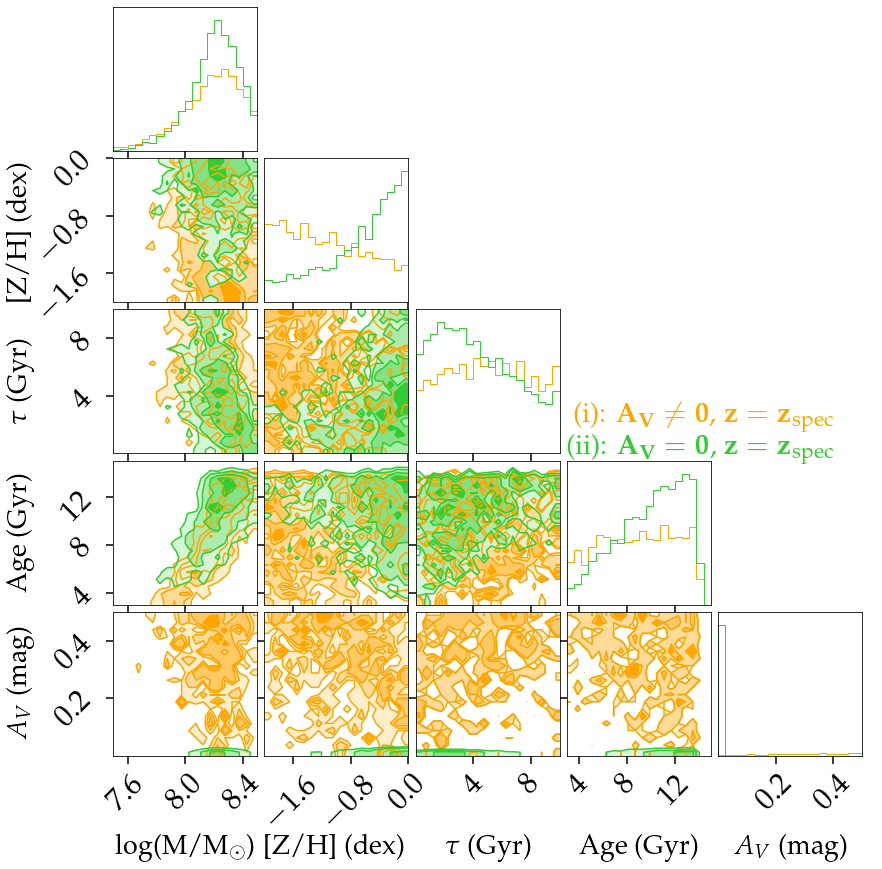}
    \caption{As Fig. \ref{fig:corner_PUDG-R24}, but for DF40.}
    \label{fig:corner_DF40}
\end{figure*}

\begin{figure*}
    \centering
    \includegraphics[width=\columnwidth]{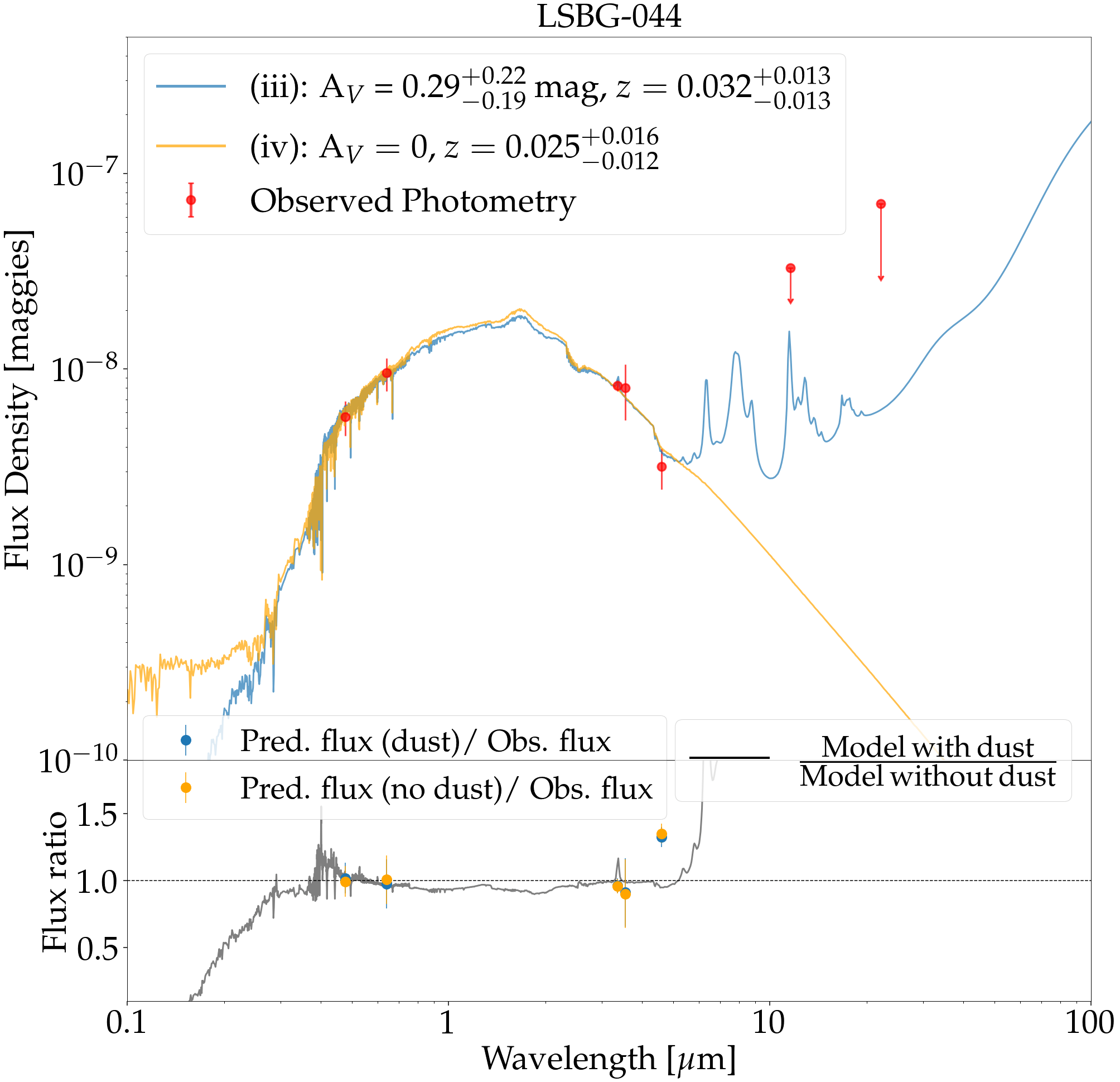}
    \includegraphics[width=\columnwidth]{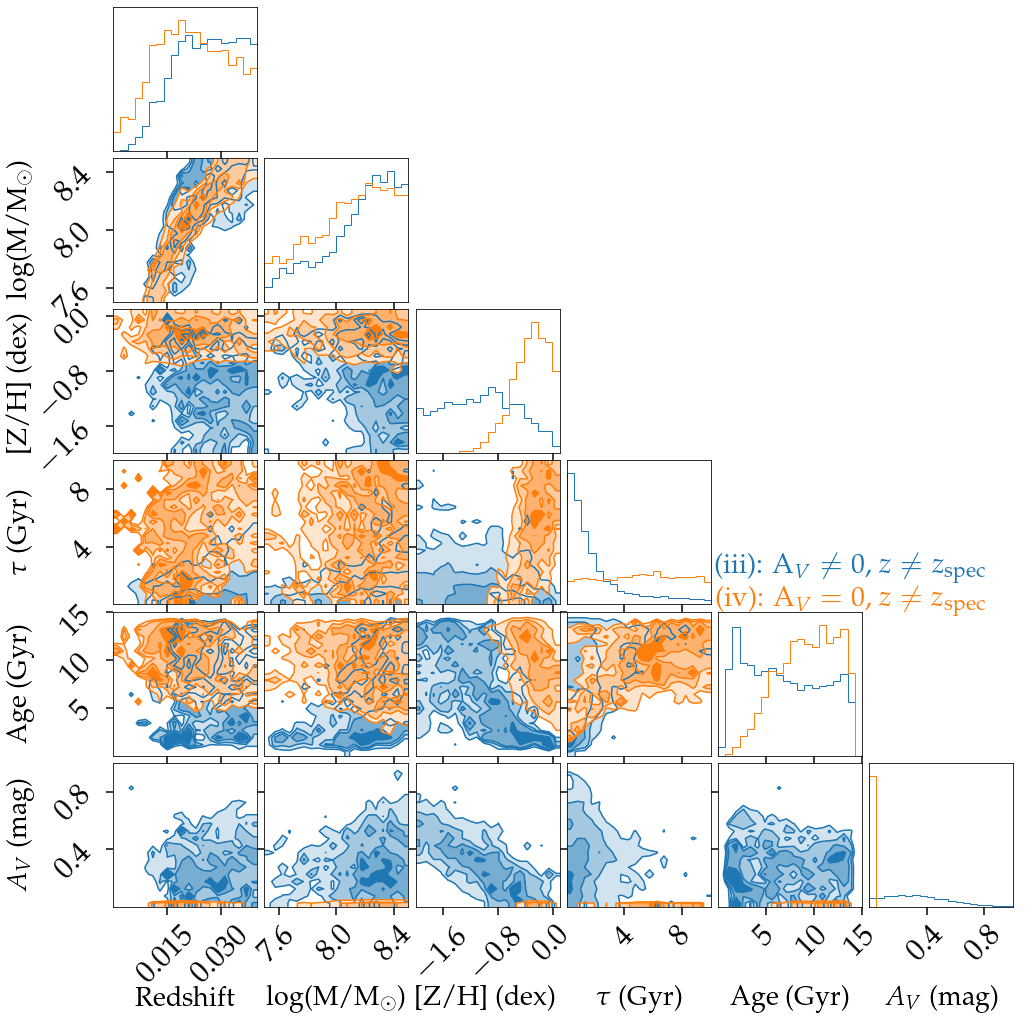}
    \caption{As Fig. \ref{fig:corner_LSBG-490}, but for LSBG-044.}
    \label{fig:corner_LSBG-044}
\end{figure*}

\begin{figure*}
    \centering
    \includegraphics[width=\columnwidth]{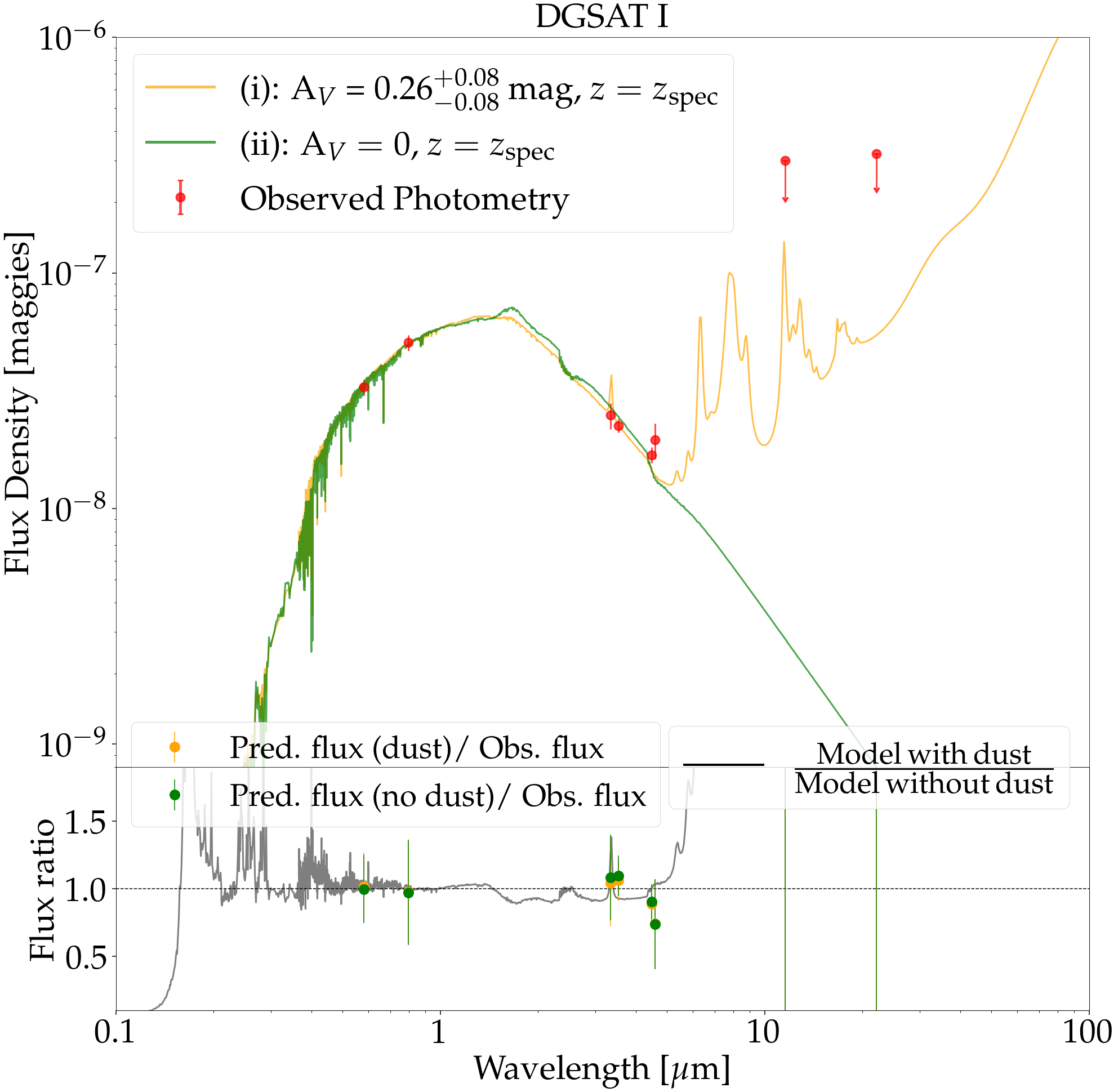}
    \includegraphics[width=\columnwidth]{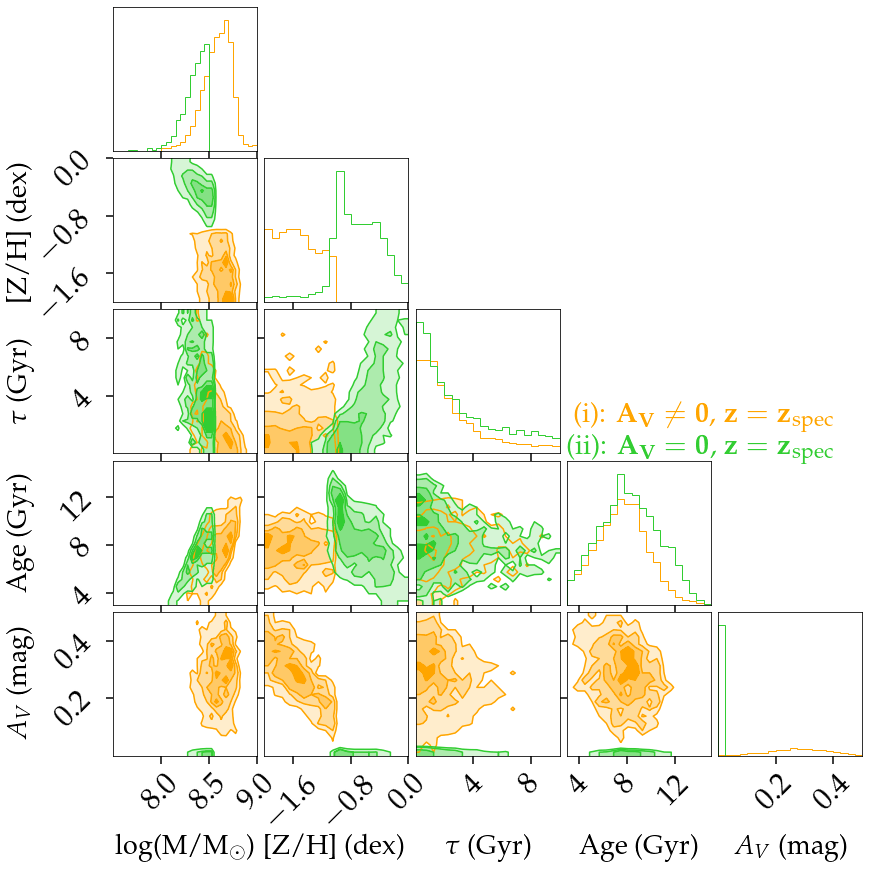}
    \caption{As Fig. \ref{fig:corner_PUDG-R24}, but for DGSAT I.}
    \label{fig:corner_DGSATI}
\end{figure*}

\begin{figure*}
    \centering
    \includegraphics[width=\columnwidth]{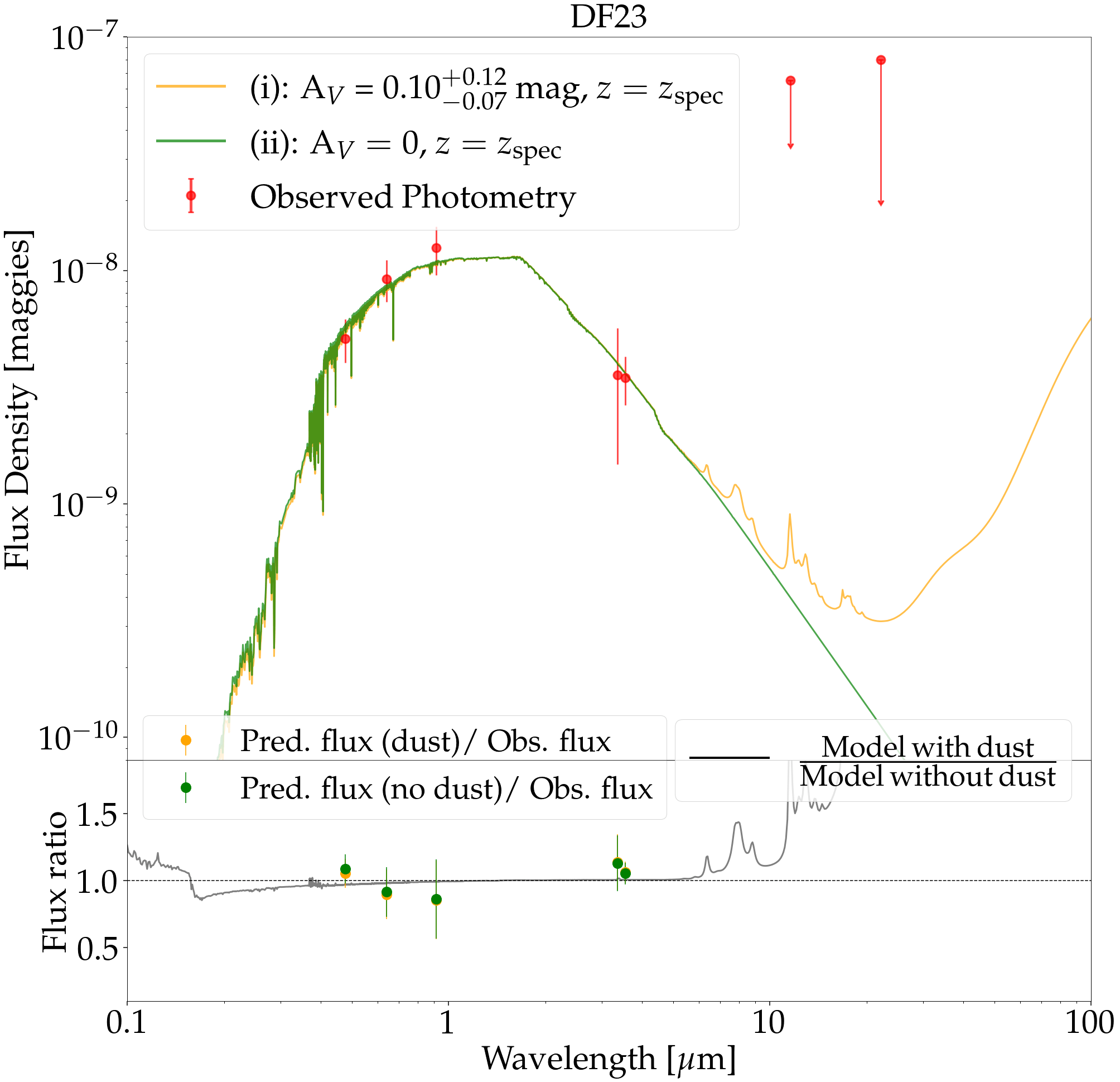}
    \includegraphics[width=\columnwidth]{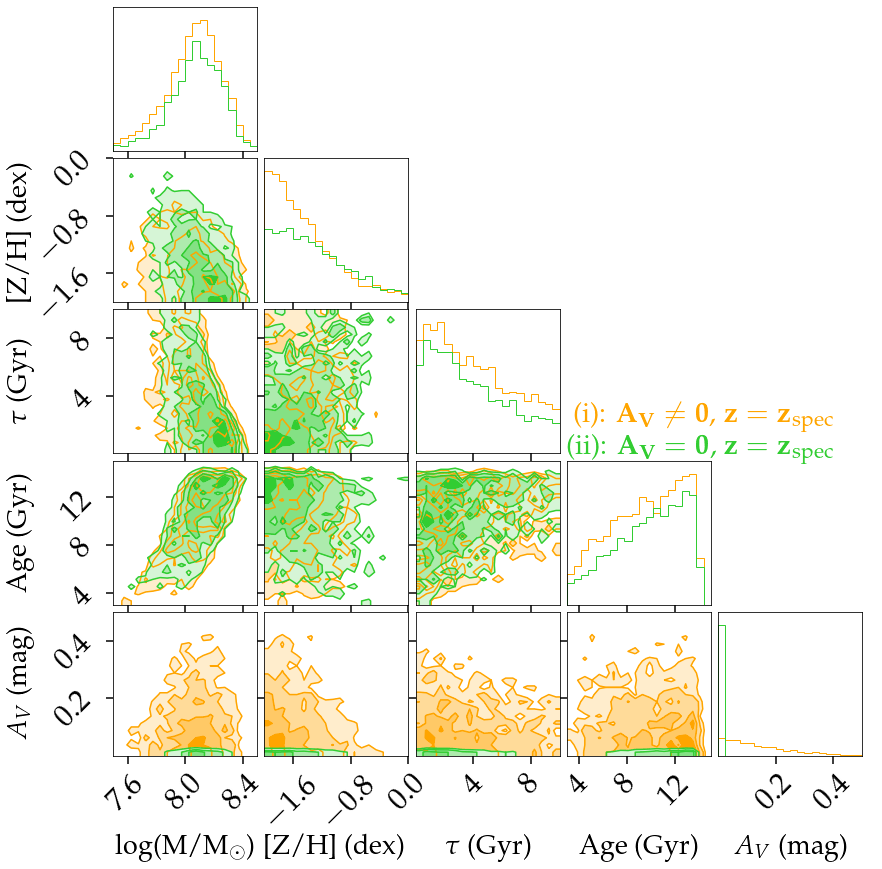}
    \caption{As Fig. \ref{fig:corner_PUDG-R24}, but for DF23.}
    \label{fig:corner_DF23}
\end{figure*}

\begin{figure*}
    \centering
    \includegraphics[width=\columnwidth]{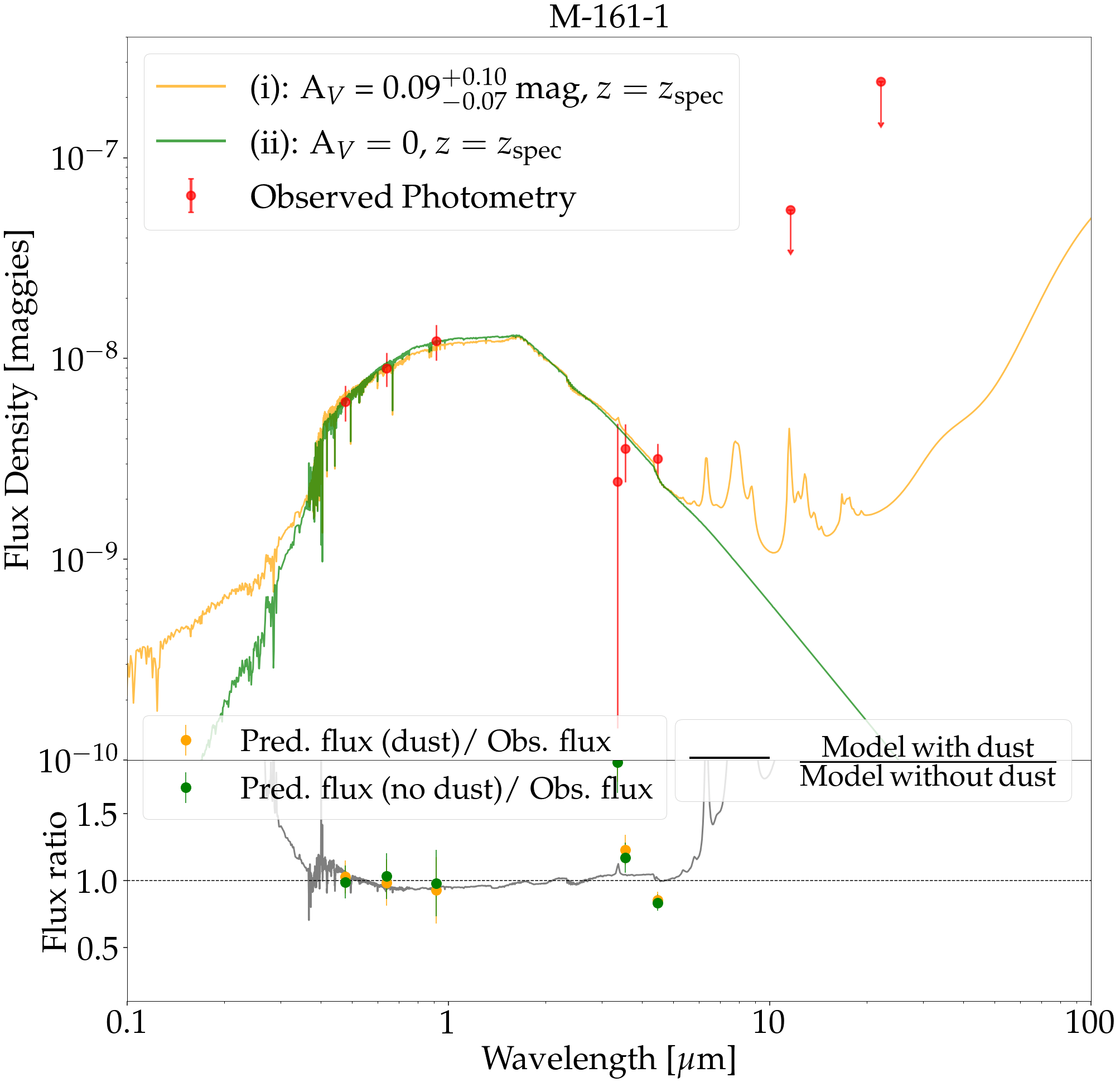}
    \includegraphics[width=\columnwidth]{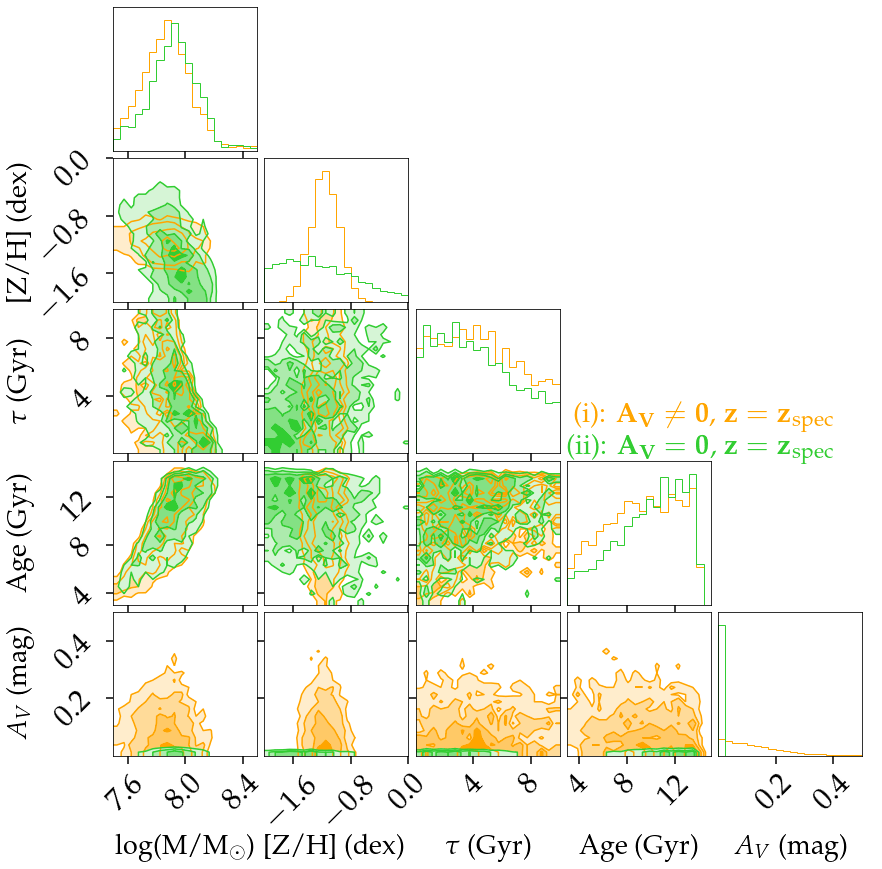}
    \caption{As Fig. \ref{fig:corner_PUDG-R24}, but for M-161-1.}
    \label{fig:corner_M-161-1}
\end{figure*}

\begin{figure*}
    \centering
    \includegraphics[width=\columnwidth]{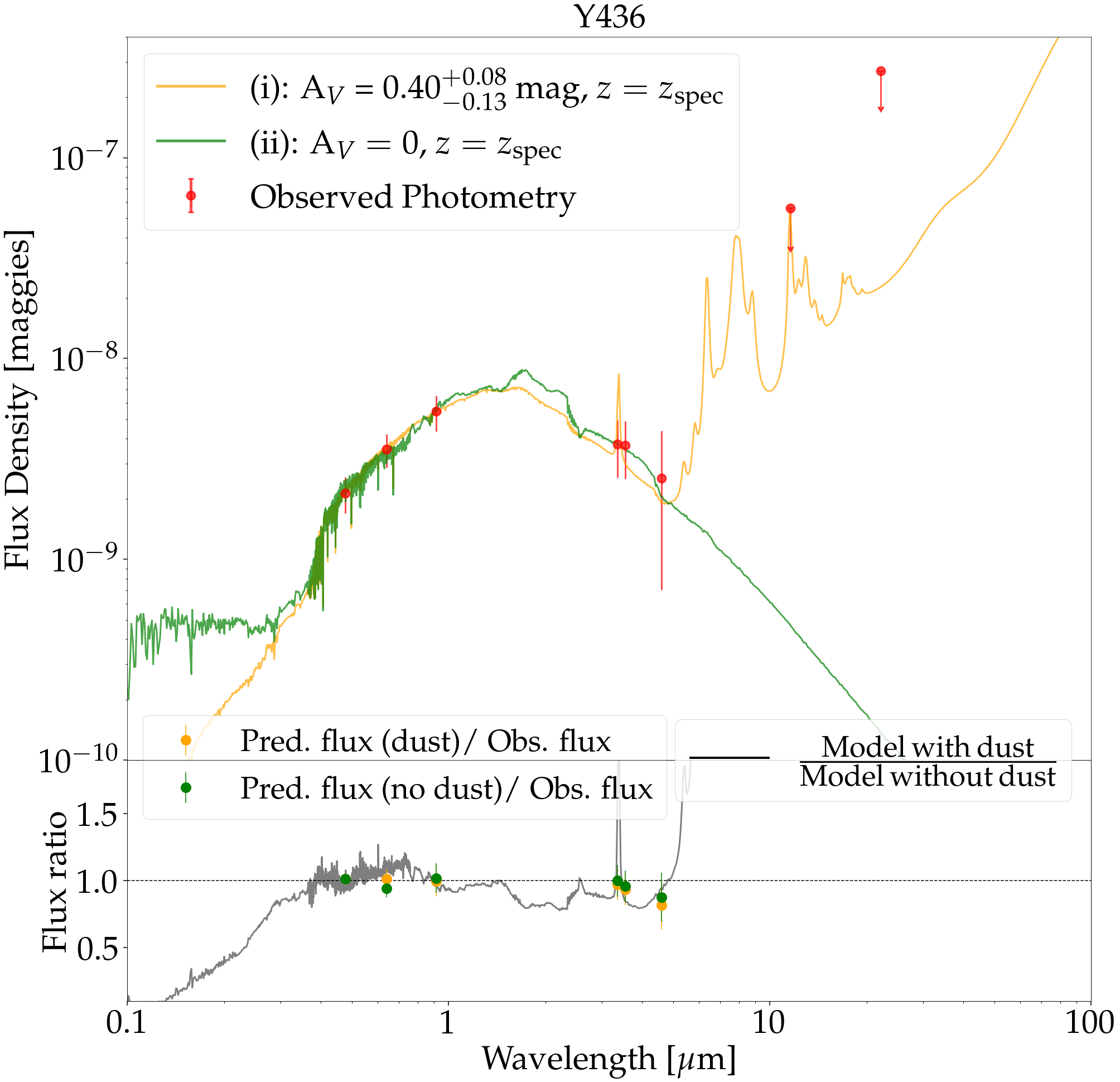}
    \includegraphics[width=\columnwidth]{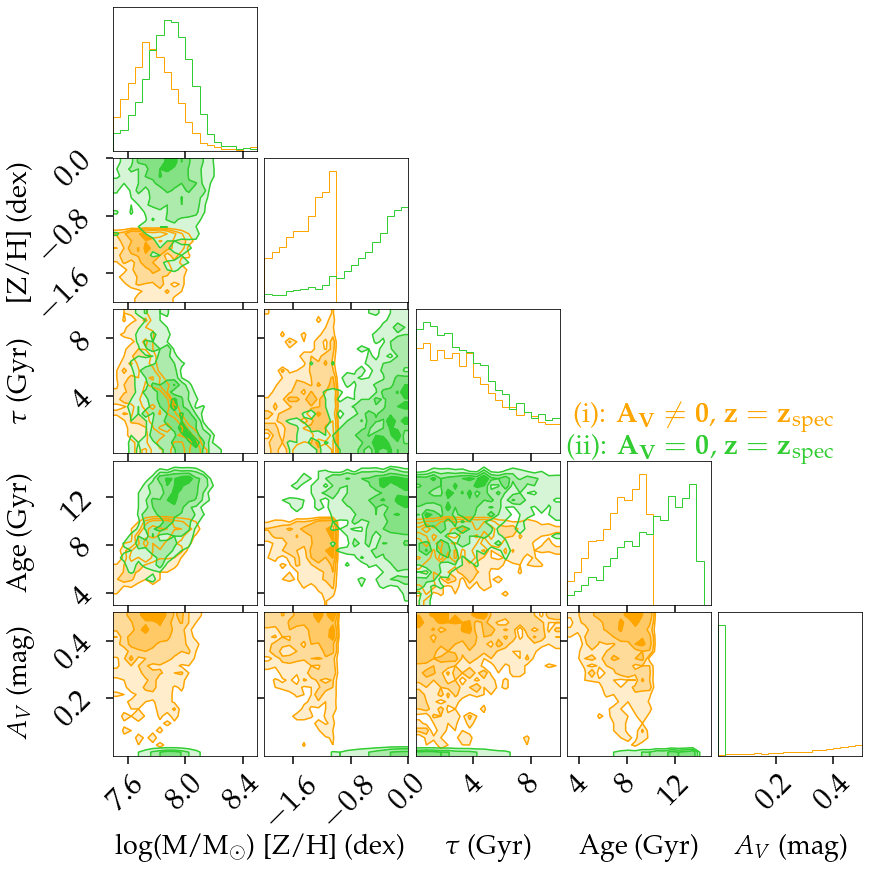}
    \caption{As Fig. \ref{fig:corner_PUDG-R24}, but for Y436.}
    \label{fig:corner_Y436}
\end{figure*}

\begin{figure*}
    \centering
    \includegraphics[width=\columnwidth]{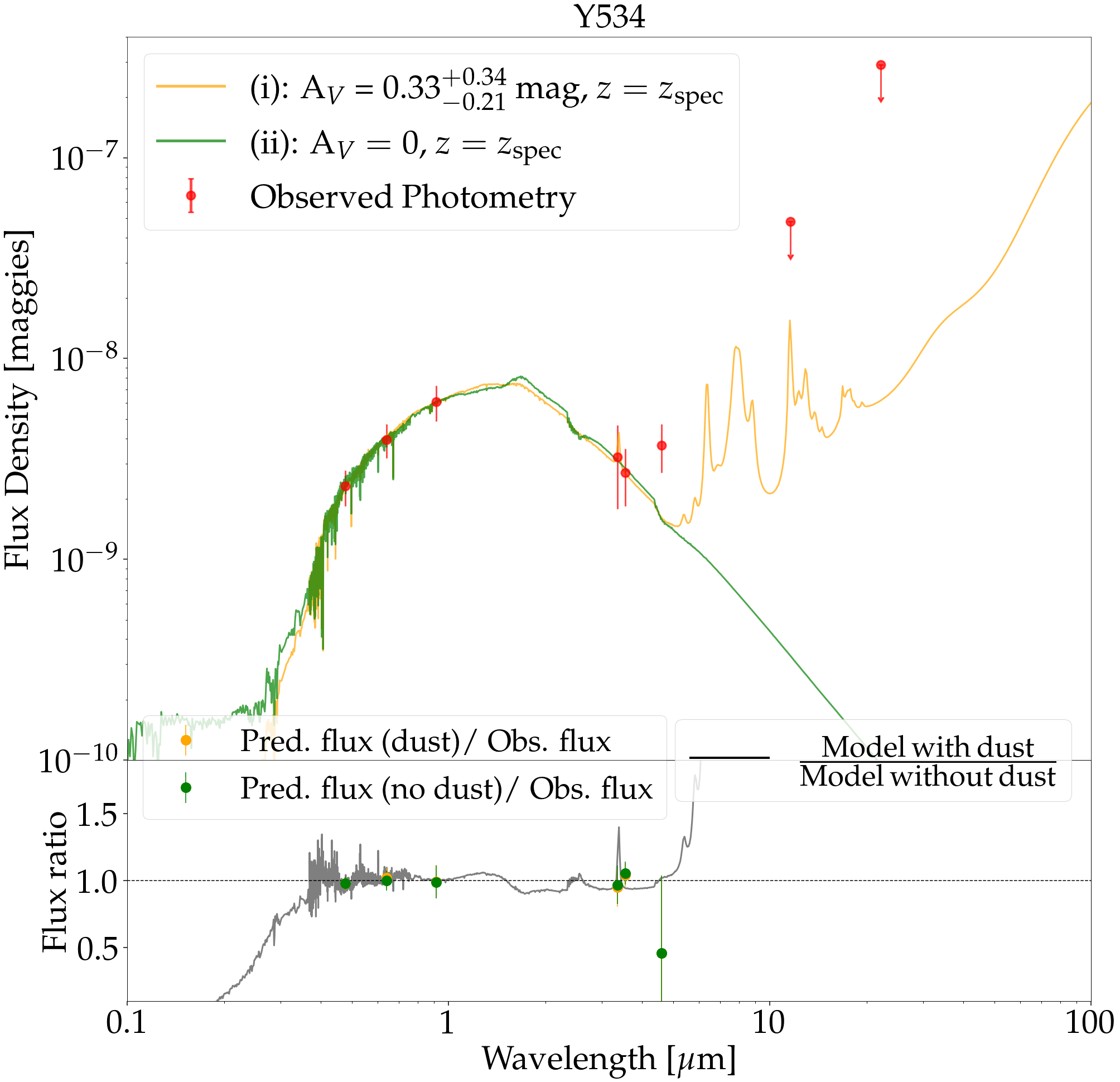}
    \includegraphics[width=\columnwidth]{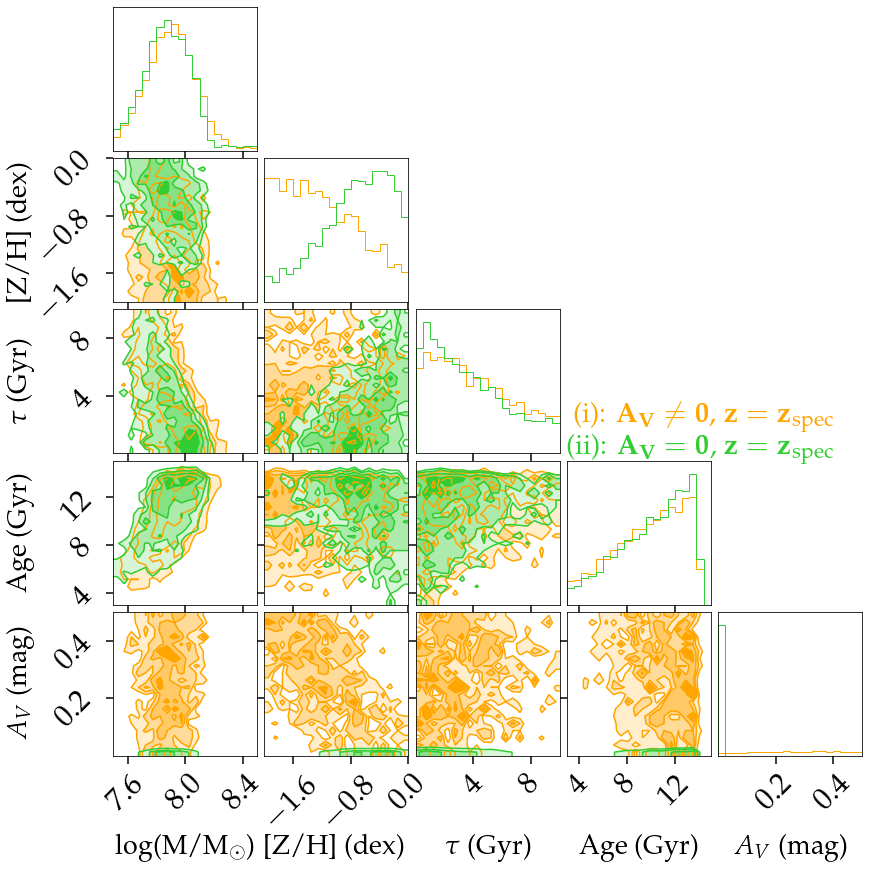}
    \caption{As Fig. \ref{fig:corner_PUDG-R24}, but for Y534.}
    \label{fig:corner_Y534}
\end{figure*}

\begin{figure*}
    \centering
    \includegraphics[width=\columnwidth]{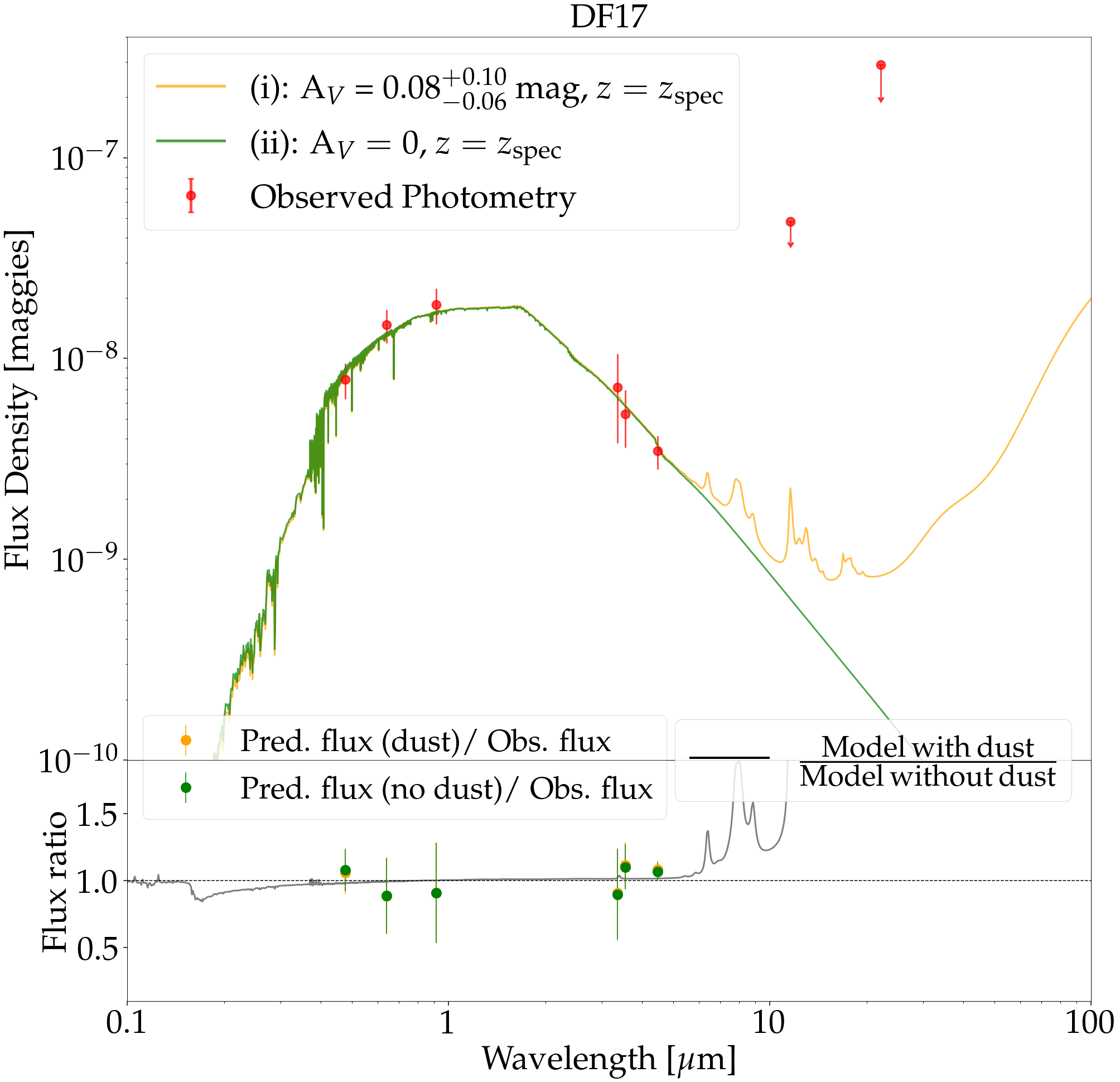}
    \includegraphics[width=\columnwidth]{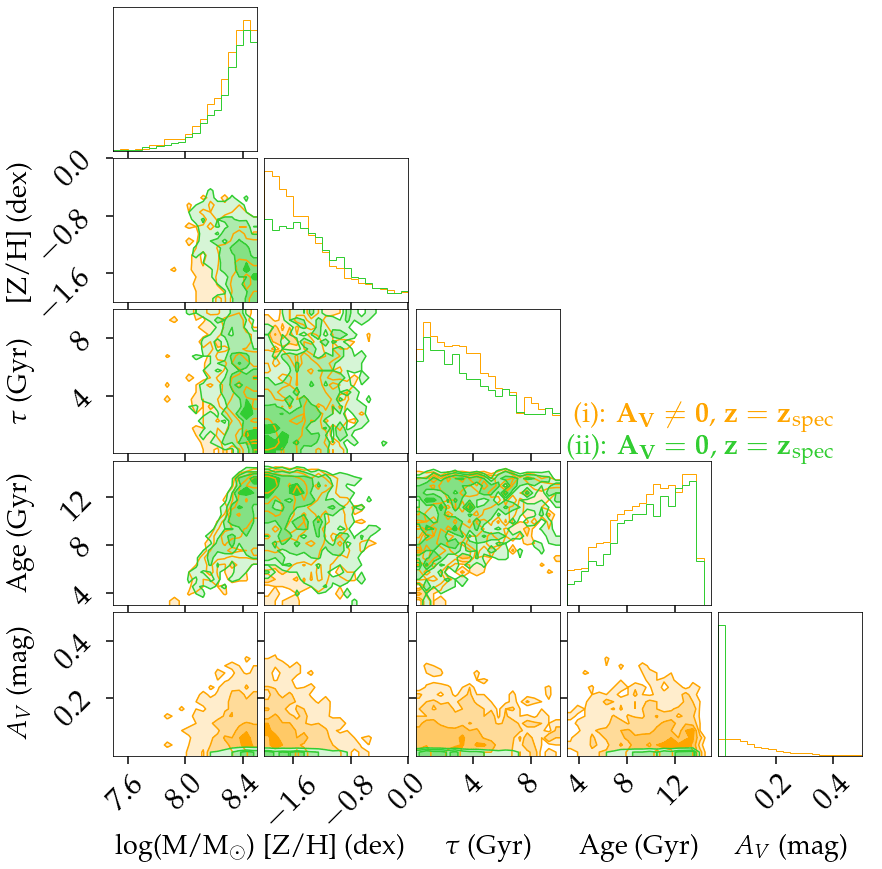}
    \caption{As Fig. \ref{fig:corner_PUDG-R24}, but for DF17.}
    \label{fig:corner_DF17}
\end{figure*}

\begin{figure*}
    \centering
    \includegraphics[width=\columnwidth]{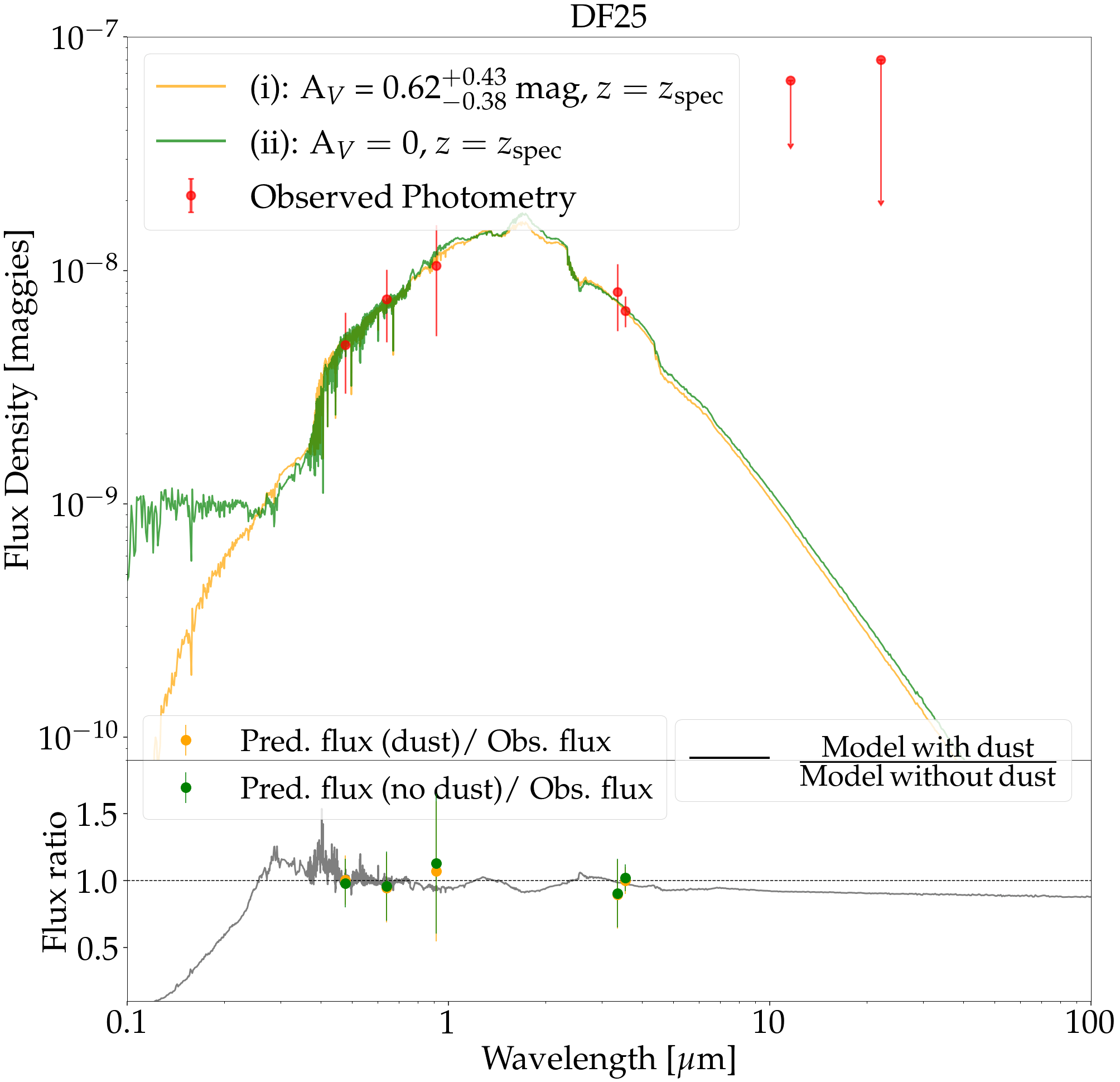}
    \includegraphics[width=\columnwidth]{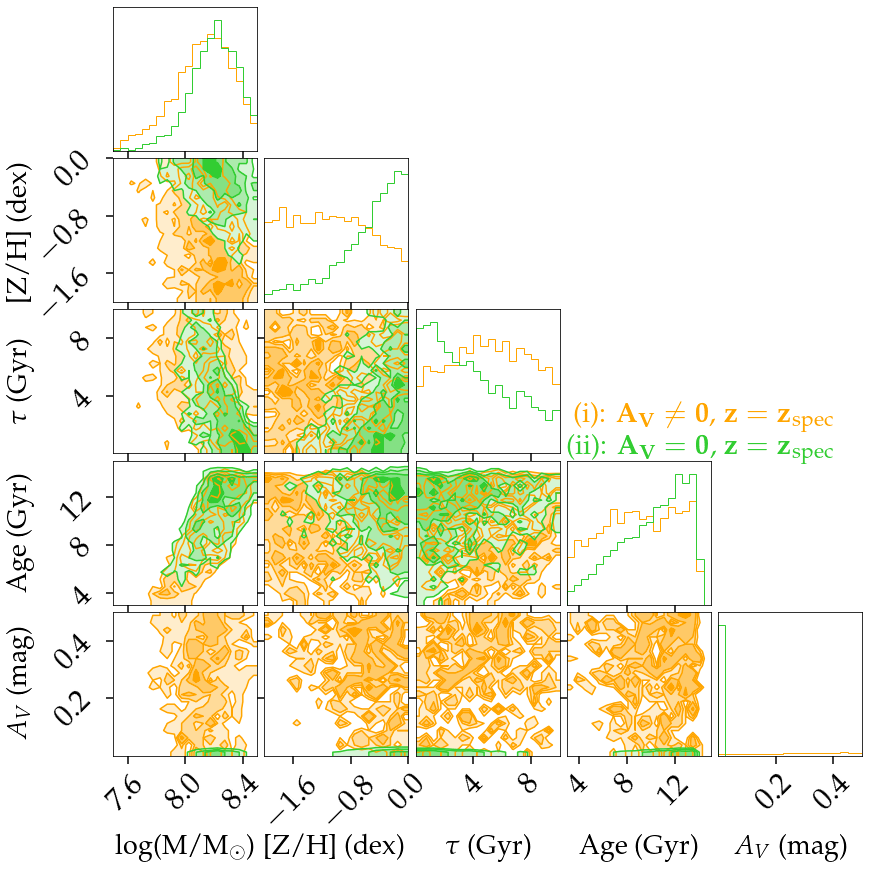}
    \caption{As Fig. \ref{fig:corner_PUDG-R24}, but for DF25.}
    \label{fig:corner_DF25}
\end{figure*}

\begin{figure*}
    \centering
    \includegraphics[width=\columnwidth]{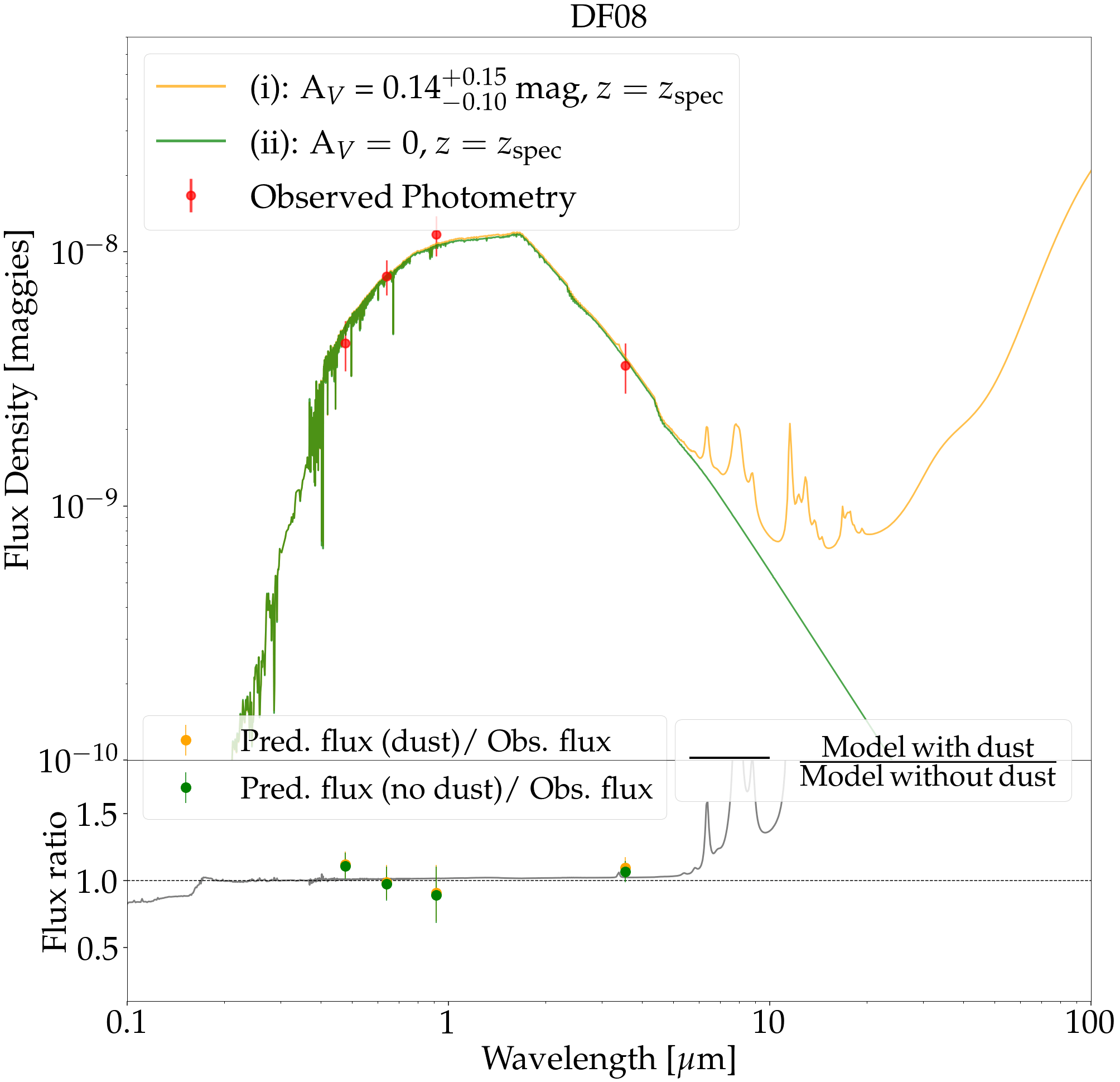}
    \includegraphics[width=\columnwidth]{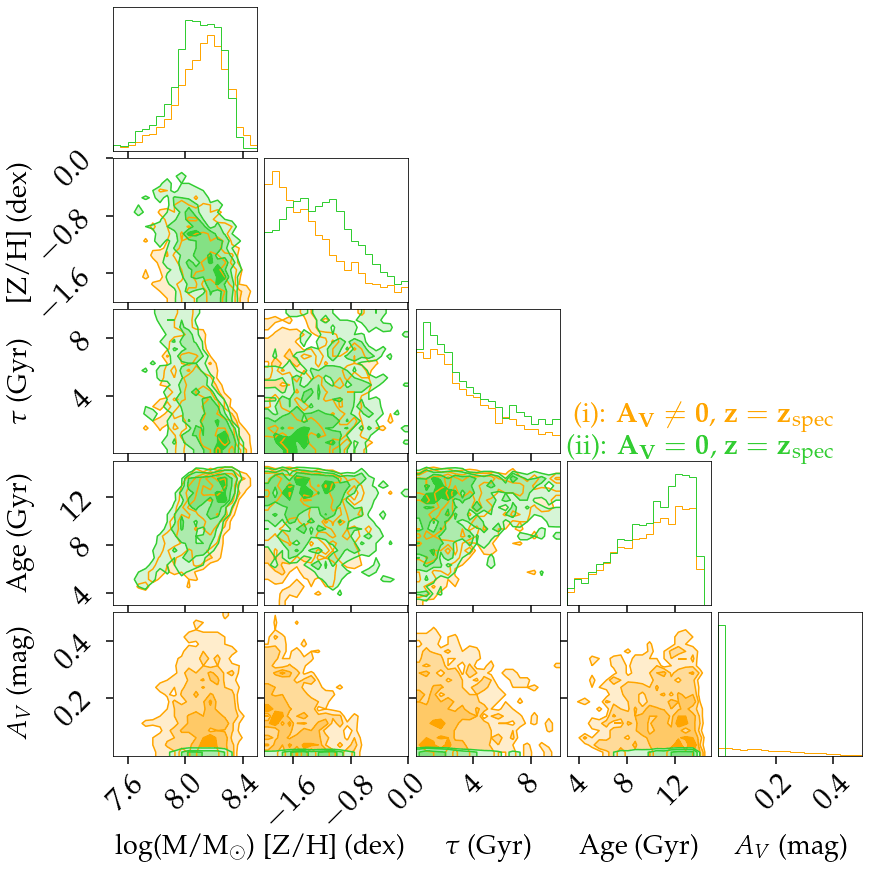}
    \caption{As Fig. \ref{fig:corner_PUDG-R24}, but for DF08.}
    \label{fig:corner_DF8}
\end{figure*}

\begin{figure*}
    \centering
    \includegraphics[width=\columnwidth]{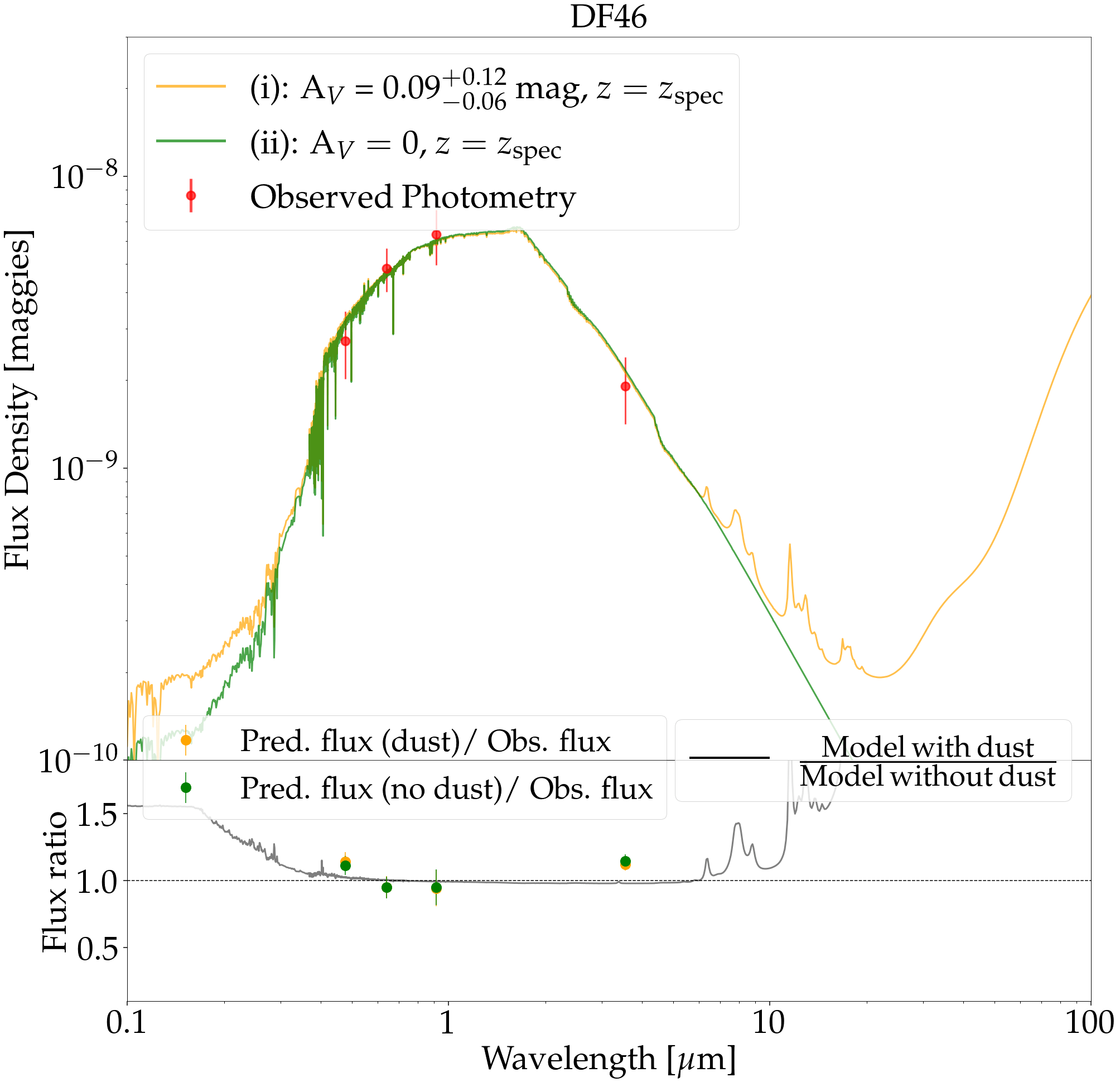}
    \includegraphics[width=\columnwidth]{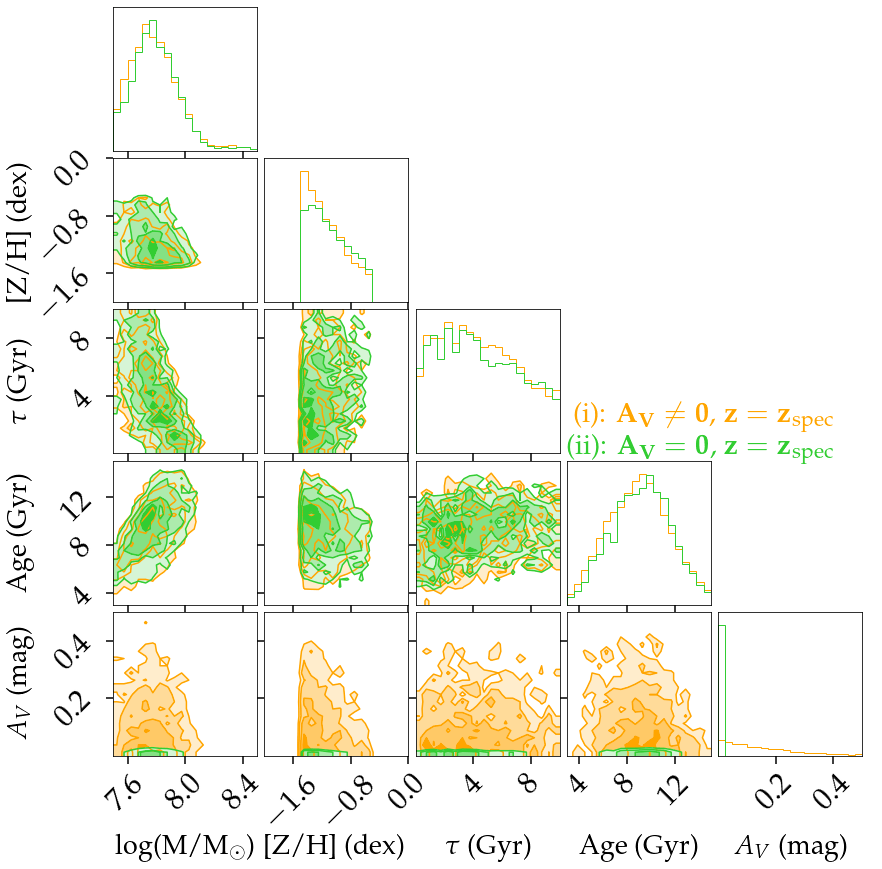}
    \caption{As Fig. \ref{fig:corner_PUDG-R24}, but for DF46.}
    \label{fig:corner_DF46}
\end{figure*}

\begin{figure*}
    \centering
    \includegraphics[width=\columnwidth]{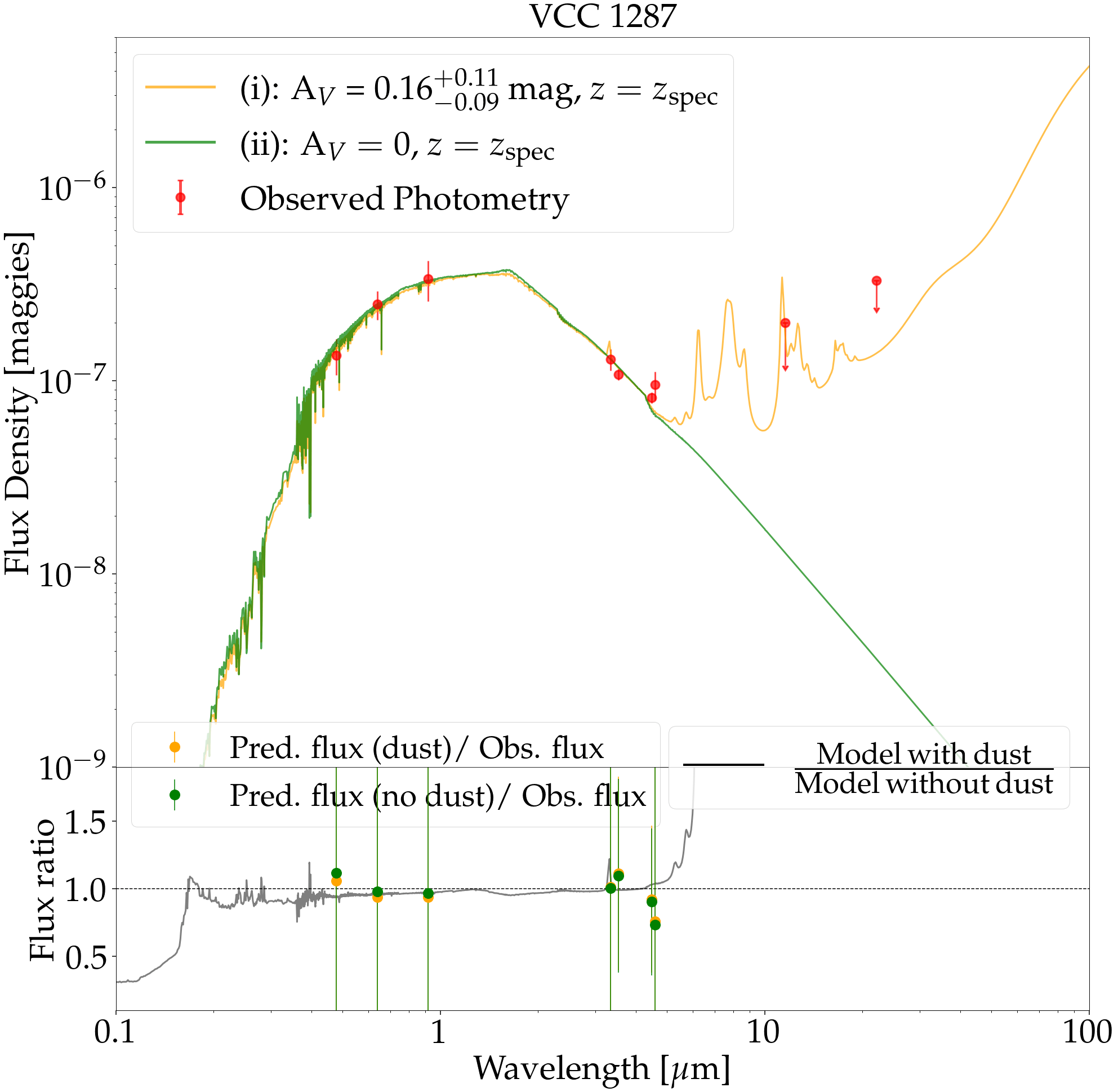}
    \includegraphics[width=\columnwidth]{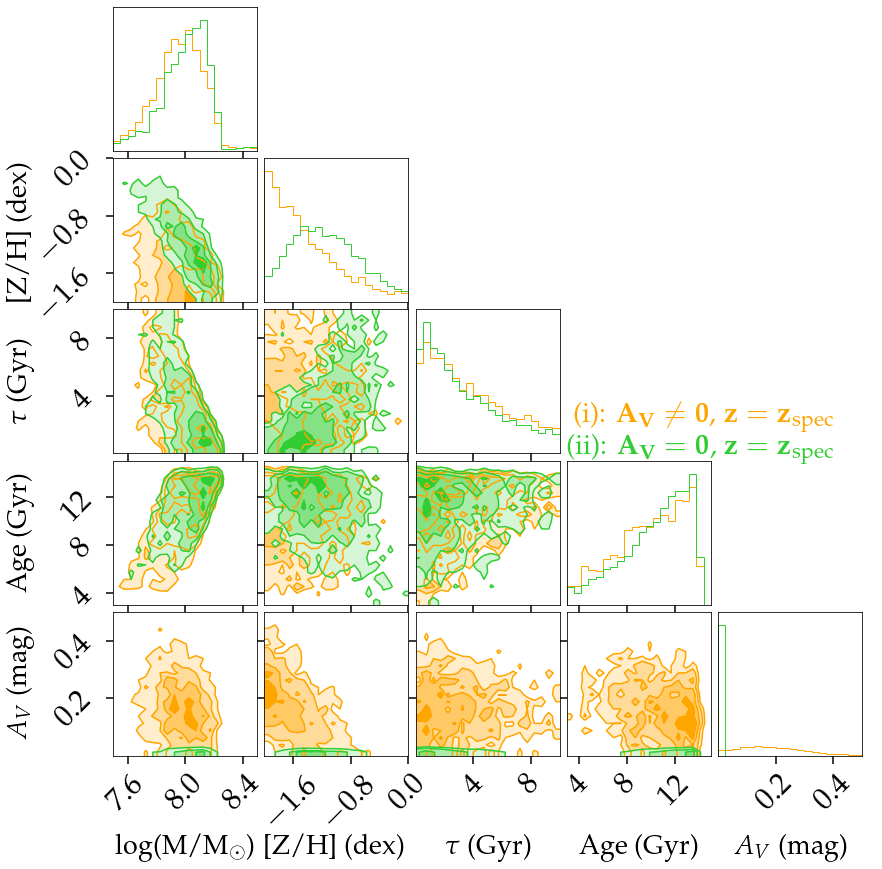}
    \caption{As Fig. \ref{fig:corner_PUDG-R24}, but for VCC 1287.}
    \label{fig:corner_VCC1287}
\end{figure*}

\begin{figure*}
    \centering
    \includegraphics[width=\columnwidth]{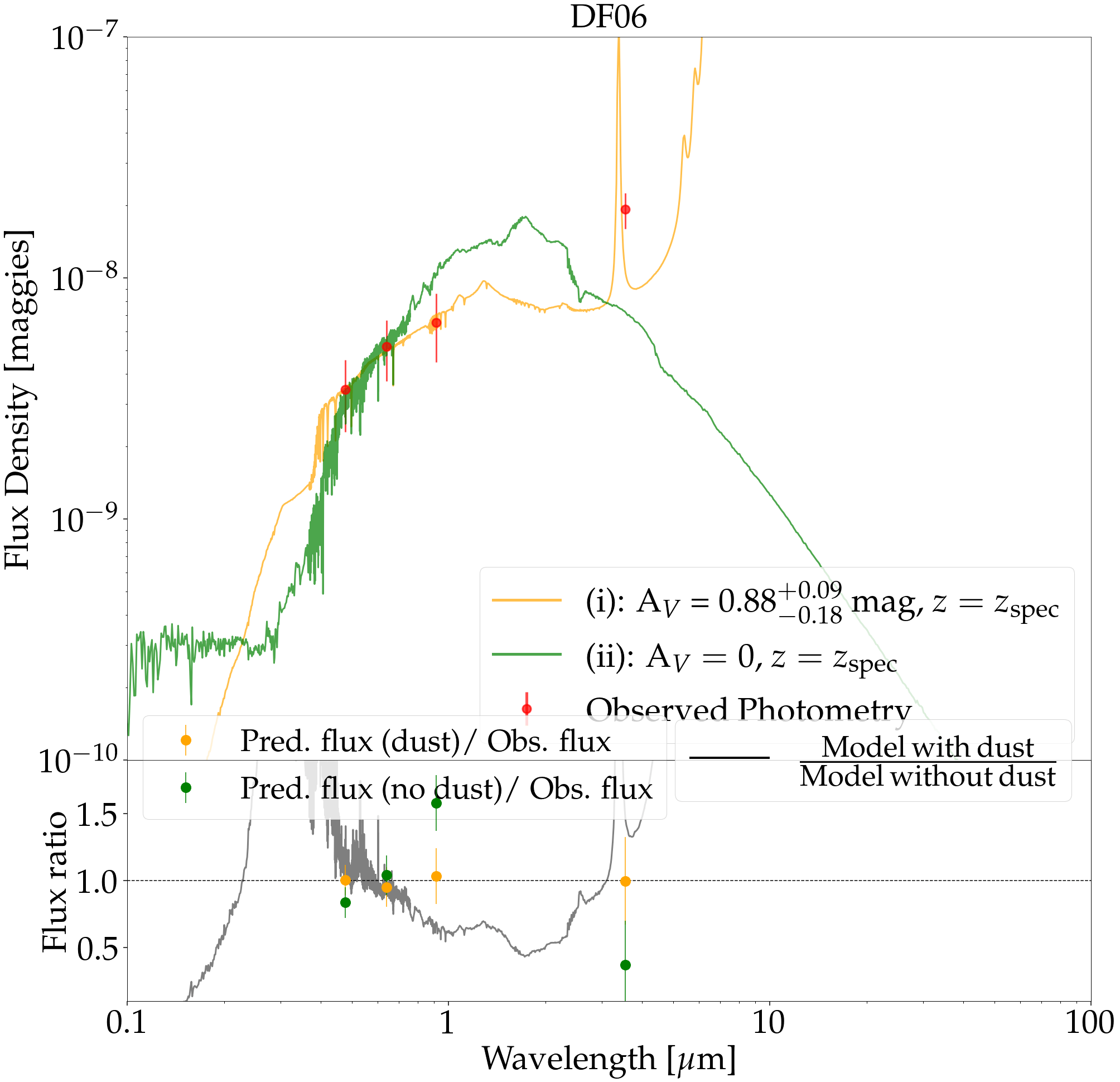}
    \includegraphics[width=\columnwidth]{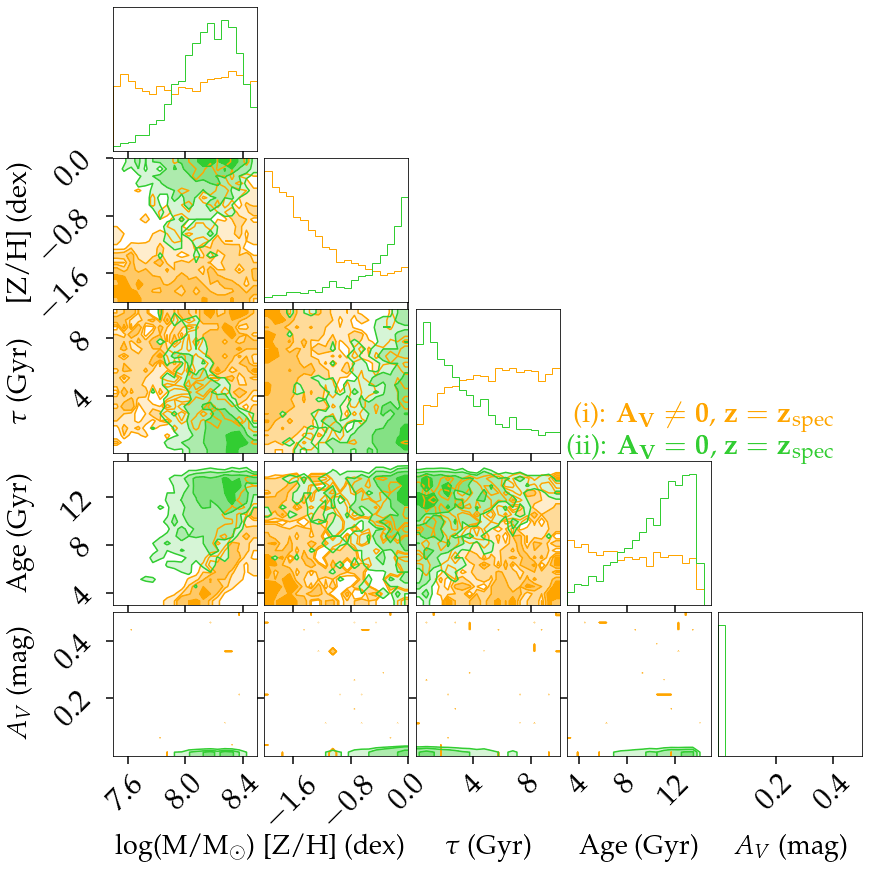}
    \caption{As Fig. \ref{fig:corner_PUDG-R24}, but for DF06.}
    \label{fig:corner_DF6}
\end{figure*}

\begin{figure*}
    \centering
    \includegraphics[width=\columnwidth]{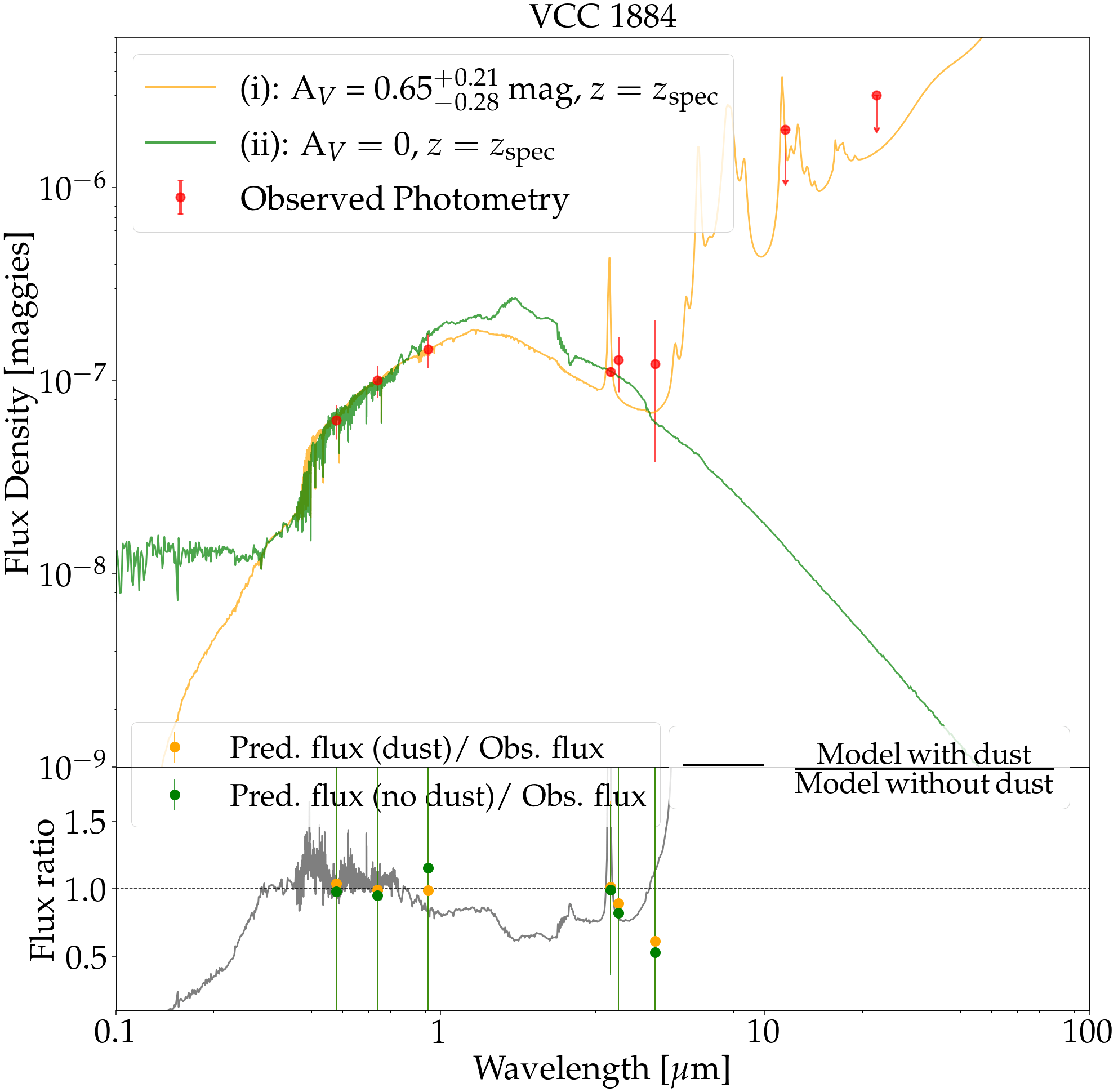}
    \includegraphics[width=\columnwidth]{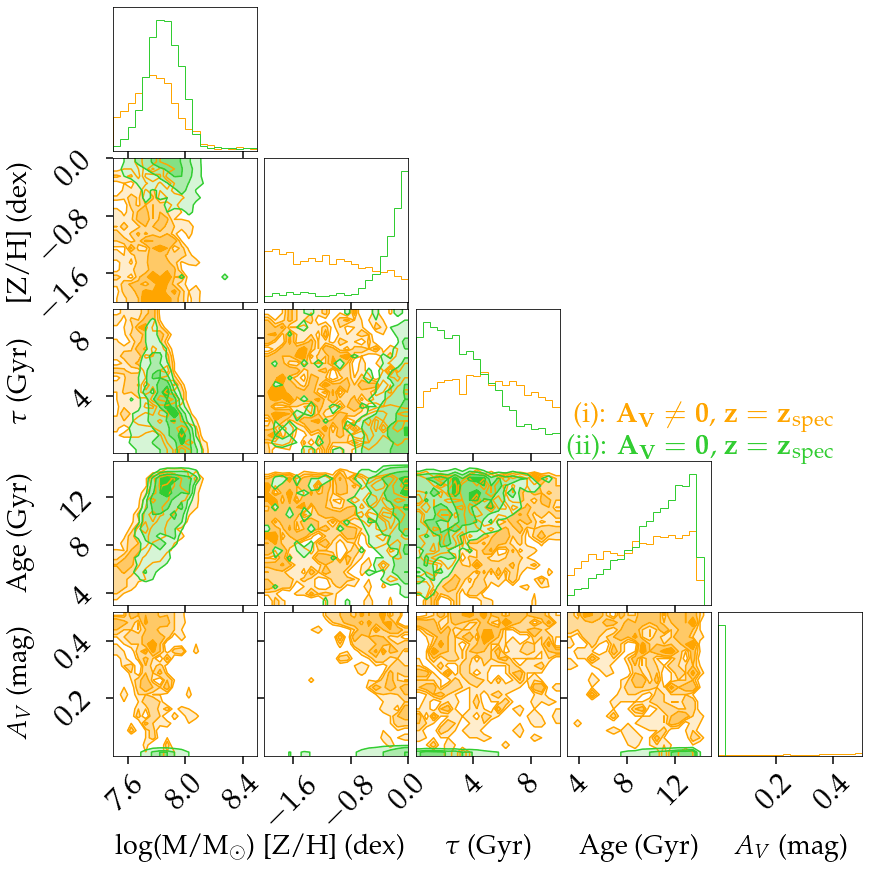}
    \caption{As Fig. \ref{fig:corner_PUDG-R24}, but for VCC 1884.}
    \label{fig:corner_VCC1884}
\end{figure*}

\begin{figure*}
    \centering
    \includegraphics[width=\columnwidth]{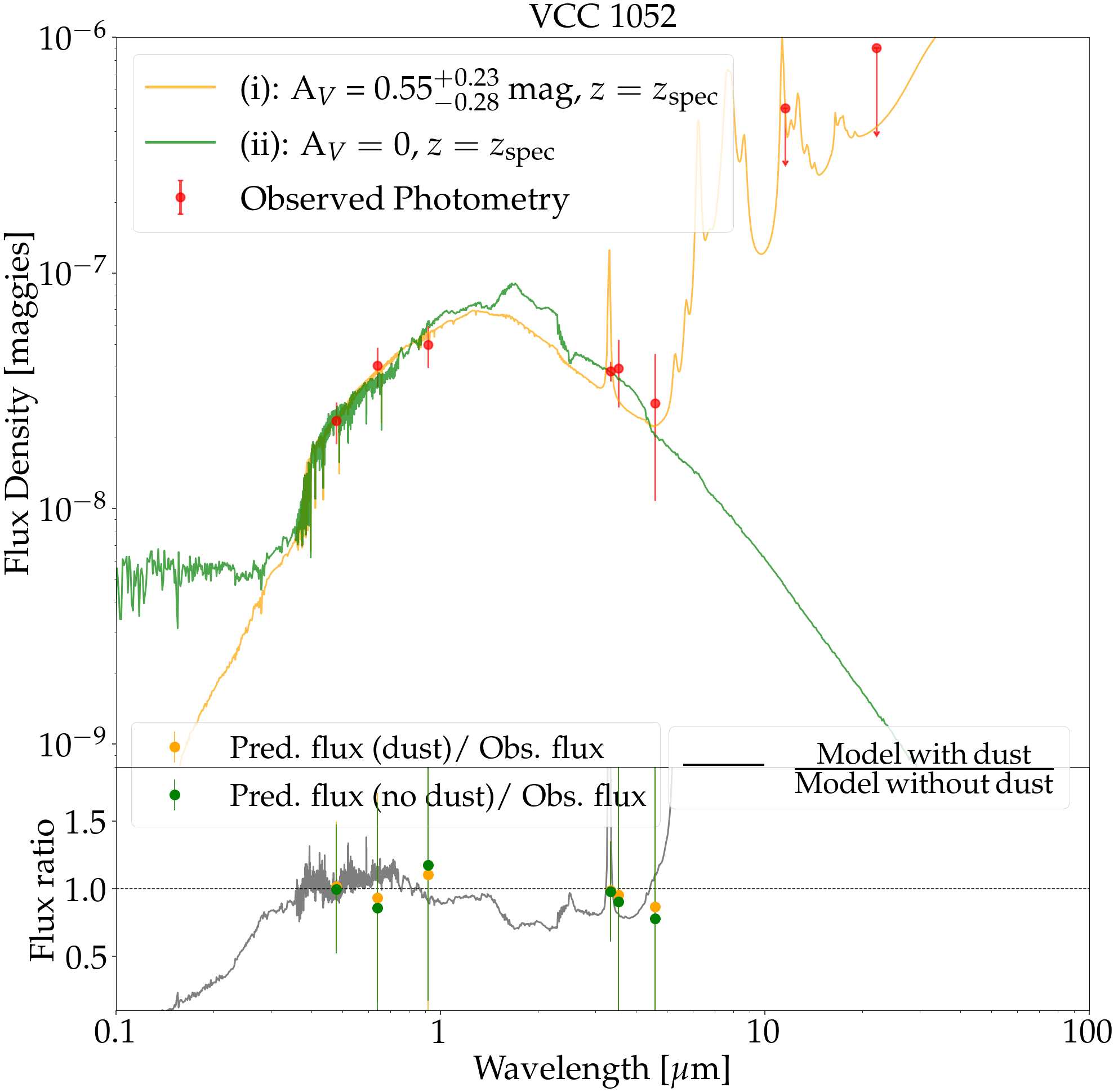}
    \includegraphics[width=\columnwidth]{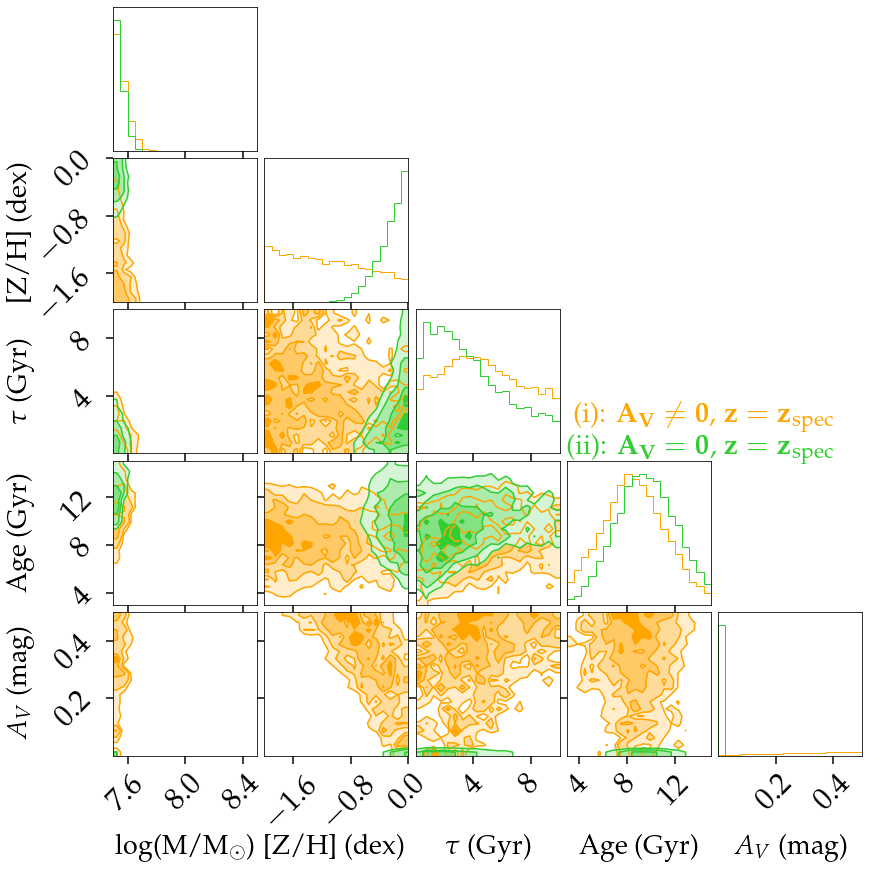}
    \caption{As Fig. \ref{fig:corner_PUDG-R24}, but for VCC 1052.}
    \label{fig:corner_VCC1052}
\end{figure*}

%% file: AppendixB.tex
\section{The effect of excluding the \textit{Spitzer}-IRAC or the \textit{WISE} bands}
\label{sec:appendixB}

\begin{figure*}
    \centering
    \includegraphics[width=\columnwidth]{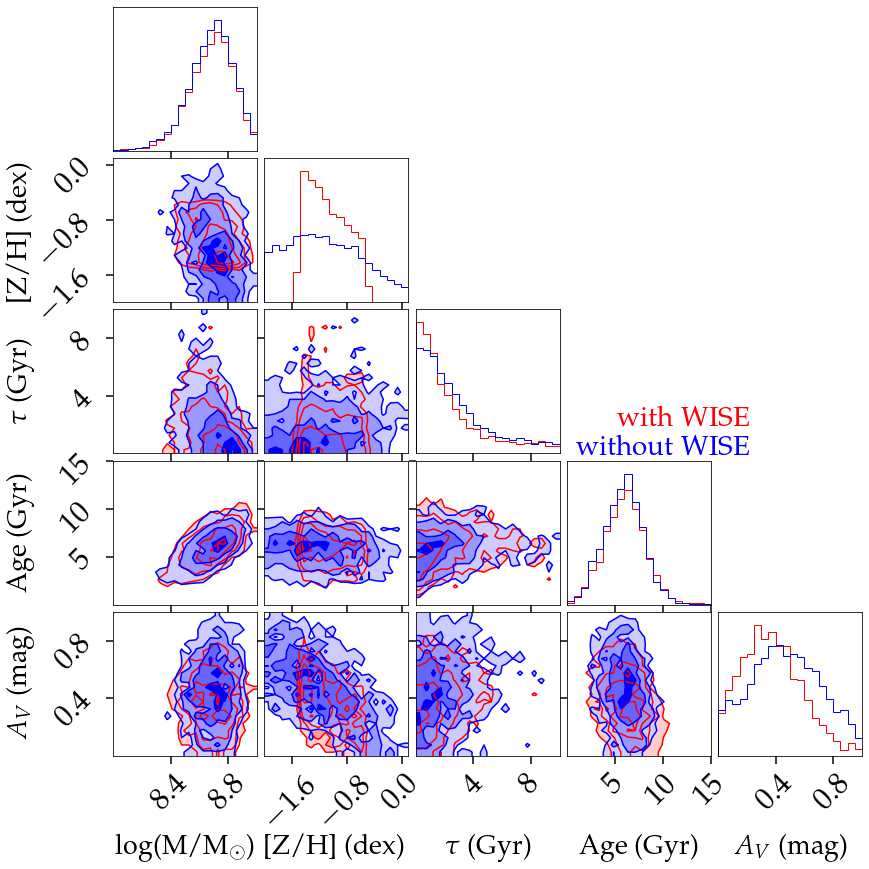}
    \includegraphics[width=\columnwidth]{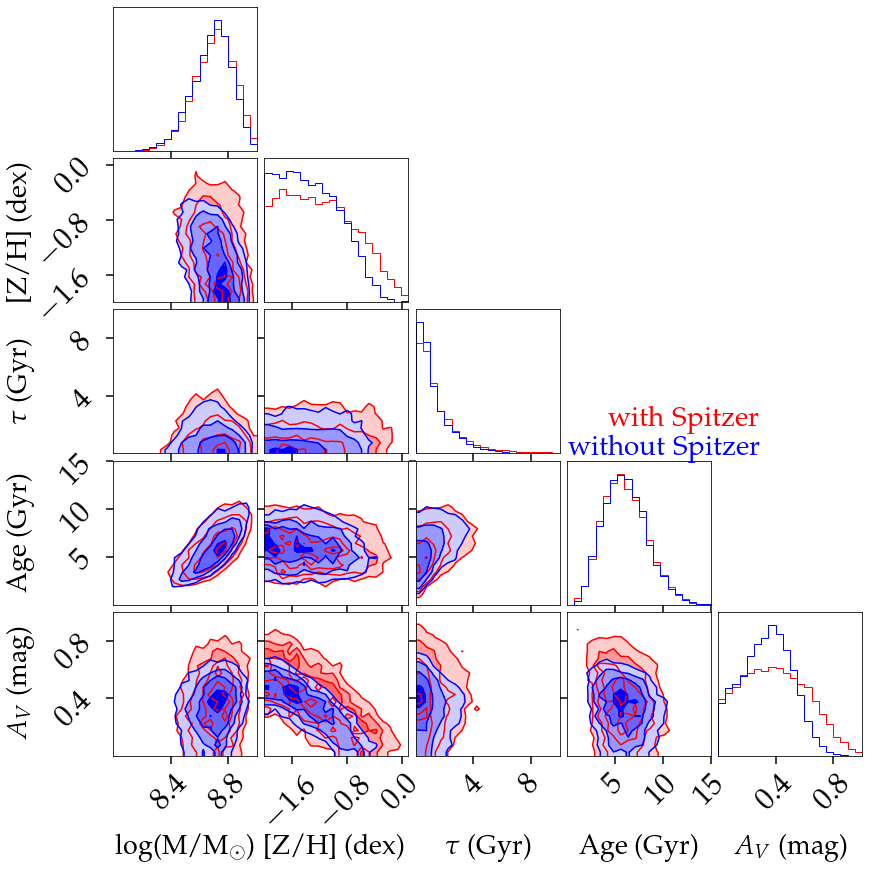}
    \caption{SED fitting results for PUDG-R24, with and without the inclusion of \textit{WISE} and \textit{Spitzer}-IRAC bands.}
    \label{fig:corner_PUDG-R24_nowise}
\end{figure*}

In this appendix, we provide a comparison between the fits performed with all of the available bands for the galaxy PUDG-R24 and for fits performed excluding the \textit{Spitzer}-IRAC bands and the \textit{WISE} bands. 
From Fig. \ref{fig:corner_PUDG-R24_nowise}, we can see that the exclusion of the \textit{Spitzer}-IRAC bands does not significantly affect the recovered stellar population properties, returning similar posterior distributions. On the other hand, we see that the exclusion of the \textit{WISE} bands has a strong effect on the estimate of the dust extinction and metallicity. With the \textit{WISE} bands, the dust posterior peaks at smaller values because the 12 and 22$\mu$m upper limits help to constrain the amount of dust found. Because of this constraint on the dust, we can also see that we have a much more constrained estimate of the metallicity.

%% file: Appendix_Table.tex
\section{SED Fitting results without dust}
\label{sec:Appendix_Table}

In this Appendix, we present in Tables \ref{tab:prospector_results_nodust} and \ref{tab:prospector_results_redshift_nodust} the results from the best-fitting SEDs without dust attenuation in the models. These tables are the complement of Tables \ref{tab:prospector_results} and \ref{tab:prospector_results_redshift}.

\renewcommand{\arraystretch}{1.5}
\begin{table*}
\scalebox{0.8}{
\begin{threeparttable}
\caption{\texttt{PROSPECTOR} SED fitting results without dust for galaxies with spectroscopic redshifts.}
\begin{tabular}{c|c|c|c|c|c|c|c} \hline
\multirow{2}{*}{} & \multirow{2}{*}{Galaxy} & \multirow{2}{*}{Configuration} & \multirow{2}{*}{$\log$(M$_\star$/M$_\odot$)} & [$Z$/H] & $\tau$ & Age & $A_{V}$ \\ 
& & & & [dex] & [Gyr] & [Gyr] & [mag] \\ 
& (1) & (2) & (3) & (4) & (5) & (6) & (7) \\ \hline

\multirow[t]{6}{*}{\textbf{Field}} & 

\textbf{PUDG-R24 \textcolor{blue}{(GC-poor)}} 
 & (ii): A$_V=0$; z = z$_{\rm spec}$ & $8.69^{+0.12}_{-0.14}$ & $-0.54^{+0.35}_{-0.41}$ & $1.20^{+1.97}_{-0.81}$ & $6.95^{+1.92}_{-1.80}$ & --   \\[0.1cm]
 \cline{2-8} 

& \textbf{DGSAT I (No GC info)} 

 & (ii): A$_V=0$; z = z$_{\rm spec}$ & $8.55^{+0.17}_{-0.18}$ & $-0.60^{+0.27}_{-0.31}$ & $2.62^{+4.09}_{-1.93}$ & $7.80^{+2.20}_{-2.28}$ & --   \\[0.1cm]  
\cline{2-8}

& \textbf{M-161-1 (No GC info)} 
& (ii): A$_V=0$; z = z$_{\rm spec}$ & $8.10^{+0.15}_{-0.18}$ & $-1.31^{+0.54}_{-0.44}$ & $3.99^{+3.41}_{-2.65}$ & $10.22^{+2.63}_{-3.81}$ & --    \\[0.1cm] \hline

\multirow[t]{8}{*}{\textbf{Group}}  &  \textbf{NGC 1052-DF4 \textcolor{blue}{(GC-poor)}}

& (ii): A$_V=0$; z = z$_{\rm spec}$ & $8.33^{+0.14}_{-0.14}$ & $-0.31^{+0.24}_{-0.25}$ & $3.79^{+3.31}_{-2.65}$ & $10.55^{+2.54}_{-3.52}$ & --    \\[0.1cm]  

 \cline{2-8}

&  \textbf{NGC 1052-DF2 \textcolor{blue}{(GC-poor)}}  
& (ii): A$_V=0$; z = z$_{\rm spec}$ & $8.45^{+0.15}_{-0.14}$ & $-0.55^{+0.35}_{-0.34}$ & $3.01^{+3.30}_{-2.11}$ & $9.86^{+2.14}_{-2.50}$ & --   \\[0.1cm]  \cline{2-8} 

& \textbf{DF03 \textcolor{blue}{(GC-poor)}}
 & (ii): A$_V=0$; z = z$_{\rm spec}$ & $8.12^{+0.16}_{-0.19}$ & $-1.22^{+0.50}_{-0.47}$ & $2.90^{+3.37}_{-1.97}$ & $10.55^{+2.53}_{-3.45}$ & --    \\[0.1cm]

 \hline

\multirow[t]{44}{*}{\textbf{Cluster}}  &
 
\textbf{DFX1 \textcolor{red}{(GC-rich)}} 
& (ii): A$_V=0$; z = z$_{\rm spec}$ & $8.36^{+0.16}_{-0.17}$ & $-1.24^{+0.47}_{-0.46}$ & $3.23^{+3.61}_{-2.27}$ & $10.60^{+2.45}_{-3.72}$ & --   \\[0.1cm] 
 \cline{2-8} 
 
& \textbf{DF26 \textcolor{red}{(GC-rich)}} 
 & (ii): A$_V=0$; z = z$_{\rm spec}$ & $8.52^{+0.14}_{-0.16}$ & $-0.62^{+0.43}_{-0.50}$ & $2.68^{+3.51}_{-1.89}$ & $11.10^{+2.06}_{-3.44}$ & --   \\[0.1cm] 
 \cline{2-8} 
 
& \textbf{DF02 \textcolor{blue}{(GC-poor)}} 
 & (ii): A$_V=0$; z = z$_{\rm spec}$ & $8.33^{+0.16}_{-0.18}$ & $-0.12^{+0.24}_{-0.49}$ & $2.70^{+2.81}_{-1.86}$ & $11.09^{+2.03}_{-3.25}$ & --    \\[0.1cm]  
 \cline{2-8} 
 
& \textbf{DF07 \textcolor{red}{(GC-rich)}}
& (ii): A$_V=0$; z = z$_{\rm spec}$ & $8.60^{+0.15}_{-0.13}$ & $-0.41^{+0.30}_{-0.31}$ & $3.64^{+3.06}_{-2.39}$ & $10.83^{+2.30}_{-3.19}$ & --    \\[0.1cm]  
 \cline{2-8} 

& \textbf{Y358 \textcolor{red}{(GC-rich)}}
 & (ii): A$_V=0$; z = z$_{\rm spec}$ & $8.46^{+0.13}_{-0.17}$ & $-0.44^{+0.39}_{-0.50}$ & $2.83^{+3.15}_{-1.98}$ & $10.81^{+2.25}_{-3.29}$ & --  \\[0.1cm]  
 \cline{2-8} 

& \textbf{DF44 \textcolor{red}{(GC-rich)}}
 & (ii): A$_V=0$; z = z$_{\rm spec}$ & $8.54^{+0.12}_{-0.11}$ & $-0.27^{+0.24}_{-0.26}$ & $3.35^{+2.68}_{-2.15}$ & $11.05^{+2.10}_{-2.96}$ & --   \\[0.1cm]
 \cline{2-8} 
 
& \textbf{DFX2 (No GC info)} 
 & (ii): A$_V=0$; z = z$_{\rm spec}$ & $8.55^{+0.06}_{-0.08}$ & $-0.54^{+0.03}_{-0.05}$ & $6.97^{+1.95}_{-2.42}$ & $10.42^{+1.72}_{-2.27}$ & --    \\[0.1cm]  
 \cline{2-8} 

& \textbf{PUDG-R16 \textcolor{blue}{(GC-poor)}} 
 & (ii): A$_V=0$; z = z$_{\rm spec}$ & $8.63^{+0.06}_{-0.08}$ & $-0.52^{+0.19}_{-0.22}$ & $0.93^{+0.91}_{-0.59}$ & $12.07^{+1.42}_{-2.44}$ & --   \\[0.1cm] 

\cline{2-8} 

& \textbf{DF40 \textcolor{blue}{(GC-poor)}}
 & (ii): A$_V=0$; z = z$_{\rm spec}$ & $8.22^{+0.16}_{-0.17}$ & $-0.30^{+0.35}_{-0.63}$ & $3.75^{+3.44}_{-2.44}$ & $10.44^{+2.50}_{-3.63}$ & --  \\[0.1cm] 
\cline{2-8} 

  &  \textbf{DF23 \textcolor{red}{(GC-rich)}} 
& (ii): A$_V=0$; z = z$_{\rm spec}$ & $8.06^{+0.16}_{-0.19}$ & $-1.41^{+0.51}_{-0.40}$ & $3.47^{+3.49}_{-2.38}$ & $10.18^{+2.69}_{-3.71}$ & --    \\[0.1cm]  
 \cline{2-8}

& \textbf{Y436 \textcolor{red}{(GC-rich)}}  & (ii): A$_V=0$; z = z$_{\rm spec}$ & $7.98^{+0.14}_{-0.16}$ & $-0.24^{+0.30}_{-0.48}$ & $3.11^{+3.24}_{-2.17}$ & $10.79^{+2.34}_{-3.36}$ & --    \\[0.1cm] 
 \cline{2-8} 
 
& \textbf{Y534 \textcolor{red}{(GC-rich)}} 
 & (ii): A$_V=0$; z = z$_{\rm spec}$ & $8.06^{+0.15}_{-0.17}$ & $-0.58^{+0.44}_{-0.58}$ & $3.12^{+3.45}_{-2.22}$ & $10.75^{+2.36}_{-3.64}$ & --   \\[0.1cm] 

 \cline{2-8}  
 
 & \textbf{DF17 \textcolor{red}{(GC-rich)}}
 & (ii): A$_V=0$; z = z$_{\rm spec}$ & $8.51^{+0.15}_{-0.17}$ & $-1.44^{+0.46}_{-0.38}$ & $3.56^{+3.53}_{-2.48}$ & $10.21^{+2.72}_{-3.45}$ & --   \\[0.1cm]  
 \cline{2-8} 
 
& \textbf{DF25 \textcolor{blue}{(GC-poor)}} 
 & (ii): A$_V=0$; z = z$_{\rm spec}$ & $8.22^{+0.15}_{-0.17}$ & $-0.28^{+0.31}_{-0.45}$ & $3.21^{+3.85}_{-2.31}$ & $10.79^{+2.29}_{-3.64}$ & --    \\[0.1cm]  
 \cline{2-8} 
 
& \textbf{DF08 \textcolor{red}{(GC-rich)}} 
 & (ii): A$_V=0$; z = z$_{\rm spec}$ & $8.18^{+0.15}_{-0.16}$ & $-1.22^{+0.49}_{-0.48}$ & $2.97^{+3.82}_{-2.09}$ & $10.70^{+2.33}_{-3.46}$ & --    \\[0.1cm]  
 \cline{2-8}

& \textbf{DF46 \textcolor{blue}{(GC-poor)}} 
 & (ii): A$_V=0$; z = z$_{\rm spec}$ & $8.05^{+0.17}_{-0.15}$ & $-1.14^{+0.35}_{-0.24}$ & $4.30^{+3.60}_{-2.77}$ & $9.31^{+2.38}_{-2.70}$ & --    \\[0.1cm]  
 \cline{2-8}

& \textbf{VCC1287 \textcolor{red}{(GC-rich)}}
& (ii): A$_V=0$; z = z$_{\rm spec}$ & $8.21^{+0.12}_{-0.16}$ & $-1.18^{+0.46}_{-0.43}$ & $2.51^{+3.41}_{-1.75}$ & $11.16^{+2.04}_{-3.29}$ & -- \\[0.1cm]     
 \cline{2-8}  

& \textbf{DF06 \textcolor{red}{(GC-rich)}} 
 & (ii): A$_V=0$; z = z$_{\rm spec}$ & $8.16^{+0.18}_{-0.21}$ & $0.08^{+0.14}_{-0.40}$ & $2.37^{+3.36}_{-1.68}$ & $11.07^{+2.04}_{-3.47}$ & --    \\[0.1cm]  
 \cline{2-8} 

& \textbf{VCC1884 \textcolor{blue}{(GC-poor)}}  
& (ii): A$_V=0$; z = z$_{\rm spec}$ & $7.95^{+0.12}_{-0.11}$ & $0.02^{+0.13}_{-0.23}$ & $2.76^{+2.64}_{-1.88}$ & $11.05^{+2.12}_{-3.21}$ & --   \\[0.1cm]  

 \cline{2-8} 
 
& \textbf{VCC1052 \textcolor{blue}{(GC-poor)}} 
 & (ii): A$_V=0$; z = z$_{\rm spec}$ & $8.03^{+0.12}_{-0.13}$ & $-0.07^{+0.19}_{-0.30}$ & $3.36^{+2.88}_{-2.20}$ & $9.92^{+2.46}_{-2.47}$ & --   \\[0.1cm]  

 \hline 
 
\end{tabular}
\begin{tablenotes}
      \small
      \item \textbf{Note.} UDGs are separated by the environment that they reside in (see Table \ref{tab:properties}). Columns are: (1) Galaxy ID with GC-richness in parentheses (i.e., rich $geq$ 20 GCs, poor < 20 GCs); (2) \texttt{PROSPECTOR} configuration; (3) Total stellar mass; (4) Metallicity; (5) Star formation time scale; (6) Mass-weighted age; (7) Dust reddening; `--' stands for fixed parameters.
\end{tablenotes}
\label{tab:prospector_results_nodust}
\end{threeparttable}}
\end{table*}

\begin{table*}
\scalebox{0.8}{
\begin{threeparttable}
\caption{\texttt{PROSPECTOR} SED fitting results without dust for galaxies without spectroscopic redshift.}
\begin{tabular}{c|c|c|c|c|c|c|c|c} \hline
\multirow{2}{*}{} & \multirow{2}{*}{Galaxy} & \multirow{2}{*}{Configuration} & \multirow{2}{*}{$\log$(M$_\star$/M$_\odot$)} & [$Z$/H] & $\tau$ & Age & $A_{V}$ & \multirow{2}{*}{z} \\ 
& & & & [dex] & [Gyr] & [Gyr] & [mag] & \\ 
& (1) & (2) & (3) & (4) & (5) & (6) & (7) & (8) \\ \hline

\multirow[t]{10}{*}{\textbf{Field}} & \textbf{LSBG-490 (No GC info)} 
 & (iv): A$_V=0$; $z \neq z_{\rm spec}$ & $8.28^{+0.39}_{-0.66}$ & $-0.15^{+0.24}_{-0.31}$ & $1.38^{+3.23}_{-1.06}$ & $7.77^{+4.57}_{-4.82}$ & -- & $0.023^{+0.012}_{-0.011}$   \\  \cline{2-9} 
 
& \textbf{LSBG-378 (No GC info)} 
 & (iv): A$_V=0$; $z \neq z_{\rm spec}$ & $8.64^{+0.42}_{-0.67}$ & $-0.47^{+0.35}_{-0.41}$ & $2.70^{+4.01}_{-2.13}$ & $8.49^{+3.90}_{-4.42}$ & -- & $0.022^{+0.013}_{-0.012}$   \\  
  
 \cline{2-9}  
 
& \textbf{LSBG-044 (No GC info)}  
 & (iv): A$_V=0$; $z \neq z_{\rm spec}$ & $8.29^{+0.39}_{-0.55}$ & $-0.28^{+0.27}_{-0.31}$ & $5.22^{+3.14}_{-3.51}$ & $9.48^{+3.14}_{-3.57}$ & -- & $0.025^{+0.016}_{-0.012}$   \\  
  \hline 

\end{tabular}
\begin{tablenotes}
      \small
      \item \textbf{Note.} UDGs are separated by the environment that they reside in (see Table \ref{tab:properties}). Columns are: (1) Galaxy ID with GC-richness in parentheses; (2) \texttt{PROSPECTOR} configuration; (3) Total stellar mass; (4) Metallicity; (5) Star formation time scale; (6) Mass-weighted age; (7) Dust reddening; (8) Redshift. `--' stands for fixed parameters.
\end{tablenotes}
\label{tab:prospector_results_redshift_nodust}
\end{threeparttable}}
\end{table*}